\catcode`\@=11
\font\tensmc=cmcsc10      
\def\smc{\tensmc}
\def\pagewidth#1{\hsize= #1 }
\def\pageheight#1{\vsize= #1 }
\def\hcorrection#1{\advance\hoffset by #1 }
\def\vcorrection#1{\advance\voffset by #1 }
\def\wlog#1{}
\newif\iftitle@
\outer\def\title{\title@true\vglue 24\p@ plus 12\p@ minus 12\p@
   \bgroup\let\\=\cr\tabskip\centering
   \halign to \hsize\bgroup\tenbf\hfill\ignorespaces##\unskip\hfill\cr}
\def\endtitle{\cr\egroup\egroup\vglue 18\p@ plus 12\p@ minus 6\p@}
\outer\def\author{\iftitle@\vglue -18\p@ plus -12\p@ minus -6\p@\fi\vglue
    12\p@ plus 6\p@ minus 3\p@\bgroup\let\\=\cr\tabskip\centering
    \halign to \hsize\bgroup\smc\hfill\ignorespaces##\unskip\hfill\cr}
\def\endauthor{\cr\egroup\egroup\vglue 18\p@ plus 12\p@ minus 6\p@}
\outer\def\heading{\bigbreak\bgroup\let\\=\cr\tabskip\centering
    \halign to \hsize\bgroup\smc\hfill\ignorespaces##\unskip\hfill\cr}
\def\endheading{\cr\egroup\egroup\nobreak\medskip}

\outer\def\endproclaim{\par\ifdim\lastskip<\medskipamount\removelastskip
  \penalty 55 \fi\medskip\rm}
\outer\def\demo#1{\par\ifdim\lastskip<\smallskipamount\removelastskip
    \smallskip\fi\noindent{\smc\ignorespaces#1\unskip:\enspace}\rm
      \ignorespaces}

\newcount\footmarkcount@
\footmarkcount@=1
\def\makefootnote@#1#2{\insert\footins{\interlinepenalty=100
  \splittopskip=\ht\strutbox \splitmaxdepth=\dp\strutbox 
  \floatingpenalty=\@MM
  \leftskip=\z@\rightskip=\z@\spaceskip=\z@\xspaceskip=\z@
  \noindent{#1}\footstrut\rm\ignorespaces #2\strut}}
\def\footnote{\let\@sf=\empty\ifhmode\edef\@sf{\spacefactor
   =\the\spacefactor}\/\fi\futurelet\next\footnote@}
\def\footnote@{\ifx"\next\let\next\footnote@@\else
    \let\next\footnote@@@\fi\next}
\def\footnote@@"#1"#2{#1\@sf\relax\makefootnote@{#1}{#2}}
\def\footnote@@@#1{$^{\number\footmarkcount@}$\makefootnote@
   {$^{\number\footmarkcount@}$}{#1}\global\advance\footmarkcount@ by 1 }

\hyphenation{man-u-script man-u-scripts ap-pen-dix ap-pen-di-ces}
\hyphenation{data-base data-bases}
\ifx\amstexloaded@\relax\catcode`\@=13 
  \endinput\else\let\amstexloaded@=\relax\fi
\newlinechar=`\^^J
\def\eat@#1{}
\def\Space@.{\futurelet\Space@\relax}
\Space@. %
\newhelp\athelp@
{Only certain combinations beginning with @ make sense to me.^^J
Perhaps you wanted \string\@\space for a printed @?^^J
I've ignored the character or group after @.}
\def\futureletnextat@{\futurelet\next\at@}
{\catcode`\@=\active
\lccode`\Z=`\@ \lowercase
{\gdef@{\expandafter\csname futureletnextatZ\endcsname}
\expandafter\gdef\csname atZ\endcsname
   {\ifcat\noexpand\next a\def\next{\csname atZZ\endcsname}\else
   \ifcat\noexpand\next0\def\next{\csname atZZ\endcsname}\else
    \def\next{\csname atZZZ\endcsname}\fi\fi\next}
\expandafter\gdef\csname atZZ\endcsname#1{\expandafter
   \ifx\csname #1Zat\endcsname\relax\def\next
     {\errhelp\expandafter=\csname athelpZ\endcsname
      \errmessage{Invalid use of \string@}}\else
       \def\next{\csname #1Zat\endcsname}\fi\next}
\expandafter\gdef\csname atZZZ\endcsname#1{\errhelp
    \expandafter=\csname athelpZ\endcsname
      \errmessage{Invalid use of \string@}}}}
\def\atdef@#1{\expandafter\def\csname #1@at\endcsname}
\newhelp\defahelp@{If you typed \string\define\space cs instead of
\string\define\string\cs\space^^J
I've substituted an inaccessible control sequence so that your^^J
definition will be completed without mixing me up too badly.^^J
If you typed \string\define{\string\cs} the inaccessible control sequence^^J
was defined to be \string\cs, and the rest of your^^J
definition appears as input.}
\newhelp\defbhelp@{I've ignored your definition, because it might^^J
conflict with other uses that are important to me.}
\def\define{\futurelet\next\define@}
\def\define@{\ifcat\noexpand\next\relax
  \def\next{\define@@}%
  \else\errhelp=\defahelp@
  \errmessage{\string\define\space must be followed by a control 
     sequence}\def\next{\def\garbage@}\fi\next}
\def\undefined@{}
\def\preloaded@{}    
\def\define@@#1{\ifx#1\relax\errhelp=\defbhelp@
   \errmessage{\string#1\space is already defined}\def\next{\def\garbage@}%
   \else\expandafter\ifx\csname\expandafter\eat@\string
         #1@\endcsname\undefined@\errhelp=\defbhelp@
   \errmessage{\string#1\space can't be defined}\def\next{\def\garbage@}%
   \else\expandafter\ifx\csname\expandafter\eat@\string#1\endcsname\relax
     \def\next{\def#1}\else\errhelp=\defbhelp@
     \errmessage{\string#1\space is already defined}\def\next{\def\garbage@}%
      \fi\fi\fi\next}
\def\famzero{\fam\z@}

\def\lim{\mathop{\famzero lim}}

\def\ln{\mathop{\famzero ln}\nolimits}

\def\textfont@#1#2{\def#1{\relax\ifmmode
    \errmessage{Use \string#1\space only in text}\else#2\fi}}
\textfont@\rm\tenrm
\textfont@\it\tenit
\textfont@\sl\tensl
\textfont@\bf\tenbf
\textfont@\smc\tensmc
\let\ic@=\/
\def\/{\unskip\ic@}
\def\textfonti{\the\textfont1 }
\def\t#1#2{{\edef\next{\the\font}\textfonti\accent"7F \next#1#2}}
\let\B=\=
\let\D=\.
\def~{\unskip\nobreak\ \ignorespaces}
{\catcode`\@=\active
\gdef\@{\char'100 }}
\atdef@-{\leavevmode\futurelet\next\athyph@}
\def\athyph@{\ifx\next-\let\next=\athyph@@
  \else\let\next=\athyph@@@\fi\next}
\def\athyph@@@{\hbox{-}}
\def\athyph@@#1{\futurelet\next\athyph@@@@}
\def\athyph@@@@{\if\next-\def\next##1{\hbox{---}}\else
    \def\next{\hbox{--}}\fi\next}
\def\.{.\spacefactor=\@m}
\atdef@.{\null.}
\atdef@,{\null,}
\atdef@;{\null;}
\atdef@:{\null:}
\atdef@?{\null?}
\atdef@!{\null!}   
\def\srdr@{\thinspace}                     
\def\drsr@{\kern.02778em}
\def\sldl@{\kern.02778em}
\def\dlsl@{\thinspace}
\atdef@"{\unskip\futurelet\next\atqq@}
\def\atqq@{\ifx\next\Space@\def\next. {\atqq@@}\else
         \def\next.{\atqq@@}\fi\next.}
\def\atqq@@{\futurelet\next\atqq@@@}
\def\atqq@@@{\ifx\next`\def\next`{\atqql@}\else\def\next'{\atqqr@}\fi\next}
\def\atqql@{\futurelet\next\atqql@@}
\def\atqql@@{\ifx\next`\def\next`{\sldl@``}\else\def\next{\dlsl@`}\fi\next}
\def\atqqr@{\futurelet\next\atqqr@@}
\def\atqqr@@{\ifx\next'\def\next'{\srdr@''}\else\def\next{\drsr@'}\fi\next}

\def\textfontii{\the\textfont2 }
\def\{{\relax\ifmmode\lbrace\else
    {\textfontii f}\spacefactor=\@m\fi}
\def\}{\relax\ifmmode\rbrace\else
    \let\@sf=\empty\ifhmode\edef\@sf{\spacefactor=\the\spacefactor}\fi
      {\textfontii g}\@sf\relax\fi}   
\def\nonhmodeerr@#1{\errmessage
     {\string#1\space allowed only within text}}
\def\linebreak{\relax\ifhmode\unskip\break\else
    \nonhmodeerr@\linebreak\fi}
\def\allowlinebreak{\relax
   \ifhmode\allowbreak\else\nonhmodeerr@\allowlinebreak\fi}
\newskip\saveskip@
\def\nolinebreak{\relax\ifhmode\saveskip@=\lastskip\unskip
  \nobreak\ifdim\saveskip@>\z@\hskip\saveskip@\fi
   \else\nonhmodeerr@\nolinebreak\fi}
\def\newline{\relax\ifhmode\null\hfil\break
    \else\nonhmodeerr@\newline\fi}
\def\nonmathaerr@#1{\errmessage
     {\string#1\space is not allowed in display math mode}}
\def\nonmathberr@#1{\errmessage{\string#1\space is allowed only in math mode}}
\def\mathbreak{\relax\ifmmode\ifinner\break\else
   \nonmathaerr@\mathbreak\fi\else\nonmathberr@\mathbreak\fi}
\def\nomathbreak{\relax\ifmmode\ifinner\nobreak\else
    \nonmathaerr@\nomathbreak\fi\else\nonmathberr@\nomathbreak\fi}
\def\allowmathbreak{\relax\ifmmode\ifinner\allowbreak\else
     \nonmathaerr@\allowmathbreak\fi\else\nonmathberr@\allowmathbreak\fi}
\def\pagebreak{\relax\ifmmode
   \ifinner\errmessage{\string\pagebreak\space
     not allowed in non-display math mode}\else\postdisplaypenalty-\@M\fi
   \else\ifvmode\penalty-\@M\else\edef\spacefactor@
       {\spacefactor=\the\spacefactor}\vadjust{\penalty-\@M}\spacefactor@
        \relax\fi\fi}
\def\nopagebreak{\relax\ifmmode
     \ifinner\errmessage{\string\nopagebreak\space
    not allowed in non-display math mode}\else\postdisplaypenalty\@M\fi
    \else\ifvmode\nobreak\else\edef\spacefactor@
        {\spacefactor=\the\spacefactor}\vadjust{\penalty\@M}\spacefactor@
         \relax\fi\fi}
\def\newpage{\relax\ifvmode\vfill\penalty-\@M\else\nonvmodeerr@\newpage\fi}
\def\nonvmodeerr@#1{\errmessage
    {\string#1\space is allowed only between paragraphs}}
\def\smallpagebreak{\relax\ifvmode\smallbreak
      \else\nonvmodeerr@\smallpagebreak\fi}
\def\medpagebreak{\relax\ifvmode\medbreak
       \else\nonvmodeerr@\medpagebreak\fi}
\def\bigpagebreak{\relax\ifvmode\bigbreak
      \else\nonvmodeerr@\bigpagebreak\fi}
\newdimen\captionwidth@
\captionwidth@=\hsize
\advance\captionwidth@ by -1.5in
\def\caption#1{}
\def\topspace#1{\gdef\thespace@{#1}\ifvmode\def\next
    {\futurelet\next\topspace@}\else\def\next{\nonvmodeerr@\topspace}\fi\next}
\def\topspace@{\ifx\next\Space@\def\next. {\futurelet\next\topspace@@}\else
     \def\next.{\futurelet\next\topspace@@}\fi\next.}
\def\topspace@@{\ifx\next\caption\let\next\topspace@@@\else
    \let\next\topspace@@@@\fi\next}
 \def\topspace@@@@{\topinsert\vbox to 
       \thespace@{}\endinsert}
\def\topspace@@@\caption#1{\topinsert\vbox to
    \thespace@{}\nobreak
      \smallskip
    \setbox\z@=\hbox{\noindent\ignorespaces#1\unskip}%
   \ifdim\wd\z@>\captionwidth@
   \centerline{\vbox{\hsize=\captionwidth@\noindent\ignorespaces#1\unskip}}%
   \else\centerline{\box\z@}\fi\endinsert}
\def\midspace#1{\gdef\thespace@{#1}\ifvmode\def\next
    {\futurelet\next\midspace@}\else\def\next{\nonvmodeerr@\midspace}\fi\next}
\def\midspace@{\ifx\next\Space@\def\next. {\futurelet\next\midspace@@}\else
     \def\next.{\futurelet\next\midspace@@}\fi\next.}
\def\midspace@@{\ifx\next\caption\let\next\midspace@@@\else
    \let\next\midspace@@@@\fi\next}
 \def\midspace@@@@{\midinsert\vbox to 
       \thespace@{}\endinsert}
\def\midspace@@@\caption#1{\midinsert\vbox to
    \thespace@{}\nobreak
      \smallskip
      \setbox\z@=\hbox{\noindent\ignorespaces#1\unskip}%
      \ifdim\wd\z@>\captionwidth@
    \centerline{\vbox{\hsize=\captionwidth@\noindent\ignorespaces#1\unskip}}%
    \else\centerline{\box\z@}\fi\endinsert}
\mathchardef\prime@="0230
\def\prime{{{}\prime@{}}}
\def\prim@s{\prime@\futurelet\next\pr@m@s}

\def\,{\relax\ifmmode\mskip\thinmuskip\else\thinspace\fi}
\def\!{\relax\ifmmode\mskip-\thinmuskip\else\negthinspace\fi}
\def\frac#1#2{{#1\over#2}}

\def\:{\nobreak\hskip.1111em{:}\hskip.3333em plus .0555em\relax}
\def\intic@{\mathchoice{\hskip5\p@}{\hskip4\p@}{\hskip4\p@}{\hskip4\p@}}
\def\negintic@
 {\mathchoice{\hskip-5\p@}{\hskip-4\p@}{\hskip-4\p@}{\hskip-4\p@}}
\def\intkern@{\mathchoice{\!\!\!}{\!\!}{\!\!}{\!\!}}
\def\intdots@{\mathchoice{\cdots}{{\cdotp}\mkern1.5mu
    {\cdotp}\mkern1.5mu{\cdotp}}{{\cdotp}\mkern1mu{\cdotp}\mkern1mu
      {\cdotp}}{{\cdotp}\mkern1mu{\cdotp}\mkern1mu{\cdotp}}}
\newcount\intno@             
\def\iint{\intno@=\tw@\futurelet\next\ints@} 
\def\iiint{\intno@=\thr@@\futurelet\next\ints@}
\def\iiiint{\intno@=4 \futurelet\next\ints@}
\def\idotsint{\intno@=\z@\futurelet\next\ints@}
\def\ints@{\findlimits@\ints@@}
\newif\iflimtoken@
\newif\iflimits@
\def\findlimits@{\limtoken@false\limits@false\ifx\next\limits
 \limtoken@true\limits@true\else\ifx\next\nolimits\limtoken@true\limits@false
    \fi\fi}
\def\multintlimits@{\intop\ifnum\intno@=\z@\intdots@
  \else\intkern@\fi
    \ifnum\intno@>\tw@\intop\intkern@\fi
     \ifnum\intno@>\thr@@\intop\intkern@\fi\intop}
\def\multint@{\int\ifnum\intno@=\z@\intdots@\else\intkern@\fi
   \ifnum\intno@>\tw@\int\intkern@\fi
    \ifnum\intno@>\thr@@\int\intkern@\fi\int}
\def\ints@@{\iflimtoken@\def\ints@@@{\iflimits@
   \negintic@\mathop{\intic@\multintlimits@}\limits\else
    \multint@\nolimits\fi\eat@}\else
     \def\ints@@@{\multint@\nolimits}\fi\ints@@@}
\def\Sb{_\bgroup\vspace@
        \baselineskip=\fontdimen10 \scriptfont\tw@
        \advance\baselineskip by \fontdimen12 \scriptfont\tw@
        \lineskip=\thr@@\fontdimen8 \scriptfont\thr@@
        \lineskiplimit=\thr@@\fontdimen8 \scriptfont\thr@@
        \Let@\vbox\bgroup\halign\bgroup \hfil$\scriptstyle
            {##}$\hfil\cr}
\def\endSb{\crcr\egroup\egroup\egroup}
\def\Sp{^\bgroup\vspace@
        \baselineskip=\fontdimen10 \scriptfont\tw@
        \advance\baselineskip by \fontdimen12 \scriptfont\tw@
        \lineskip=\thr@@\fontdimen8 \scriptfont\thr@@
        \lineskiplimit=\thr@@\fontdimen8 \scriptfont\thr@@
        \Let@\vbox\bgroup\halign\bgroup \hfil$\scriptstyle
            {##}$\hfil\cr}
\def\endSp{\crcr\egroup\egroup\egroup}
\def\Let@{\relax\iffalse{\fi\let\\=\cr\iffalse}\fi}
\def\vspace@{\def\vspace##1{\noalign{\vskip##1 }}}
\def\aligned{\,\vcenter\bgroup\vspace@\Let@\openup\jot\m@th\ialign
  \bgroup \strut\hfil$\displaystyle{##}$&$\displaystyle{{}##}$\hfil\crcr}
\def\endaligned{\crcr\egroup\egroup}
\def\matrix{\,\vcenter\bgroup\Let@\vspace@
    \normalbaselines
  \m@th\ialign\bgroup\hfil$##$\hfil&&\quad\hfil$##$\hfil\crcr
    \mathstrut\crcr\noalign{\kern-\baselineskip}}
\def\endmatrix{\crcr\mathstrut\crcr\noalign{\kern-\baselineskip}\egroup
                \egroup\,}
\newtoks\hashtoks@
\hashtoks@={#}
\def\format{\crcr\egroup\iffalse{\fi\ifnum`}=0 \fi\format@}
\def\format@#1\\{\def\preamble@{#1}%
  \def\c{\hfil$\the\hashtoks@$\hfil}%
  \def\r{\hfil$\the\hashtoks@$}%
  \def\l{$\the\hashtoks@$\hfil}%
  \setbox\z@=\hbox{\xdef\Preamble@{\preamble@}}\ifnum`{=0 \fi\iffalse}\fi
   \ialign\bgroup\span\Preamble@\crcr}

\def\cases{\left\{\,\vcenter\bgroup\vspace@
     \normalbaselines\openup\jot\m@th
       \Let@\ialign\bgroup$##$\hfil&\quad$##$\hfil\crcr
      \mathstrut\crcr\noalign{\kern-\baselineskip}}
\def\endcases{\endmatrix\right.}
\newif\iftagsleft@
\tagsleft@true
\def\TagsOnRight{\global\tagsleft@false}
\def\tag#1$${\iftagsleft@\leqno\else\eqno\fi
 \hbox{\def\pagebreak{\global\postdisplaypenalty-\@M}%
 \def\nopagebreak{\global\postdisplaypenalty\@M}\rm(#1\unskip)}%
  $$\postdisplaypenalty\z@\ignorespaces}
\interdisplaylinepenalty=\@M
\def\allowdisplaybreak@{\def\allowdisplaybreak{\noalign{\allowbreak}}}
\def\displaybreak@{\def\displaybreak{\noalign{\break}}}
\def\align#1\endalign{\def\tag{&}\vspace@\allowdisplaybreak@\displaybreak@
  \iftagsleft@\lalign@#1\endalign\else
   \ralign@#1\endalign\fi}
\def\ralign@#1\endalign{\displ@y\Let@\tabskip\centering\halign to\displaywidth
     {\hfil$\displaystyle{##}$\tabskip=\z@&$\displaystyle{{}##}$\hfil
       \tabskip=\centering&\llap{\hbox{(\rm##\unskip)}}\tabskip\z@\crcr
             #1\crcr}}
\def\lalign@
 #1\endalign{\displ@y\Let@\tabskip\centering\halign to \displaywidth
   {\hfil$\displaystyle{##}$\tabskip=\z@&$\displaystyle{{}##}$\hfil
   \tabskip=\centering&\kern-\displaywidth
        \rlap{\hbox{(\rm##\unskip)}}\tabskip=\displaywidth\crcr
               #1\crcr}}
\def\overrightarrow{\mathpalette\overrightarrow@}
\def\overrightarrow@#1#2{\vbox{\ialign{$##$\cr
    #1{-}\mkern-6mu\cleaders\hbox{$#1\mkern-2mu{-}\mkern-2mu$}\hfill
     \mkern-6mu{\to}\cr
     \noalign{\kern -1\p@\nointerlineskip}
     \hfil#1#2\hfil\cr}}}
\def\overleftarrow{\mathpalette\overleftarrow@}
\def\overleftarrow@#1#2{\vbox{\ialign{$##$\cr
     #1{\leftarrow}\mkern-6mu\cleaders\hbox{$#1\mkern-2mu{-}\mkern-2mu$}\hfill
      \mkern-6mu{-}\cr
     \noalign{\kern -1\p@\nointerlineskip}
     \hfil#1#2\hfil\cr}}}
\def\overleftrightarrow{\mathpalette\overleftrightarrow@}
\def\overleftrightarrow@#1#2{\vbox{\ialign{$##$\cr
     #1{\leftarrow}\mkern-6mu\cleaders\hbox{$#1\mkern-2mu{-}\mkern-2mu$}\hfill
       \mkern-6mu{\to}\cr
    \noalign{\kern -1\p@\nointerlineskip}
      \hfil#1#2\hfil\cr}}}
\def\underrightarrow{\mathpalette\underrightarrow@}
\def\underrightarrow@#1#2{\vtop{\ialign{$##$\cr
    \hfil#1#2\hfil\cr
     \noalign{\kern -1\p@\nointerlineskip}
    #1{-}\mkern-6mu\cleaders\hbox{$#1\mkern-2mu{-}\mkern-2mu$}\hfill
     \mkern-6mu{\to}\cr}}}
\def\underleftarrow{\mathpalette\underleftarrow@}
\def\underleftarrow@#1#2{\vtop{\ialign{$##$\cr
     \hfil#1#2\hfil\cr
     \noalign{\kern -1\p@\nointerlineskip}
     #1{\leftarrow}\mkern-6mu\cleaders\hbox{$#1\mkern-2mu{-}\mkern-2mu$}\hfill
      \mkern-6mu{-}\cr}}}
\def\underleftrightarrow{\mathpalette\underleftrightarrow@}
\def\underleftrightarrow@#1#2{\vtop{\ialign{$##$\cr
      \hfil#1#2\hfil\cr
    \noalign{\kern -1\p@\nointerlineskip}
     #1{\leftarrow}\mkern-6mu\cleaders\hbox{$#1\mkern-2mu{-}\mkern-2mu$}\hfill
       \mkern-6mu{\to}\cr}}}
\def\sqrt#1{\radical"270370 {#1}}
\def\dots{\relax\ifmmode\let\next=\ldots\else\let\next=\tdots@\fi\next}
\def\tdots@{\unskip\ \tdots@@}
\def\tdots@@{\futurelet\next\tdots@@@}
\def\tdots@@@{$\mathinner{\ldotp\ldotp\ldotp}\,
   \ifx\next,$\else
   \ifx\next.\,$\else
   \ifx\next;\,$\else
   \ifx\next:\,$\else
   \ifx\next?\,$\else
   \ifx\next!\,$\else
   $ \fi\fi\fi\fi\fi\fi}
\def\text{\relax\ifmmode\let\next=\text@\else\let\next=\text@@\fi\next}
\def\text@@#1{\hbox{#1}}
\def\text@#1{\mathchoice
 {\hbox{\everymath{\displaystyle}\def\textfonti{\the\textfont1 }%
    \def\textfontii{\the\textfont2 }\textdef@@ T#1}}
 {\hbox{\everymath{\textstyle}\def\textfonti{\the\textfont1 }%
    \def\textfontii{\the\textfont2 }\textdef@@ T#1}}
 {\hbox{\everymath{\scriptstyle}\def\textfonti{\the\scriptfont1 }%
   \def\textfontii{\the\scriptfont2 }\textdef@@ S\rm#1}}
 {\hbox{\everymath{\scriptscriptstyle}\def\textfonti{\the\scriptscriptfont1 }%
   \def\textfontii{\the\scriptscriptfont2 }\textdef@@ s\rm#1}}}
\def\textdef@@#1{\textdef@#1\rm \textdef@#1\bf
   \textdef@#1\sl \textdef@#1\it}

\def\textdef@#1#2{\def\next{\csname\expandafter\eat@\string#2fam\endcsname}%
\if S#1\edef#2{\the\scriptfont\next\relax}%
 \else\if s#1\edef#2{\the\scriptscriptfont\next\relax}%
 \else\edef#2{\the\textfont\next\relax}\fi\fi}
\scriptfont\itfam=\tenit \scriptscriptfont\itfam=\tenit
\scriptfont\slfam=\tensl \scriptscriptfont\slfam=\tensl
\mathcode`\0="0030
\mathcode`\1="0031
\mathcode`\2="0032
\mathcode`\3="0033
\mathcode`\4="0034
\mathcode`\5="0035
\mathcode`\6="0036
\mathcode`\7="0037
\mathcode`\8="0038
\mathcode`\9="0039
\def\Cal{\relax\ifmmode\let\next=\Cal@\else
     \def\next{\errmessage{Use \string\Cal\space only in math mode}}\fi\next}
\def\Cal@#1{{\fam2 #1}}
\def\bold{\relax\ifmmode\let\next=\bold@\else
   \def\next{\errmessage{Use \string\bold\space only in math
      mode}}\fi\next}\def\bold@#1{{\fam\bffam #1}}
\mathchardef\Gamma="0000
\mathchardef\Delta="0001
\mathchardef\Theta="0002
\mathchardef\Lambda="0003
\mathchardef\Xi="0004
\mathchardef\Pi="0005
\mathchardef\Sigma="0006
\mathchardef\Upsilon="0007
\mathchardef\Phi="0008
\mathchardef\Psi="0009
\mathchardef\Omega="000A
\mathchardef\varGamma="0100
\mathchardef\varDelta="0101
\mathchardef\varTheta="0102
\mathchardef\varLambda="0103
\mathchardef\varXi="0104
\mathchardef\varPi="0105
\mathchardef\varSigma="0106
\mathchardef\varUpsilon="0107
\mathchardef\varPhi="0108
\mathchardef\varPsi="0109
\mathchardef\varOmega="010A

\def\fontlist@{\\{\tenrm}\\{\sevenrm}\\{\fiverm}\\{\teni}\\{\seveni}%
 \\{\fivei}\\{\tensy}\\{\sevensy}\\{\fivesy}\\{\tenex}\\{\tenbf}\\{\sevenbf}%
 \\{\fivebf}\\{\tensl}\\{\tenit}\\{\tensmc}}
\def\dodummy@{{\def\\##1{\global\let##1=\dummyft@}\fontlist@}}
\newif\ifsyntax@
\newcount\countxviii@
\def\newtoks@{\alloc@5\toks\toksdef\@cclvi}
\def\nopages@{\output={\setbox\z@=\box\@cclv \deadcycles=\z@}\newtoks@\output}
\def\syntax{\syntax@true\dodummy@\countxviii@=\count18
\loop \ifnum\countxviii@ > \z@ \textfont\countxviii@=\dummyft@
   \scriptfont\countxviii@=\dummyft@ \scriptscriptfont\countxviii@=\dummyft@
     \advance\countxviii@ by-\@ne\repeat
\dummyft@\tracinglostchars=\z@
  \nopages@\frenchspacing\hbadness=\@M}
\def\magstep#1{\ifcase#1 1000\or
 1200\or 1440\or 1728\or 2074\or 2488\or 
 \errmessage{\string\magstep\space only works up to 5}\fi\relax}
{\lccode`\2=`\p \lccode`\3=`\t 
 \lowercase{\gdef\tru@#123{#1truept}}}

\def\scaletype#1{\mag=#1\relax
 \hsize=\expandafter\tru@\the\hsize
 \vsize=\expandafter\tru@\the\vsize
 \dimen\footins=\expandafter\tru@\the\dimen\footins}

\def\scalefont#1#2\andcallit#3{\edef\font@{\the\font}#1\font#3=
  \fontname\font\space scaled #2\relax\font@}
\def\Mag@#1#2{\ifdim#1<1pt\multiply#1 #2\relax\divide#1 1000 \else
  \ifdim#1<10pt\divide#1 10 \multiply#1 #2\relax\divide#1 100\else
  \divide#1 100 \multiply#1 #2\relax\divide#1 10 \fi\fi}
\def\scalelinespacing#1{\Mag@\baselineskip{#1}\Mag@\lineskip{#1}%
  \Mag@\lineskiplimit{#1}}
\def\wlog#1{\immediate\write-1{#1}}
\catcode`\@=\active

\def\oversetbrace#1\to#2{\overbrace{#2}^{#1}}
\def\undersetbrace#1\to#2{\underbrace{#2}_{#1}}

\magnification=1200
\baselineskip=16pt
\nopagenumbers

\pagewidth{12.15cm}
\pageheight{18,3cm}
\voffset .3cm 

\def\noi{\noindent}
\def\sq{\hbox{\vrule\vbox{\hrule\phantom{o}\hrule}\vrule}}
\def\wh{\widehat}
\def\wt{\widetilde}
\def\ovr{\overrightarrow}

\def\ov{\overline}
\def\noi{\noindent}
\def\ep{\varepsilon}
\def\Bbb#1{\text{\bf{#1}}}

\font\ninerm=cmr9

\font\docebf=cmbx10 scaled 1728
\font\catorcerm=cmr10 scaled 1728
\font\catorcebf=cmbx10 scaled 1728
\font\catorce=cmr10 scaled 1728
\font\dseisbf=cmr10 scaled 2073
\def\bigt{\bigtriangleup}
\TagsOnRight
\mathsurround=1pt

\font\biggbf=cmr10 scaled 1728
\font\bigbf=cmr10 scaled 1200
\font\bigitbf=cmti10 scaled 1440
\font\catorce=cmr10 scaled 1728
\font\catbf=cmbx10 scaled 1728
\font\dseis=cmr10 scaled 2073

$\ $

\pageno=-1

\headline={\ifnum\pageno=-1\hfil\else\hss\tenrm \folio\ \fi}

\hskip 8cm hep-th/9803083
\vskip 1cm

\centerline{\bf{\biggbf UNIVERSIDAD SIM\'ON BOL\'IVAR}}
\vskip 2.5cm
\centerline{\bf{\biggbf SPIN$\ \ $2$\ \ $EN$\ \ $DIMENSI\'ON$\ \ $2+1}}
\vskip 1cm
\bigbf
\centerline{Trabajo final presentado a la Universidad Sim\'on Bol\'{\i}var 
por el}
\vskip 5mm
\centerline{Lic. P\'{\i}o J. Arias}
\vskip 5mm
\centerline{como requisito parcial para optar al t\'{\i}tulo de}
\vskip 3mm
\centerline{Doctor en F\'{\i}sica}
\vskip 5mm
\centerline{Realizado con la asesor\'{\i}a del}
\vskip 3mm
\centerline{Prof. Carlos Aragone}
\vskip 4cm
\centerline{Febrero, 1994}
 
\newpage

$\ $

\pageno=-2 

\vskip 5cm
{\bigitbf
\hskip 7cm {\it A la memoria de mi padre,}

\hskip 7cm {\it P\'{\i}o Jos\'e Arias B.}

\hskip 7cm {\it A Zulay y Dafne}
}

\newpage

$\ $

\vskip 3cm

\centerline{\bf{\biggbf RESUMEN}}

\vskip 2cm

En este trabajo se analizan din\'amicamente distintas teor\'{\i}as masivas de
spin 1 y spin 2, mostrando su equivalencia (entre teor\'{\i}as de igual spin) 
y su analog\'{\i}a (entre teor\'{\i}as de spin distinto). Se muestra una 
equivalencia can\'onica entre la teor\'{\i}a de spin 2 autodual y la 
teor\'{\i}a de gravedad Masiva Vectorial de Chern-Simons linealizada. Se 
introduce una teor\'{\i}a gravitatoria masiva curva, que no es invariante 
Lorentz en el espacio tangente, pero s\'{\i} ante difeomorfismos. Se analiza la
posibilidad de rotura de simetr\'{\i}a en la teor\'{\i}a de spin 1
Topol\'ogica Masiva, y en las teor\'{\i}as de spin 2 masivas
Topol\'ogica Masiva y  vectorial de Chern-Simons linealizadas. Se estudia el
comportamiento  any\'onico de las distintas teor\'{\i}as de spin 1 y spin 2,
consiguiendo el  an\'alogo gravitacional de la teor\'{\i}a de Chern-Simons
vectorial pura. 

\newpage

$\ $
\vskip 1cm
\bigbf
{\bf \'INDICE GENERAL}
\vskip 1cm 
\noi
{\bf I $\ \ $INTRODUCCI\'ON}\hskip 8.1cm 1
\vskip 1cm
\noi
{\bf II $\ $TEOR\'IAS DE SPIN 1 MASIVO}\hskip 5.6cm 7
\vskip 5mm
{\bf
\item{1.}La acci\'on autodual}\dotfill 7
{\bf
\item{2.}La acci\'on Topol\'ogica Masiva Vectorial}\dotfill 10
{\bf
\item{3.}La acci\'on de Hagen}\dotfill 12
\vskip 1cm
\noi
{\bf III TEOR\'IAS DE SPIN 2 MASIVO}\hskip 5.3cm 15
\vskip 5mm
{\bf
\item{1.}La acci\'on de Fierz-Pauli y la teor\'{\i}a
\item{}de spin 2 autodual}\dotfill 16
\itemitem{1.1}La acci\'on de Fierz-Pauli y la condici\'on 
\itemitem{}de autodualidad \dotfill 16
\itemitem{1.2}La acci\'on autodual \dotfill 18
\itemitem{1.3}An\'alisis can\'onico de la teor\'{\i}a  
\itemitem{}de spin 2 autodual \dotfill 20
{\bf
\item{2.}La acci\'on de gravedad Topol\'ogica 
\item{}Masiva, linealizada}\dotfill 24
{\bf
\item{3.}La acci\'on intermedia o la gravedad masiva
\item{}Vectorial de Chern-Simons linealizada}\dotfill 29
\itemitem{3.1}An\'alisis covariante \dotfill 29
\itemitem{3.2}Descomposici\'on 2+1 y la forma invariante
\itemitem{}de calibre de $S_{VCS}^l$ \dotfill 30
\itemitem{3.3}An\'alisis can\'onico de $S_{VCS}^l$ \dotfill 35
{\bf
\item{4.}Equivalencia can\'onica entre $S_{AD}^2$ y $S_{VCS}^l$}\dotfill 39
\itemitem{4.1}El conjunto com\'un de v\'{\i}nculos \dotfill 39
\itemitem{4.2}El hamiltoniano invariante de calibre \dotfill 41
\itemitem{4.3}La extensi\'on ``invariante de calibre'' de $H_0^{(VCS)}$
\dotfill 44 
\vskip 1cm
\noi
{\bf IV LA GRAVEDAD MASIVA VECTORIAL}

\noi{\bf \ \ \ \ \ DE CHERN-SIMONS}\hskip 7.1cm 47
\vskip 5mm
{\bf
\item{1.}La acci\'on dentro de marco
\item{}jer\'arquico de simetr\'{\i}as}
\dotfill 47
\vskip 1cm
\noi
{\bf V ROTURA DE SIMETR\'IA}\hskip 6.6cm 54
\vskip 5mm
{\bf
\item{1.}Teor\'{\i}a de Proca-Chern-Simons}\dotfill 54
\itemitem{1.1}La acci\'on como producto de un proceso de rotura 
\itemitem{}espont\'anea de simetr\'{\i}a \dotfill 54
\itemitem{1.2}An\'alisis covariante \dotfill 56 
\itemitem{1.3}Descomposici\'on 2+1 y le energ\'{\i}a \dotfill 57
{\bf
\item{2.}Teor\'{\i}a de Einstein autodual}\dotfill 59
\itemitem{2.1}La acci\'on. An\'alisis covariante \dotfill 59
\itemitem{2.2}Descomposici\'on 2+1 \dotfill 64
{\bf
\item{3.}La no viabilidad de romper la simetr\'{\i}a 
\item{}para la teor\'{\i}a $TM$}\dotfill 65
\itemitem{3.1}La teor\'{\i}a con los dos t\'erminos de $CS$ \dotfill 65
\itemitem{3.2}La teor\'{\i}a $TM$ con todas sus simetr\'{\i}as rotas 
\dotfill 70
\vskip 1cm
\noi
{\bf VI COMPORTAMIENTO ANY\'ONICO EN TEOR\'IAS}
{\bf VEC\-TO\-RIA\-LES Y DE GRAVEDAD LINEALIZADA}\hskip 4.5cm 75
\vskip 5mm
{\bf
\item{1.}Spin y estad\'{\i}stica en dimensi\'on 2+1}\dotfill 75
{\bf
\item{2.}Implementaci\'on din\'amica de estad\'{\i}stica fraccionaria}
\dotfill 80
{\bf 
\item{3.}Comportamiento any\'onico en teor\'{\i}as vectoriales}\dotfill 83
\itemitem{3.1}Teor\'{\i}a de $CS$ pura y la $TM$ vectorial \dotfill 83
\itemitem{3.2}Las teor\'{\i}as autodual y $TM$, y el problema de los 
\itemitem{}acoplamientos no minimales \dotfill 86
\itemitem{3.3}La teor\'{\i}a de Hagen \dotfill 88
{\bf
\item{4.}Comportamiento any\'onico en teor\'{\i}as 
\item{}de gravedad linealizada}\dotfill 89
\itemitem{4.1}La posibilidad de tener anyones gravitacionales \dotfill 89
\itemitem{4.2}El par\'ametro de comportamiento any\'onico para  
\itemitem{}la teor\'{\i}a $VCS$ linealizada y la $TM$ linealizada \dotfill 92
\itemitem{4.3}Par\'ametro de comportamiento any\'onico de  
\itemitem{}la teor\'{\i}a $AD$ y de la teor\'{\i}a de Einstein autodual
\dotfill 95 
\itemitem{4.4}La teor\'{\i}a $AD$ con acoplamiento no minimal
\dotfill 97
\vskip 1cm
\noi
{\bf VII CONCLUSIONES}\hskip 7.5 cm 100
\vskip 5mm
{\bf
\item{}Referencias} \dotfill 103
{\bf
\item{}Ap\'endice A} \dotfill 107
{\bf
\item{}Ap\'endice B} \dotfill 114

\newpage

$\ $

\pageno=1
\headline={\ifnum\pageno=1\hfil\else\hss\tenrm \folio\ \fi}
\vskip 1cm

\centerline{\biggbf Cap\'{\i}tulo {\biggbf I}}

\vskip 1cm

\centerline{\biggbf INTRODUCCI\'ON}

\vskip 1.5cm

El estudio de teor\'{\i}as vectoriales y tensoriales en dimensi\'on 2+1 (2 
espaciales + 1 temporal) estuvo, inicialmente motivado por su conexi\'on con 
el comportamiento de modelos, en 3+1 dimensiones, a altas temperaturas [1]. 
Sin embargo, en la actualidad, la f\'{\i}sica planar (en 2+1 dimensiones) 
posee un inter\'es real intr\'{\i}nseco, ya que presenta caracter\'{\i}sticas 
propias que hacen sumamente atractivo su estudio y an\'alisis en el contexto 
de la teor\'{\i}a cu\'antica de campos. Inclusive, al nivel mas fundamental 
se ha propuesto que a una escala planckiana los grados f\'{\i}sicos de 
libertad observables podr\'{\i}an ser mejor descritos por una red bidimenional 
que evoluciona con el tiempo [2]. En esta dimensi\'on espacio-temporal aparecen 
naturalmente y exclusivamente los ``anyones" [3], o part\'{\i}culas con spin 
y estad\'{\i}stica distintos a los que estamos acostumbrados en 3+1 
dimensiones. Estas part\'{\i}culas podr\'{\i}an tener aplicaciones en f\'{\i}sica
de la  materia condensada [4,5]. En otro orden de ideas, la gravitaci\'on en 2+1 
dimensiones es claro que tiene que jugar un papel importante en el estudio de 
fen\'omenos donde esten involucradas cuerdas c\'osmicas [6].

Las diferencias entre 2+1 y 3+1 dimensiones, empiezan a notarse si miramos 
las representaciones del grupo inhomog\'eneo de Lorentz, o grupo de 
Poincar\'e, en 2+1 dimensiones [7]. Si $P^m$ representa al generador de las 
traslaciones y $M^{mn}(= - M^{nm})$ al generador de las transformaciones de 
Lorentz, el \'algebra de Poincar\'e, en 2+1 dimensiones, es 

$$
\align
[P^m, P^n] &= 0, \tag 1,a\\
[J^m, P^n] &= - i\ep^{mnl}P_l, \tag 1,b\\
[J^m, J^n] &= i\ep^{mnl}J_l, \tag 1,c
\endalign
$$
donde $J^{m}\equiv(1/2)\ep^{mnl}M_{nl}$, se introduce por la antisimetr\'{\i}a 
de $M^{mn}$. Mirando el sector del grupo de Lorentz $L^\uparrow_+$ (grupo 
propio, ortocrono), observamos que $J^0$ es el generador de las rotaciones 
espaciales y $-\ep_{ij}J^j$ $(\ep_{ij}=\ep^{ij}=\ep^{oij},i,j=1,2)$ genera 
los ``boosts" en la direcci\'on $X^i$. (1,c) representa al \'algebra del 
grupo de Lorentz SO(2,1) y es isomorfa a la del grupo SL(2,$\Bbb{R}$) por lo 
que estos grupos admiten el mismo grupo de recubrimiento.

En 3+1 dimensiones las representaciones irreducibles estan caracterizadas por 
los autovalores de los casimires $P_mP^m$ y $W_mW^m$, donde 
$W_m\equiv (1/2)\ep_{mnrl}$ $P^nM^{rl}$ es el vector de Pauli-Lubanski. En 
cambio para 2+1 dimensiones el invariante de Pauli-Lubanski es un escalar 
$(\ep_{mnl}P^mM^{nl}=2P^mJ_m)$. As\'{\i}, las representaciones masivas 
est\'an caracterizadas por los autovalores de $P_mP^m$ y $P_mJ^m$ [8]

$$
\align 
(P_{m}P^m+m^2)\psi &=0, \tag 2,a \\
(P_{m}J^m-mS)\psi &=0. \tag 2,b \\
\endalign
$$
En (2,b) no hay ninguna restricci\'on sobre el valor de S, que representa el 
spin o ``helicidad" de la representaci\'on. Este hecho marca una gran 
diferencia con la contraparte en 3+1 dimensiones donde el autovalor del 
invariante de Pauli-Lubanski es $-{m}^{2}S(S+1)$ con 
$S\in ``\frac{1}{2}{\Bbb{Z}}''$ (conjunto de los m\'ultiplos enteros de 
(1/2)), debido a la no abelianilidad del grupo de rotaciones en 3 dimensiones 
espaciales. En cambio, el grupo de rotaciones en el plano es abeliano, por lo 
que no hay restricciones para los autovalores del momento angular y, por 
ende, del spin.

En (2) observamos que para representaciones con masa nula, $S$ no est\'a bien 
definido. Sin embargo, existen dos realizaciones de estas representaciones con 
masa nula. Estas son la teor\'{\i}a de Maxwell, que es equivalente a la de un 
campo escalar, y la de Dirac, con espinores de dos componentes [7,9]. Para 
teor\'{\i}as con tensores de mas alto rango no hay excitaciones con masa 
nula. Existen an\'alisis rigurosos donde se encuentra que las part\'{\i}culas 
creadas por campos localizables en regiones compactas poseen spin 
$S\in$``$\frac{1}{2}\Bbb{Z}$'', y se cumple el teorema spin-estad\'{\i}stica. 
Esta restricci\'on no existe para campos no localizables en regiones 
compactas, ni se obtiene informaci\'on precisa respecto a la estad\'{\i}stica 
[10,11].

En 2+1 dimensiones las transformaciones impropias de paridad, $P$, e 
inversi\'on temporal $T$ estan definidas como 

$$
P:(t,x,y)\longrightarrow (t,-x,y)\ \ ;\ \ T:(t,x,y)\longrightarrow (-t,x,y),
\tag 3 
$$
donde observamos que $P$ no corresponde a la inversi\'on espacial, como en 
3+1 dimensiones. Esto se debe a que en el plano hacer inversi\'on espacial es 
equivalente a una rotaci\'on. $J^0$ es sensible a $P$ y $T$, por lo que bajo 
estas transformaciones $S\to -S$, y puede probarse que el spin en 2+1 viola 
$P$ y $T$ [12]. Si quisieramos tener una teor\'{\i}a invariante bajo estas 
transformaciones discretas debemos tener presentes pares de part\'{\i}culas de
igual masa y spines  opuestos. 

En vista de lo expuesto anteriormente, si queremos una teor\'{\i}a de spin 
$S\ne0$, \'esta debe ser necesariamente masiva, y si describe una sola 
excitaci\'on masiva, la teor\'{\i}a violar\'a $P$ y $T$. Para los casos de 
spin 1 y 2, sucede que podemos sumarle a las acciones de Maxwell, $S_M$, y 
de Einstein, $S_E$, respectivamente, un t\'ermino $S_{TM}$, de car\'acter 
topol\'ogico, y as\'{\i} obtener una teor\'{\i}a masiva, consistente, con 
spin 1 \'o 2 seg\'un el caso. Estos t\'erminos estan relacionados con las 
clases caracter\'{\i}sticas de Chern-Simons, de ah\'{\i} la denominaci\'on de 
t\'erminos de Chern-Simons; adem\'as, violan $P$ y $T$. A las teor\'{\i}as 
resultantes se les llama teor\'{\i}as topol\'ogicas masivas $(TM)$ del spin 
correspondiente [13,14]. Estas teor\'{\i}as tienen, respectivamente, las 
mismas simetr\'{\i}as que $S_M$ y $S_E$.

Para la teor\'{\i}a $TM$ vectorial (spin 1) y la gravedad $TM$ linealizada 
(spin 2) tenemos una formulaci\'on equivalente de caracter autodual $(AD)$. 
Para el caso de spin 1, la ecuaci\'on de movimiento de la teor\'{\i}a 
autodual es como la ``ra\'{\i}z" de las ecuaciones de Proca para el campo 
vectorial [15,16]. Para el caso de spin 2 [17,18,31] las ecuaciones son como 
la ``ra\'{\i}z" de las ecuaciones de Fierz-Pauli. Estas teor\'{\i}as 
autoduales son equivalentes como teor\'{\i}as libres a las 
corres\-pon\-dien\-tes teor\'{\i}as TM. Para spin 3 y 4 tambi\'en es posible 
construir teor\'{\i}as masivas de caracter autodual [19,20]. Estas 
formulaciones autoduales estan relacionadas con una formulaci\'on relativista 
reciente de anyones [8]. Es posible, tambi\'en cons\-truir ``acciones maestras" 
para spin 1 [16] y spin 2 [18] las cuales sobre las ecuaciones de movimiento 
(``on shell") se reducen a las acciones $AD$ o $TM$, respectivas.

Volviendo a los anyones, aunque todav\'{\i}a no existe una teor\'{\i}\-a de 
campos definitiva que describa part\'{\i}culas con estad\'{\i}stica y spin 
frac\-cionario, re\-sul\-ta in\-tere\-sante estu\-diar teor\'{\i}as donde las 
part\'{\i}culas bos\'onicas o fermi\'onicas adquieren spin y estad\'{\i}stica 
fraccionaria a trav\'es de un mecanismo externo. Un ejemplo de esto es el 
modelo que resulta de acoplar minimalmente part\'{\i}culas cargadas con un 
campo electromagn\'etico y tener adicionalmente un t\'ermino de Chern-Simons 
$(CS)$ [10,21,22]. En este caso el t\'ermino de $CS$ hace que las 
part\'{\i}culas se ``vean" como puntos singulares del flujo magn\'etico, y 
el generador de las rotaciones adquiere una contribuci\'on determinada 
por el acoplamiento con el t\'ermino de $CS$ [23]. Este cuadro es igual al 
que aparece cuando se estudia la mec\'anica cu\'antica de dos anyones con 
par\'ametros estad\'{\i}stico $\theta$ [65]. La fase del efecto 
Aharanov-Bohm resulta ser proporcional a $\theta$. Se dice, 
entonces, que la estad\'{\i}stica y spin del modelo han sido implementados 
din\'amicamente, y se hipotiza que este efecto pueda proveer las bases de una 
explicaci\'on te\'orica de la superconductividad a altas temperaturas 
cr\'{\i}ticas [24,25].

En el caso de acoplamiento con teor\'{\i}as de gravedad linea\-li\-zada, 
bas\'andonos en la similitud que existe entre la acci\'on no re\-la\-ti\-vis\-ta de 
una part\'{\i}cula cargada en un campo electromagn\'etico y la acci\'on no 
relativista de una part\'{\i}cula masiva en un campo gravitatorio d\'ebil 
[26], podemos definir el an\'alogo gravitacional del punto de flujo 
magn\'etico y tendremos lo que se conoce como efecto Aharanov-Bohm 
gravitacional [27,28]. Este an\'alogo gra\-vi\-ta\-cio\-nal fue analizado 
recientemente [29,30], en el contexto de la gravedad $TM$ linealizada. Sin 
embargo, la analog\'{\i}a con la teor\'{\i}a $TM$ vectorial se obtiene en 
determinados casos. Queda abierto, en la actualidad, el problema de poder 
implementar estad\'{\i}stica y spin fraccionarios a nivel exacto, curvo.

Relacionado con la implementaci\'on din\'amica de spin y estad\'{\i}stica 
fraccionarios, tenemos el hecho de que cuando hay rotura espont\'anea de 
simetr\'{\i}a, no es posible esta implementaci\'on [32,33]. De manera que, 
resulta importante, entre otros aspectos, ver la viabilidad f\'{\i}sica de 
poder romper las simetr\'{\i}as de las acciones de spin 2 masivo, y su 
repercusi\'on en el comportamiento any\'onico de las teor\'{\i}as, viables, 
as\'{\i} obtenidas. En el caso vectorial al romper espont\'aneamente la 
teor\'{\i}a de $CS$ pura nos queda en la parte vectorial la acci\'on 
autodual, para la teor\'{\i}a $TM$ nos queda, en cambio, una teor\'{\i}a de 

Proca con un t\'ermino de $CS$ que describe dos excitaciones autoduales con 
masas distintas [34]. Para teor\'{\i}as de spin 3 esto se ha hecho, 
encontrando estrecha similitud con los casos vectoriales [20].

En este trabajo, se estudian teor\'{\i}as masivas de spin 1 y 2 desde un 
punto de vista comparativo, mostrando que los distintos fen\'omenos 
f\'{\i}sicos que ocurren en las teor\'{\i}as de spin 1 tienen una 
analog\'{\i}a uniforme en spin 2. Presentamos la teor\'{\i}a curva que 
llamamos Gravedad Vectorial masiva de Chern-Simons $(VCS)$ cuya teor\'{\i}a 
linealizada aparece como an\'alogo de la teor\'{\i}a $TM$ vectorial. 
Adem\'as, encontramos el an\'alogo gravitacional de la teor\'{\i}a de 
Chern-Simons pura vectorial en el contexto de la implementaci\'on din\'amica 
de anyones. Por \'ultimo mostramos que s\'olo en la teor\'{\i}a de gravedad 
$VCS$ linealizada podemos romper consistentemente las simetr\'{\i}as que esta 
presenta, a diferencia de la teor\'{\i}a de gravedad $TM$ linealizada donde 
no es posible hacerlo.

La organizaci\'on es como sigue en el cap\'{\i}tulo II analizamos 
las teor\'{\i}as de spin 1 masivo y las posibles equivalencias entre ellas. 
En el cap\'{\i}tulo III realizamos un an\'alisis an\'alogo, mas extenso, de 
las teor\'{\i}as de spin 2 masivo. Parte de los an\'alisis 
din\'amicos hechos para spin 1 y spin 2, son hechos heur\'{\i}sticamente 
u\-san\-do los proyectores que se introducen en el ap\'endice B. Adem\'as, 
mostramos una equivalencia can\'onica entre la acci\'on autodual de spin 2 
y la acci\'on de la gravedad $VCS$ linealizada, se analiza su posible
conexi\'on can\'onica con la teor\'{\i}a $TM$ linealizada. En el cap\'{\i}tulo
IV  presentamos la acci\'on curva de la gravedad $VCS$. En el cap\'{\i}tulo V, 
miramos las teor\'{\i}as con rotura de simetr\'{\i}a correspondientes a la 
acci\'on $TM$ vectorial, $VCS$ linealizada y $TM$ li\-nea\-li\-za\-da. 
Encontramos que en este \'ultimo caso, este proceso no es consitente. Por 
\'ultimo, en el cap\'{\i}tulo VI estudiamos el comportamiento any\'onico en 
teor\'{\i}as vectoriales y de gravedad linealizada. Observamos que  permitiendo
acoplamientos no minimales los par\'ametros de comportamiento any\'onico son 
radicalmente distintos. 

\newpage

$\ $

\pageno=7
\headline={\ifnum\pageno=7\hfil\else\hss\tenrm \folio\ \fi}
\vskip 1cm

\centerline{\catorce Cap\'{\i}tulo {\catbf{II}}}

\vskip 1cm

\centerline{\dseis TEOR\'IAS DE SPIN 1 MASIVO}

\vskip 2cm

Las teor\'{\i}as de spin 1 masivo, por ser mas tratables que las de spin 2 
masivo, resultan un laboratorio muy conveniente antes de abordar los 
problemas que querramos estudiar. En este cap\'{\i}tulo analizamos tres 
teor\'{\i}as de spin 1 masivo que son equivalentes como teor\'{\i}as libres. 
Estas ya han sido presentadas y analizadas anteriormente [13-16,35,36], sin 
embargo damos aqu\'{\i} un an\'alisis personal de ellas, el cual nos 
servir\'a de base al abordar el an\'alisis de las teor\'{\i}as de spin 2 
masivo.

\noi
{\bf 1.- La acci\'on autodual}

\vskip 3mm
La acci\'on de Proca, es la usual

$$
S_P=<-\frac{1}{2}F_{mn}F^{mn}-\frac{1}{2}m^2a_ra^r>,\tag 1.1
$$
con ecuaciones de movimiento

$$
\partial^mF_{mn}-m^2a_n=0,\tag 1.2
$$
las cuales aseguran que se cumple la condici\'on de Lorentz $\partial^ma_m=0$. 
Las dos componentes independientes que quedan corresponden a las excitaciones 
que se propagan en esta teor\'{\i}a.

En D = 2+1 es posible hallar la ``ra\'{\i}z cuadrada" de la ecuaci\'on (1.2), 
\'esta es [15]

$$
\pm \ep_m{}^{ln}\partial_l a_n-ma_m=0.\tag 1.3
$$
La ecuaci\'on (1.3) implica, igualmente, que $\partial^m a_m=0$. 
Covariantemente, usando los proyectores\footnote"*"{\ninerm{Ver ap\'endice B}} 
$P_{\pm m}{}^n=\frac{1}{2}(P_m{}^n\pm \xi_m{}^n)$, la reescribiremos como

$$
[\pm \sq^{1/2}(P_{+m}{}^n-P_{-m}{}^n)-m(P_{+m}{}^n+P_{-m}{}^n)]a_n=0,\tag 1.4
$$
donde hemos usado el hecho de que $a_m$ es transverso. Esta \'ultima 
ecuaci\'on, debido a que los $P_\pm$ son ``ortogonales", es equivalente a

$$\align
\pm (\sq^{1/2}\mp m)P_{+m}{}^na_n & =0,\tag 1.5,a\\
\pm (\sq^{1/2}\pm m)P_{-m}{}^na_n & =0,\tag 1.5,b
\endalign
$$
de donde observamos que el signo ``mas" (``menos") en (1.3) co\-rres\-pon\-de a una 
excitaci\'on f\'{\i}sica de masa $m$ y spin +1(-1).

Decimos que un campo vectorial $a_m$ que verifica (1.3) es autodual. A la 
acci\'on que tiene a la condici\'on de autodualidad como ecuaci\'on de 
movimiento se le denomina acci\'on Autodual $(AD)$ de spin 1. Esta es [15]

$$
S_{AD}=\frac{m}{2}<a_r\ep^{rmn}\partial_ma_n-ma^ra_r>.\tag 1.6
$$
La acci\'on $S_{AD}$, a diferencia de la de Proca, es sensible a paridad 
$(P)$ e inversi\'on temporal $(T)$ debido a la presencia del t\'ermino de 
$CS$ [14]. El signo del t\'ermino $-m^2a_ra^r$ debe ser el que tiene en la 
acci\'on, ya que si lo cambiamos dar\'{\i}a origen a un hamiltoniano no 
definido positivo. Sin embargo el t\'ermino de $CS$ puede tener cualquier 
signo, \'este determina la helicidad de la propagaci\'on, tal como mostramos 
en (1.5). Una manera de  construir un sistema invariante bajo $P$ y $T$ es 
tomar una acci\'on, con dos campos de calibre $a_{1m}$ y $a_{2m}$, donde el 
t\'ermino de $CS$ para cada campo de calibre tenga signo distinto y definir 
$P$ y $T$ de forma que incluya un cambio de los 2 campos [35].

Podemos mirar, r\'apidamente, la acci\'on reducida de $S_{AD}$. Des\-com\-ponemos 
$a_m$ en

$$\align
a_0 & = a,\tag 1.7,a\\
a_i & =\ep_{ij}\partial_ja^T+\partial_ia^L.\tag 1.7,b
\endalign
$$
Al hacer la descomposici\'on 2+1 de $S_{AD}$, aparece un v\'{\i}nculo 
cuadr\'a\-ti\-co asociado a $a_0$ que resolvemos

$$
a=\frac{1}{m}\Delta a^T,\tag 1.8
$$
al sustituirlo nos lleva a la forma

$$
S=\frac{1}{2}<-2m\Delta a^L\dot{a}^T-\Delta a^T\Delta a^T+m^2(a^T\Delta a^T+
a^L\Delta a^L)>.\tag 1.9
$$
Haciendo las redefiniciones

$$\align
Q &\equiv (-\Delta )^{1/2}a^T\tag 1.10,a\\
\pi &=m(-\Delta )^{1/2}a^L\tag 1.10,b
\endalign
$$
llegamos a la forma final de la acci\'on reducida

$$
S^{(red)}_{AD}=<\pi \dot{Q}-\frac{1}{2}\pi \pi-\frac{1}{2}Q(-\Delta +m^2)Q>.
\tag 1.11
$$
De esta \'ultima relaci\'on observamos que la excitaci\'on es masiva, con 
masa $m$. Adem\'as distinguimos la densidad hamiltoniana $\Cal{H}$, en
funci\'on de las variables independientes, pues 
${\Cal{L}}^{(red)}_{AD}\sim p\dot{q}-\Cal{H}$. Notamos,  entonces, que: 1) $\Cal{H}$
es definida positiva, 2) si cambiamos el signo de $m^2$  en $S_{AD}$
terminaremos con una $\Cal{H}$ no definida positiva, 3) el t\'ermino de $CS$
puede tener cualquier signo ya que $\Cal{H}$ es invariante si cambio $m$  por $-m$.

\vskip 5mm
\noi
{\bf 2.- La acci\'on Topol\'ogica Masiva vectorial}
\vskip 3mm
Esta teor\'{\i}a de spin 1 masivo es el centro de las otras teor\'{\i}as que 
aqu\'{\i} presentamos, ya que estas teor\'{\i}as por ser equivalentes a ella 
son tomadas como teor\'{\i}as alternas para la descripci\'on de spin 1 
masivo. La denominaci\'on de Topol\'ogica proviene del hecho que el 
t\'ermino de $CS$ tiene un significado topol\'ogico. Ade\-m\'as, tiene la 
particularidad que el campo de calibre adquiere masa sin perder la 
invariancia de calibre usual. La acci\'on $TM$ es [13,14]

$$
S_{TM}=\frac{1}{2}<-f_rf^r+2f_r\ep^{rmn}\partial_ma_n-\mu a_r\ep^{rmn}
\partial_ma_n>.\tag 2.1
$$
Esta acci\'on es invariante bajo las transformaciones de calibre 
$\delta a_m=\partial_m\xi$, $\delta f_m=0$. Sus ecuaciones de movimiento

$$\align
& f^r=\ep^{rmn}\partial_ma_n,\tag 2.2,a\\
& \ep^{rmn}\partial_mf_n-\mu \ep^{rmn}\partial_ma_n=0,\tag 2.2,b
\endalign
$$
nos conducen al sistema de ecuaciones, de segundo orden, para $a_m$

$$
\partial_m F^{mr}-\mu \ep^{rmn}\partial_ma_n=0.\tag 2.3
$$
La helicidad de la excitaci\'on se obtiene si reescribimos (1.6) como

$$
[\sq (P_{+m}{}^r+P_{-m}{}^r)-\mu \sq^{1/2}(P_{+m}{}^r-P_{-m}{}^r)]a_r=0.
\tag 2.4
$$
Si tomamos el calibre de Lorentz $\partial^ma_m=0$, es inmediato notar que la 
excitaci\'on tiene masa $\mu$ y spin +1. Un tratamiento mas riguroso [14] nos 
muestra que el spin de la excitaci\'on es $\mu /|\mu |$.

Si hacemos la des\-com\-po\-si\-ci\'on 2+1 de $S_{TM}$, tendremos que $f_0$ y $a_0$ 
son mul\-ti\-pli\-ca\-do\-res de Lagrange asociados a los v\'{\i}nculos

$$\align
f_0 & = \Delta a^T, \tag 2.5,a\\
f^T & = \mu a^T,\tag 2.5,b
\endalign
$$
donde (2.5,a) es el v\'{\i}nculo cuadr\'atico asociado a $f_0$ y $f^T$ en 
(2.5,b) es la parte transversa de $f_i$ en el mismo esp\'{\i}ritu de (1.7). 
Al sustituir (2.5) en la acci\'on llegamos a

$$
S^{(red)}_{TM}=\frac{1}{2}<-2f^L\Delta \dot{a}^T-\Delta a^T\Delta a^T+
f^L\Delta f^L+\mu^2 a^T\Delta a^T>.\tag 2.6
$$
En (2.6) no aparece $a^L$, que representa la componente de $a_i$ sensible a 
los cambios de calibre. Haciendo las redefiniciones

$$\align
Q &\equiv (-\Delta )^{1/2}a^T\tag 2.7,a\\
\pi &\equiv (-\Delta )^{1/2}f^L\tag 2.7,b
\endalign
$$
llegamos a

$$
S^{(red)}_{TM}=<\pi \dot{Q}-\frac{1}{2}\pi \pi-\frac{1}{2}
Q(\Delta +\mu^2)Q>,\tag 2.8
$$
que es id\'entica a (1.11). Observamos que la excitaci\'on es masiva, con 
masa $\mu$ y tiene energ\'{\i}a definida positiva. El signo de $\mu$ no 
afecta el resultado.

La estructura de (2.6) es expl\'{\i}citamente invariante de calibre. Podemos 
escoger el calibre\footnote"$^\dagger$"{\ninerm{Si partimos de (2.6) con el 
procedimiento can\'onico, resulta que el v\'{\i}nculo que genera las 
transformaciones de calibre es $\zeta =\pi^L$ donde $\pi^L$ es el momento 
conjugado de $a^L$.}}

$$
a^L=\frac{1}{\mu}f^L,\tag 2.9
$$
y, entonces, la acci\'on (2.6) es id\'entica a (1.9). Esta equivalencia entre 
la acci\'on autodual y la $TM$ fu\'e reportada con anterioridad en el 
contexto de las acciones reducidas [16] y tambi\'en en el contexto del 
formalismo can\'onico donde se observa que la acci\'on autodual corresponde a 
la $TM$ con el calibre fijado [36].

\newpage

\vskip 5mm
\noi
{\bf 3.- La acci\'on de Hagen}
\vskip 3mm
Otra forma alterna para describir spin 1 con invariancia de calibre 
constituye la acci\'on propuesta por Hagen [37]

$$
S_H=<-\frac{1}{2}f^rf_r+\ep^{rst}f_r\partial_sa_t-\frac{\mu}{2}\ep^{rst}a_r
\partial_sa_t-\frac{\lambda}{2\mu}\ep^{rst}f_r\partial_sf_t>,\tag 3.1
$$
donde $a_r$ y $f^r$ son independientes, $\lambda$ es un par\'ametro 
num\'erico. Observamos que la particularidad 
que tiene $S_H$ es que se agrega un t\'ermino,parecido al de $CS$, construido 
con el objeto $f_r$, que puede pensarse como el dual de 
$F_{mn}=\partial_ma_n-\partial_na_m$, aunque las ecuaciones de movimiento 
indican que no es exactamente as\'{\i}

$$\align
&-f^r+\ep^{rst}\partial_sa_t-\frac{\lambda}{\mu}\ep^{rst}\partial_sf_t=0,
\tag 3.2,a\\
&\ep^{rst}\partial_sf_t-\mu \ep^{rst}\partial_sa_t=0.\tag 3.2,b
\endalign
$$
Cuando $\lambda =0$ $f_r$ es el dual de $F_{mn}$, pero este es el l\'{\i}mite 
donde $S_H|_{\lambda =0}=S_{TM}$. La acci\'on $S_H$ es invariante bajo las 
transformaciones de calibre $\delta a_m=\partial_m \xi$, $\delta f_m=0$. 
Analizando covariantemente las ecuaciones (3.2), observamos que $f_{r}$ es 
transverso. Tomamos el calibre $\partial_ra^r=0$, ya que la parte 
longuitudinal de $a_{r}$ no aparece en las ecuaciones. As\'{\i} podemos 
reescribir (3.2) como

$$\align
-(P_++P_-)_m{}^rf_r+&\sq^{1/2}(P_+-P_-)_m{}^ra_r+\\
-&{\lambda \over \sq^{1/2}}{\mu}(P_+-P_-)_m{}^rf_r=0,\tag 3.3,a\\
 \sq^{1/2}(P_+-P_-)_m{}^rf_r-\mu &\sq^{1/2}(P_+-P_-)_m{}^ra_r=0.\tag 3.3,b
\endalign
$$
De estas ecuaciones obtenemos

$$
P_{\pm m}{}^rf_r=\pm (1-\lambda )\sq^{1/2}P_{\pm m}{}^ra_r,\tag 3.4
$$
con lo que queda $f_{r}$ completamente determinado. Volviendo a (3.3,b)

$$
\sq^{1/2}[(1-\lambda )(P_{+m}{}^r+P_{-m}{}^r)\sq^{1/2}-\mu (P_{+m}{}^r-
P_{-m}{}^r)]a_r=0,\tag 3.5
$$ 
de donde observamos que tenemos una excitaci\'on masiva de masa 
$\mu /(1-\lambda )$ y spin + 1. Adem\'as, el signo de $(1-\lambda )$ influye 
en la determinaci\'on de la helicidad que se propaga (aqu\'{\i} estamos 
suponiendo $1>\lambda $). Este signo no influye en la positividad de la 
energ\'{\i}a. $\lambda =1$ constituye un valor singular.

La acci\'on reducida se obtiene an\'alogamente a los dos casos ya presentados. 
$f_0$ y $a_0$ son multiplicadores asociados a los v\'{\i}nculos

$$\align
f^T & =\mu a^T,\tag 3.6,a\\
f &= (1-\lambda )\Delta a^T,\tag 3.6,b
\endalign
$$
Al sustituirlos en la acci\'on, \'esta queda expl\'{\i}citamente invariante 
de calibre (no depende de $a^L$)

$$
S^{(red)}_H=\frac{1}{2}<-(1-\lambda )^2\Delta a^T\Delta a^T+f^L\Delta f^L+
+\mu^2a^T\Delta a^T-2(1-\lambda )f^L\Delta \dot{a}^T>,\tag 3.7
$$
y definiendo

$$\align
Q&=(1-\lambda )(-\Delta )^{1/2}a^T,\tag 3.8,a\\
\pi &= (-\Delta )^{1/2}f^L,\tag 3.8,b
\endalign
$$
llegamos a 

$$
S^{(red)}_H=<\pi \dot{Q}-\frac{1}{2}\pi \pi -\frac{1}{2}Q(-\Delta +
(\frac{\mu}{(1-\lambda )})^2)Q>,\tag 3.9
$$
que es igual a $S^{(red)}_{TM}$ con masa $\mu /(1-\lambda )$.

Si $\lambda \to 1$ en (3.7), tendremos que la acci\'on resultante no aporta 
ninguna din\'amica a los campos. En la acci\'on (3.9) observamos que la 
energ\'{\i}a es definida positiva. Esto no depende del signo de 
${\mu \over (\lambda -1)}$ que s\'olo determina la helicidad de la excitaci\'on. La 
equivalencia con $S_{TM}$ se observa si en el proceso de reducci\'on 
cambiamos ${\mu \over (1-\lambda)}$ por $\mu$ y $(1-\lambda )a^T$ por $a^T$. La 
equivalencia con la autodual es inmediata.

\newpage

$\ $

\pageno=15
\headline={\ifnum\pageno=15\hfil\else\hss\tenrm \folio\ \fi}

\vskip 1cm

\centerline{\catorcerm Cap\'{\i}tulo {\docebf III}}

\vskip 1cm

\centerline{\dseisbf TEOR\'IAS DE SPIN 2 MASIVO}

\vskip 2cm

En dimensiones mayores que tres la teor\'{\i}a que cl\'asicamente se utiliza 
para describir al gravit\'on es la acci\'on de Einstein. En el caso de 
$D=2+1$ no sucede igual que con la acci\'on de Maxwell, ya que la acci\'on de 
Einstein en $2+1$ no tiene grados din\'amicos de libertad [40][41][42]. Sin 
embargo, podemos darle din\'amica sum\'andole distintos t\'erminos los cuales 
adem\'as de proporcionarle masa a la teor\'{\i}a se combinan para dar 
teor\'{\i}as de spin 2 puro.

En este cap\'{\i}tulo nos ocuparemos de revisar los modelos de spin 2 masivo 
(gravedad linealizada masiva) y la equivalencia entre ellos. Todas estas 
teor\'{\i}as corresponden a la din\'amica de un campo tensorial $h_{mn}$ que 
no es sim\'etrico en $m$ y $n$, ya que estamos pensando en la linealizaci\'on 
del dreibein, $e_m{}^a=\delta_m{}^a+kh_m{}^a$. La conexi\'on con el 
$h_{mn}^{(s)}$, sim\'etrico, que corresponde a linearizar la m\'etrica, 
$g_{mn}=\eta_{mn}+kh_{mn}^{(s)}$, es $h^{(s)}_{mn}=h_{mn}+h_{nm}$. Todas las 
teor\'{\i}as estar\'an en un fondo (background) plano. Observaremos que hay una
estrecha analog\'{\i}a entre las teor\'{\i}as de spin 1 y 2. En esta
analog\'{\i}a la teor\'{\i}a $TM$ linealizada no aparece como el an\'alogo a la
$TM$ vectorial. A esta \'ultima le corresponde la teor\'{\i}a de gravedad
masiva vectorial de $CS$ linealizada.

\newpage
\vskip 5mm
\noi
{\bf 1. Acci\'on de Fierz-Pauli y la teor\'{\i}a de spin 2 autodual}
\vskip 3mm
\noi
{\bf 1.1 La acci\'on de Fierz-Pauli y la condici\'on de autodualidad}
\vskip 3mm
La manera que usualmente se utiliza para proporcionar masa a la 
teor\'{\i}a de Einstein linealizada es sum\'andole el t\'ermino de 
Fierz-Pauli. La correspondiente acci\'on es la de Fierz-Pauli (FP)

$$
S_{FP}=S_E-\frac{m^2}{2}<h_{pa}h^{ap}-h_p{}^ph_a{}^a>,\tag 1.1
$$
donde $S_E$ es la acci\'on de Einstein linealizada\footnote"*"
{\ninerm{Ver ap\'endice A}}

$$
S_E=-\frac{1}{2}<-2h_{pa}\ep^{pmn}\partial_m\omega_n{}^a+\
\ep^a{}_{bc}\ep^{pmn}\eta_{pa}\omega_m{}^b\omega_n{}^c>.\tag 1.2
$$
Las ecuaciones de movimiento asociadas a $S_{FP}$ son

$$\align
\ep^{pmn}\partial_ph_{ma}+\delta_a{}^n\omega_m{}^m-\omega_a{}^n &=0,
\tag 1.3,a\\ 
\ep^{pmn}\partial_m\omega_n{}^a-m^2(h^{ap}-\eta^{pa}h_l{}^l) &=0.\tag 1.3,b
\endalign
$$
De (1.3,a) se obtiene la expresi\'on de $\omega_m{}^a$ en funci\'on de 
$h_m{}^a$ 

$$
\omega_m{}^a=-\frac{1}{2}\delta_m{}^a\ep^{pls}\partial_ph_{ls}+
\ep^{apl}\partial_ph_{lm},\tag 1.4
$$
que al sustituir en (1.3,b) nos da el sistema de ecuaciones para $h_{mn}$

$$
[(\ep^{pmb}\ep^{rsa}-\frac{1}{2}\ep^{pma}\ep^{rsb})\partial_m\partial_r-
m^2(\eta^{sa}\eta^{pb}-\eta^{pa}\eta^{sb})]h_{sb}=0.\tag 1.5
$$
Esto puede reescribirse como

$$
[-\frac{1}{2}(\ep^{pms}\ep^{arb}+\ep^{pmb}\ep^{ars})\partial_m\partial_r-
m^2(\eta^{sa}\eta^{pb}-\eta^{pa}\eta^{sb})]h_{sb}=0,\tag 1.6
$$
donde, notamos que, el primer t\'ermino es sim\'etrico en $p$ y $a$, y 
adem\'as s\'olo contribuye la parte sim\'etrica de $h_{sb}$. Esto debe ser 
as\'{\i}, ya que este t\'ermino representa al tensor de Einstein linealizado.

Utilizando los proyectores de las distintas partes irreducibles de 
$h_{sb}$\footnote"${}^\dagger$"{\ninerm{Ver ap\'endice B}}, escribimos la
ecuaci\'on  (1.6) como

$$
[(\sq -m^2)(P^2_S-P^0_S)-m^2(P^1_S-P_E^1-P_B^0-\sqrt{2}P^0_{SW}-
\sqrt{2}P^0_{WS})]h=0,\tag 1.7
$$
donde $Ph$, en cada caso, se sobreentiende que significa 
$P_{mn}{}^{ls}h_{ls}$.

Aplicando $P^1_S,P^1_E,P_B^0$ y $P^0_{SW}$, sobre (1.7), obtenemos que 

$$
P^1_Sh=P^1_Eh=P^0_Sh=P^0_Bh=0.\tag 1.8
$$
De (1.8) tenemos que $P_{WS}^0h=0$, lo que utilizamos al aplicar $P_{WS}^0$ 
sobre (1.7), y nos dice que

$$
P^0_Wh=0.\tag 1.9
$$
Nos queda, entonces, que la propagaci\'on es de spin 2 puro con masa $m$

$$
(\sq -m^2)P^2_Sh=0,\tag 1.10
$$
y, ya que $P_S^2=P_{+S}^2+P_{-S}^2$ tendremos que las 2 helicidades $+2$ y 
$-2$ se propagan en la teor\'{\i}a de Fierz-Pauli. Corroboramos, entonces, el 
hecho de que para tener una teor\'{\i}a que conserve $P$ y $T$ debemos tener 
las dos helicidades presentes con igual masa.

La ecuaci\'on (1.10) es la condici\'on que cumple la parte trans\-ver\-sa, 
sim\'etrica y sin traza, $h^{Tt}_{mn}$, de 
$h_{mn}$ (i.e. $h_{mn}^{Tt}=h_{nm}^{Tt}$, 
$\partial^mh_{mn}^{Tt}=\eta^{mn}h_{mn}^{Tt}=0$). Esta ecuaci\'on puede 
factorizarse como en el caso de la acci\'on de Proca

$$
(\sq -m^2)P_S^2h=[\sq^{1/2}(P^2_{+S}-P^2_{-S})-mP^2_S][-\sq^{1/2}(P^2_{+S}-
P^2_{-S})-mP^2_S]h, \tag 1.11
$$
y cuando $h_{mn}^{Tt}$ cumple una ecuaci\'on homog\'enea con alguno de estos 
factores, decimos que verifica la condici\'on de autodualidad

$$
[\pm \sq^{1/2}(P^2_{+S}-P^2_{-S})-mP^2_S]h=0.\tag 1.12
$$
Para el caso de spin 1 el problema era mas sencillo, y afirmabamos que la 
condici\'on de autodualidad constitu\'{\i}a la ``ra\'{\i}z'' de la ecuaci\'on 
de Proca. Aqu\'{\i} vemos que en todo caso es la ``ra\'{\i}z'' de la 
ecuaci\'on de Fierz-Pauli sobre la parte transversa, sim\'etrica y sin traza.

\vskip 5mm
\noi
{\bf 1.2 La acci\'on autodual}
\vskip 3mm
La acci\'on que conduce a la condici\'on (1.12) para $h_{mn}^{Tt}$ y que 
adem\'as describe una excitaci\'on masiva de spin 2 puro es la 
correspondiente a la teor\'{\i}a de spin 2 autodual [17,18,31]

$$
S^2_{AD}=\frac{m}{2}<h_{pa}\ep^{prs}\partial_rh_s{}^a-m(h_{pa}h^{ap}-
h_p{}^ph_a{}^a)>.\tag 1.13
$$
En (1.13), el signo de $m$, como veremos, determina la helicidad de la 
excitaci\'on. El primer t\'ermino de $S_{AD}$ se obtiene al linearizar el 
t\'ermino de $CS$ tri\'adico [43] (TCS) que presentaremos en el cap\'{\i}tulo 
siguiente, y el segundo t\'ermino es el de Fierz-Pauli usual.

Las ecuaciones de movimiento que surgen al hacer variaciones en $S_{AD}$ son

$$
\ep^{prs}\partial_rh_s{}^a-m(h^{ap}-\eta^{pa}h_l{}^l)=0.\tag 1.14
$$
En funci\'on de proyectores el t\'ermino $TCS$ linealizado se escribe como

$$\align
h_{pa}\ep^{prs}\partial_rh_s{}^a =& h_{pa}\sq^{1/2}[P^2_{+S}-P^2_{-S}-
P^0_{BS}-P^0_{SB} +\\
& +\frac{1}{2}(P^1_{+S}-P^1_{-S}+P^1_{+E}-P^1_{-E})+\\ 
& +\frac{1}{2}(P^1_{+SE}-P^1_{-SE}+P^1_{+ES}-P^1_{-ES})]^{pa,sb}h_{sb},
\tag 1.15
\endalign
$$
y el de $FP$ como

$$
(h_{pa}h^{ap}-h_p{}^ph_l{}^l)=[P^2_S+P^1_S-P_E^1-P_S^0-P_B^0-\sqrt{2}
(P^0_{SW}+P^0_{WS})]^{pa,sb}h_{sb}.\tag 1.16
$$
As\'{\i}, la ecuaci\'on (1.14) puede escribirse en t\'erminos de proyectores 
como

$$\align
[(&\sq^{1/2}-m)P^2_{+S}-(\sq^{1/2}+m)P^2_{-S}+\frac{1}{2}(\sq^{1/2}-2m)
(P^1_{+S}-P^1_{-E})\\
+&\frac{1}{2}(\sq^{1/2}+2m)(P^1_{+E}-P^1_{-S})+\frac{\sq^{1/2}}{2}
(P^1_{+SE}+P^1_{+ES}-P^1_{-SE}-P^1_{-ES})\\
-&\sq^{1/2}(P^0_{BS}+P^0_{SB})+m(P^0_S+P^0_B+\sqrt{2}(P^0_{WS}+P^0_{SW})]h=0.
\tag 1.17
\endalign
$$

Aplicamos $P_{\pm S}^1$ y $P^1_{\pm SE}$ a (1.17) y obtenemos

$$\align
(\pm \frac{1}{2}(\sq^{1/2}\mp 2m)P^1_{\pm S}\pm \frac{\sq^{1/2}}{2}
P^1_{\pm SE})h & =0, \tag 1.18,a\\ 
(\pm \frac{1}{2}(\sq^{1/2}\pm 2m)P^1_{\pm SE}\pm \frac{\sq^{1/2}}{2}
P^1_{\pm S})h & = 0, \tag 1.18,b
\endalign
$$
que al restarlas, nos dice que $(P^1_{\pm S}+P^1_{\pm SE})h=0$. Si 
sustituimos esto en (1.18,a) tendremos que

$$
P^1_{\pm S}h=0.\tag 1.19
$$
Hacemos un procedimiento an\'alogo aplicando $P^1_{\pm E}$ y $P^1_{\pm ES}$ 
sobre (1.17) y llegaremos a 

$$
P^1_{\pm E}h=0, \tag 1.20
$$
con lo que aseguramos que no hay propagaci\'on de las partes de spin 1. 
Aplicamos ahora $P^0_W$, $P^0_{SB}$, $P^0_S$ sobre (1.17), obteniendo 
respectivamente

$$\align
& \sqrt{2}mP^0_{WS} h  = 0,\tag 1.21,a\\ 
& (mP^0_{SB}-\sq^{1/2}P^0_S)h =0, \tag 1.21,b\\ 
& [m(P^0_S+\sqrt{2}P^0_{SW})-\sq^{1/2}P^0_{SB}]h=0.\tag 1.21,c
\endalign
$$
De (1.21,a) (con $P^0_{SW}$) es inmediato ver que $P^0_S h=0$. Siguiendo con 
(1.21,b) tendremos que $P^0_Bh=0$. Por \'ultimo con (1.21,c) concluimos que

$$
P^0_Wh=P^0_Bh=P^0_Sh=0.\tag 1.22
$$

Tenemos, entonces, que el sistema (1.17) es equivalente a 

$$
[(\sq^{1/2}-m)P^2_{+S}-(\sq^{1/2}+m)P^2_{-S}]h=0,\tag 1.23
$$
que es justamente la condici\'on de autodualidad para $h_{mn}^{Tt}$. Adem\'as 
(1.23) muestra que la excitaci\'on f\'{\i}sica de masa $m$ corresponde a la 
parte de helicidad +2. Si cambiamos $m$ por $-m$, se propagar\'a la parte de 
helicidad -2. Concluimos que la helicidad de la excitaci\'on es $2m/|m|$, 
an\'alogo al caso de spin 1. La positividad de la energ\'{\i}a puede verse si 
hacemos la decomposici\'on 2+1 de $S^2_{AD}$ y obtenemos la acci\'on reducida 
[18]. No presentamos esto aqu\'{\i} ya que en la siguiente subsecci\'on 
hacemos el an\'alisis can\'onico de $S^2_{AD}$ que nos servir\'a para 
estudiar su equivalencia con la acci\'on linealizada de la gravedad masiva 
vectorial de Chern-Simons (VCS).

\vskip 5mm
\noi
{\bf 1.3 An\'alisis can\'onico de la teor\'{\i}a de spin 2 autodual}
\vskip 3mm
Pasamos, ahora, al an\'alisis can\'onico de la teor\'{\i}a autodual. Esto 
ser\'a utilizado para mostrar la equivalencia entre esta teor\'{\i}a y la 
intermedia [44] que ser\'a presentada en la secci\'on 3. Hacemos, entonces, 
la decomposici\'on 2+1 de $S^2_{AD}$ llegando a la expresi\'on 
$(i,j,k\text{ y }l=1,2)$

$$\align
S^2_{AD} = \frac{m}{2}<& -2h_{00}(\ep_{ij}\partial_ih_{j0}+mh_{ii})+2h_{0k}
(\ep_{ij}\partial_ih_{jk}+mh_{k0})+\\
& -h_{ik}\ep_{ij}\dot{h}_{jk}+h_{i0}\ep_{ij}\dot{h}_{j0}+m(h_{ii}h_{jj}-h_{ij}
h_{ji})>.\tag 1.24
\endalign
$$
Ahora hacemos las redefiniciones

$$\align
n & = h_{00},\tag 1.25,a\\
N_i & = h_{i0},\tag 1.25,b\\ 
M_i & = h_{0i},\tag 1.25,c\\ 
H_{ij} & = \frac{1}{2}(h_{ij}+h_{ji}),\tag 1.25,d\\
V & = \frac{1}{2}\ep_{ij}h_{ij},\tag 1.25,e
\endalign
$$
donde hemos distinguido las partes sim\'etrica antisim\'etrica de $h_{ij}$. 
$S^2_{AD}$ toma la forma

$$\align
S^{(2)}_{AD} = \frac{m}{2}<&-2n(\ep_{ij}\partial_iN_j+H_{ii})+2M_k
(\ep_{ij}\partial_iH_{jk}+mN_k-\partial_kV)+\\ 
& -m(H_{ij}H_{ij}-H_{ii}H_{jj})-2\dot{H}_{ij}[\delta_{ij}V-\frac{1}{4}
(\ep_{ik}H_{kj}+\ep_{jk}H_{ki})]\\
&\hskip 3cm -\dot{N}_i\ep_{ij}N_j-2mVV>.\tag 1.26
\endalign
$$
$n$ y $M_k$ son multiplicadores asociados a los v\'{\i}nculos

$$\align
\psi & \equiv \ep_{ij}\partial_iN_j+mH_{ii},\tag 1.27,a\\ 
\psi_k & \equiv \ep_{ij}\partial_iH_{jk}+mN_k-\partial_kV.\tag 1.27,b
\endalign
$$
Debido a que $S^{2,l}_{AD}$ es de primer orden, la definici\'on de los momentos 
conjugados, asociados a las variables din\'amicas, no permite despejar ninguna 
de las velocidades. As\'{\i} que tenemos los v\'{\i}nculos primarios

$$\align
\varphi & \equiv \pi - \frac{\partial \Cal{L}}{\partial \dot{V}}=\pi ,
\tag 1.28,a\\ 
\varphi_i & = \pi_i-\frac{\partial \Cal{L}}{\partial \dot{N}_i}=\pi_i+
\frac{m}{2}\ep_{ij}N_j, \tag 1.28,b\\ 
\varphi_{ij} & = \pi_{ij}-\frac{\partial \Cal{L}}{\partial \dot{H}_{ij}}=
\pi_{ij}+m\delta_{ij}V-\frac{m}{4}(\ep_{ik}H_{kj}+\ep_{jk}H_{ki}). 
\tag 1.28,c
\endalign
$$

La densidad hamiltoniana sobre los v\'{\i}nculos es

$$
\Cal{H}_0=\frac{m^2}{2}(H_{ij}H_{ij}-H_{ii}H_{jj}+VV), \tag 1.29,a
$$
as\'{\i}, definimos el hamiltoniano can\'onico $H_c=\int dx^2\Cal{H}_c$, con

$$
\Cal{H}_c=\Cal{H}_0+\mu n\psi -\mu M_i\psi_i +\lambda\varphi 
+\lambda_i\varphi_i +\lambda_{ij}\varphi_{ij}. \tag 1.30
$$

Definimos los corchetes de Poisson a tiempos iguales entre las variables y 
sus momentos conjugados 

$$\align
\{ V(x), \pi(y)\} & \equiv \delta^{(2)}(\ovr{x}-\ovr{y}), \tag 1.31,a\\ 
\{ N_i(x),\pi_j(y)\} & \equiv \delta_{ij}\delta^{(2)}(\ovr{x}-\ovr{y}), 
\tag 1.31,b\\ 
\{ H_{ij}(x),\pi_{kl}(y)\} & \equiv \frac{1}{2}(\delta_{ik}\delta_{jl}+
\delta_{il}\delta_{jk})\delta^{(2)}(\ovr{x}-\ovr{y}), \tag 1.31,c
\endalign
$$
donde se sobreentiende que $x^p=(t,\ovr{x})$ y $y^p=(t,\ovr{y})$. Podemos, 
as\'{\i}, obtener los v\'{\i}nculos primarios

$$
\{ \Omega_A(x),\Omega_B(y)\} = M_{AB}(x)\delta^{(2)}(\ovr{x}-\ovr{y}), 
\tag 1.32,a
$$
donde

$$\align
\Omega_A(x) &=(\psi(x),\psi_i(x),\varphi (x),\varphi_i(x),\varphi_{ij}(x)),
\tag 1.32,b\\ 
\Omega_B(y) &=(\psi(y),\psi_k(y),\varphi (y),\varphi_k(y),\varphi_{kl}(x)).
\tag 1.32,c
\endalign
$$
$M_{AB}(x)$ tiene la forma 

$$
M_{AB}(x)=\left( \matrix
M_1^{(3\times 3)} & M_3^{(3\times 2)}\\
M_2^{(2\times 3)} & M_4^{(2\times 2)}
\endmatrix \right),\tag 1.33,a
$$
con

$$
M_1^{(3\times 3)}(x)=\left( \matrix
0 & 0 & 0 \\
0 & 0 & -\partial_i \\ 
0 & -\partial_k & 0  
\endmatrix \right),\tag 1.33,b
$$
\vskip 3mm

$$
M_2^{(2\times 3)}(x)=\left( \matrix
-\ep_{im}\partial_m & -m\delta_{ik} & 0 \\
-m\delta_{ij} &-\frac{1}{2}(\delta_{ik}\ep_{jm}\partial_m+
\delta_{jk}\ep_{im}\partial_m) & m\delta_{ij}
\endmatrix \right),\tag 1.33,c
$$

\vskip 3mm

$$
M_3^{(3\times 2)}(x)=\left( \matrix
-\ep_{km}\partial_m & m\delta_{kl}\\
m\partial_{ik} & -\frac{1}{2}(\delta_{il}\ep_{km}
\partial_m+\delta_{ik}\ep_{lm}\partial_m)\\ 
0 & -m\delta_{kl}
\endmatrix \right),\tag 1.33,d
$$

\vskip 3mm

$$
M_4^{(2\times 2)}(x)=\left( \matrix
m\ep_{ik} & 0 \\
0 & -\frac{m}{4}(\ep_{ik}\delta_{jl}+\ep_{il}\delta_{jk}+
\ep_{jk}\delta_{il}+\ep_{jl}\delta_{ik})
\endmatrix \right).\tag 1.33,e
$$
Miramos la conservaci\'on de los v\'{\i}nculos 

$$\align
\dot{\psi} &= \ep_{ij}\partial_i\lambda_j+m\lambda_{ii}, \tag 1.34,a\\ 
\dot{\psi}_i & = -\partial_i\lambda + m\lambda_i + \ep_{kl}\partial_k
\lambda_{li}, \tag 1.34,b\\ 
\dot{\varphi} & = -2m^2V+m(\partial_iM_i-\lambda_{ii}), \tag 1.34,c\\ 
\dot{\varphi}_i & = m^2M_i+m\ep_{ik}\lambda_k-m\ep_{ij}\partial_jn, 
\tag 1.34,d\\ 
\dot{\varphi} & = m^2 \delta_{ij}(H_{kk}-n)-m^2H_{ij}+\lambda m\delta_{ij}+
\frac{m}{2}(\ep_{jm}\partial_mM_i+\ep_{im}\partial_mM_j)+\\
&\ \ \ \ \ \ \ -\frac{m}{2}(\ep_{ik}\lambda_{kj}+\ep_{jk}\lambda_{ki}), 
\tag 1.34,e
\endalign
$$
donde pareciera que no hay v\'{\i}nculos adicionales. Sin embargo, si hacemos 
la combinaci\'on

$$
\theta = \varphi +\psi -\frac{1}{m}\partial_i\varphi_i,\tag 1.35
$$
resulta que $\dot{\theta}=-2m^2V$. $\theta$ conmuta con todos los v\'{\i}nculos 
primarios y, por tanto, surge el v\'{\i}nculo secundario

$$
\wt{\theta}=V. \tag 1.36
$$
La conservaci\'on de $\wt{\theta}$ no proporciona ning\'un v\'{\i}nculo adicional. La 
consistencia $\dot{\wt{\theta}}=0$ implica que $\lambda =0$. As\'{\i} que podemos 
eliminar a $V$ y $\pi$ de la teor\'{\i}a.

Tomamos a $\theta$ por $\psi$. El sistema, finalmente, estar\'a descrito por 

$$
\Cal{H}_0 = \frac{m^2}{2}(H_{ij}H_{ij}-H_{ii}H_{jj}), \tag 1.37,a
$$
sometido a los v\'{\i}nculos

$$\align
\theta & = -\frac{1}{m}\partial_i\pi_i+mH_{ii}+\frac{1}{2}\ep_{ij}
\partial_iN_j, \tag 1.37,b\\ 
\varphi_i & = \pi_i+\frac{m}{2}\ep_{ij}N_j, \tag 1.37,c\\ 
\psi_i & = \ep_{kl}\partial_kH_{li}-mN_i, \tag 1.37,d\\
\varphi_{ij} &= \pi_{ij}-\frac{m}{4}(\ep_{ik}H_{kj}+\ep_{jk}H_{ki}), 
\tag 1.37,e
\endalign
$$
donde ya hemos eliminado a $\pi$ y $V$. El conteo de grados de libertad nos 
da correcto: 10 variables $(N_i,\pi_i,H_{ij},\pi_{ij})$ sometidas a 8
v\'{\i}nculos  $(\theta ,\varphi_i,\psi_i,\varphi_{ij})$ quedando 2 grados de
libertad  correspondientes a la \'unica variable din\'amica, mas su momento
conjugado.

\vskip 5mm
\noi
{\bf 2. La acci\'on de gravedad Topol\'ogica Masiva, linealizada}
\vskip 3mm
La aci\'on de gravedad TM linealizada es

$$
S^{2,l}_{TM}=\frac{1}{4\mu}<-\ep^{lmr}\partial_lh_{mp}^{(s)}G^{pa}(h^{(s)})
\eta_{ra}+\mu h_{pa}^{(s)}G^{pa}(h^{(s)})>,\tag 2.1
$$
donde el tensor de Einstein linealizado\footnote"${}^\star$"
{\ninerm{Ver ap\'endice A}}

$$
G^{pa}(h^{(s)})=-\frac{1}{2}\ep^{prs}\ep^{alp}\partial_r\partial_l
h^{(s)}_{sb}, \tag 2.2
$$
ha sido escrito en forma conveniente para los c\'alculos. Observamos que el 
signo del t\'ermino correspondiente a la acci\'on de Einstein es contrario al 
que tendr\'{\i}a en la acci\'on $S_E$ (ver 1.2) si sustituyeramos (1.4). 
$S^{2,l}_{TM}$ es invariante bajo las transformaciones de calibre

$$
\delta h_s{}^b=\partial_s\xi^b\ \ ;\ \ 
\delta h^{(s)}_{pm}=\partial_p\xi_m+\partial_m\xi_p. \tag 2.3
$$

Reescribimos, ahora, $S^{2,l}_{TM}$ en funci\'on de los proyectores de las 
distintas partes de $h_s{}^b$

$$
S^{2,l}_{TM}=\frac{1}{8\mu}<h^{(s)}\sq \ \ \sq^{1/2}(P^2_{+S}-P^2_{-S})h^{(s)}
-\mu h^{(s)}\sq (P^2_S-P^0_S)h^{(s)}>.\tag 2.4,a
$$
As\'{\i}, las ecuaciones de movimiento de esta acci\'on son

$$
\frac{1}{\mu}[\sq \ \ \sq^{1/2}(P^2_{+S}-P^2_{-S})h^{(s)}-\mu \sq 
(P^2_S-P^0_S)h^{(s)}]=0. \tag 2.4,b
$$

Observamos que de (2.4,b) $P_W^0h^{(s)}=P^1_Sh^{(s)}=0$ y 

$$\align
\sq P^0_S h^{(s)} & = 0,\tag 2.5,a\\ 
\frac{1}{\mu}\sq (\sq^{1/2}-\mu )P^2_{+S}h^{(s)} & = 0, \tag 2.5,b\\ 
-\frac{1}{\mu}\sq (\sq^{1/2}+\mu )P^2_{-S}h^{(s)} & = 0, \tag 2.5,c
\endalign
$$
de donde no podemos concluir, inmediatamente, que no hayan excitaciones de 
masa cero.

En el calibre arm\'onico

$$
\partial_mh^{mn}=0,\tag 2.6
$$
i.e. $h^{(s)}=(P^2_S+P^0_S)h^{(s)}$, el inverso del operador diferencial de 
las ecuaciones de mo\-vi\-mien\-to es

$$
{\bold \bigt}=-\frac{1}{\sq}(P^0_S+\frac{\mu}{\sq^{1/2}-\mu}P^2_{+S}-
\frac{\mu}{\sq^{1/2}+\mu}P^2_{-S}),\tag 2.7
$$

donde ${\bold \bigt}[\sq^{3/2}(P^2_{+S}-P^2_{-S})-\mu\sq (P^2_S-P^0_S)]=
-\mu (P^2_S+P^0_S)$. Podemos reescribir $\bigt$ como

$$
{\bold \bigt}=\frac{1}{\sq}(P^2_S-P^0_S)-\frac{1}{\sq^{1/2}}
(\frac{1}{\sq^{1/2}-\mu}P^2_{+S}+\frac{1}{\sq^{1/2}+\mu}P^2_{-S}).\tag 2.8
$$
Los proyectores $P^2_{\pm S}$ pueden escribirse, convenientemente, como

$$\align
P^2_{\pm S}{}_{mn}{}^{ls} & = \frac{1}{2}[P^2_{S}{}_{mn}{}^{ls}\pm \frac{1}{4}
(P_m{}^l\xi_m{}^s+P_n{}^l\xi_m{}^s+P_m{}^s\xi_m{}^l+P_n{}^s\xi_m{}^l)]\\
& \equiv \frac{1}{2} [P^2_S\pm \frac{1}{4}``P\xi\text{''}],\tag 2.9
\endalign
$$
de esta forma llegamos a la forma final de ${\bold \bigt}$

$$
{\bold \bigt}=\frac{1}{\sq}(P^2_S-P^0_S)-\frac{1}{\sq -\mu^2}P^2_S-
\frac{1}{4(\sq -\mu^2)}\frac{\mu}{\sq^{1/2}}``P\xi\text{''}.\tag 2.10
$$
En componentes

$$\align
\bigt_{mn}{}^{ls} = & \frac{1}{2\sq}(P_m{}^lP_n{}^s+P_m{}^sP_n{}^l-
2P_{mn}P^{ls})+\\ 
- & \frac{1}{2(\sq -\mu^2)}(P_m{}^lP_n{}^s+P_m{}^sP_n{}^l-P_{mn}P^{ls})+\\
- & \frac{\mu}{4(\sq -\mu^2)\sq}(P_m{}^l\ep_n{}^{rs}+P_n{}^l\ep_m{}^{rs}+
P_m{}^s\ep_n{}^{rl}+P_n{}^s\ep_m{}^{rl})\partial_r, \tag 2.11
\endalign
$$
donde el primer t\'ermino es justamente el propagador de Einstein con el 
signo opuesto [40]. Cuando acoplamos con una fuente externa conservada (i.e. 
$T^{mn}$ con $\partial_mT^{mn}=0$), al despejar $h^{(s)}_{mn}$ en funci\'on 
de $T^{mn}$ y sustituirlo en la acci\'on, la contribuci\'on de este t\'ermino 
es $\sim <T^{mn}\frac{1}{\sq}T_{mn}-T^m{}_m\frac{1}{\sq}T^n{}_n>$. Esta 
contribuci\'on es de contacto. Esto puede verse si descomponemos $T^{mn}$ en 
sus com\-po\-nen\-tes independientes 
$T^{00},T^{0i}=\ep_{ij}\partial_jT^T+\partial_i((-\Delta )^{-1}
\dot{T}^{00})$ y $T^{ij}=(\delta_{ij}\Delta -\partial_i
\partial_j)T+\partial_i\partial_j(-\Delta )^{-2}\ddot{T}^{00}+
(\ep_{ik}\partial_k\partial_j+\ep_{jk}\partial_k\partial_i)(-\Delta )
\dot{T}^T$, en cuyo caso

$$
<T^{mn}\frac{1}{\sq}T_{mn}-T^m{}_m\frac{1}{\sq}T^n{}_n>\sim 
<T^TT^T+T^{00}T>.\tag 2.12
$$
Vemos, ahora, que no hay polos de masa cero en el propagador. El signo de 
la acci\'on de Einstein cobra relevancia cuando miramos la positividad de 
la energ\'{\i}a.

Hacemos la descomposici\'on 2+1 de $S^{2,l}_{TM}$, empezando con (2.1) en 
funci\'on de $h^{(s)}_{mn}$ es explicitamente as\'{\i}

$$\align
S^{2,l}_{TM}=-\frac{1}{8\mu}<(\sq h^{(s)n}{}_p  -\partial_t&\partial^n
h^{(s)t}{}_p)\ep^{psq}\partial_sh_{qn}^{(s)}+\\ 
&+\mu \partial_mh_{pa}^{(s)}\ep^{mpr}\ep^{als}\partial_lh_{sr}^{(s)}>. 
\tag 2.13
\endalign
$$
En un primer paso llegamos a 

$$\align
S^{2,l}_{TM}=-\frac{1}{8\mu}<& 2\ep_{ij}\partial_ih_{j0}^{(s)}\partial_k
\dot{h}_{k0}^{(s)}+\Delta h_{0i}^{(s)}\ep_{ij}h_{j0}^{(s)}-\Delta 
h_{ik}^{(s)}\ep_{ij}\dot{h}^{(s)}_{jk}+\\ 
& -\partial_lh_{li}^{(s)}\ep_{ij}\partial_kh^{(s)}_{kj}-2\partial_k
\dot{h}_{ki}^{(s)}\ep_{ij}\dot{h}^{(s)}_{j0}-\mu \dot{h}^{(s)}_{ii}
h^{(s)}_{jj}+\\ 
& +2\ep_{ij}\partial_i\dot{h}^{(s)}_{jk}\dot{h}^{(s)}_{k0}+\ep_{ij}\partial_i
h^{(s)}_{jk}\partial_k\dot{h}_{00}^{(s)}+\mu\dot{h}^{(s)}_{ij}\dot{h}^{(s)}_{ij}+\\ 
& +4\mu\dot{h}^{(s)}_{i0}(\partial_ih_{jj}^{(s)}-\partial_jh_{ij}^{(s)})+
\ddot{h}^{(s)}_{ik}\ep_{ij}\dot{h}_{jk}^{(s)}+\\
& +2\ep_{ij}\partial_i h^{(s)}_{jk}(\Delta
h^{(s)}_{k0}-\partial_k\partial_lh^{(s)}_{l0})- 2\mu h_{i0}^{(s)}\Delta
h_{i0}^{(s)}+\\  
& +2\mu h_{00}[\Delta 
h_{ii}^{(s)}-\partial_i\partial_jh_{ij}^{(s)}-\frac{\Delta}{\mu}\ep_{ij}
\partial_ih_{j0}^{(s)}]+\\
&- 2\mu \partial_ih_{i0}^{(s)}\partial_j^nh^{(s)}_{j0}>.
\tag 2.14 
\endalign
$$

Hacemos la descomposici\'on

$$
\align
h_{00}^{(s)} & =2n, \tag 2.15,a\\
h^{(s)}_{0i} & =2\ep_{ij}\partial_jn^T+2\partial_in^L,\tag 2.15,b\\
h^{(s)}_{ij} & =2(\delta_{ij}\Delta -\partial_i\partial_j)h^T+
2\partial_i\partial_jh^L+2(\ep_{ij}\partial_k\partial_j+\ep_{jk}\partial_k
\partial_i)h^{TL},\tag 2.15,c
\endalign
$$
donde los factores de 2 surgen al aplicar la equivalencia entre 
$h^{(s)}_{mn}$ y $h_{mn}$ no sim\'etrico, cuya parte antisim\'etrica (de la 
forma $\sim \ep_{mnl}V^l$) no aparece en (2.15). Las componentes de 
$h_{mn}^{(s)}$ ante transformaciones de calibre, con par\'ametros 
$\xi_m\ \ \ (\xi_0=\xi ,\xi_i=\ep_{ij}\partial_j\xi^T+\partial_i\xi^L)$, 
cambian como

$$\align
\delta h^T=0\ \ &,\ \ \delta h^{TL}=\frac{1}{2}\xi^T\ \ ,\ \ 
\delta h^L=\xi^L,\\ 
\delta n=\dot{\xi}\ \ &,\ \ \delta n^T=\frac{1}{2}\dot{\xi}^T\ \ ,\ \ 
\delta n^L = \frac{1}{2}(\xi +\dot{\xi}^L),\tag 2.16
\endalign
$$
donde observamos que $h^T$ es la \'unica componente de $h^{(s)}_{mn}$ 
invariante de calibre. Sin embargo, algunas combinaciones de ellas 
podr\'{\i}an, tambi\'en, ser invariantes.

$S^{2,l}_{TM}$ en funci\'on de las distintas componentes de $h^{(s)}_{mn}$ 
queda como

$$\align
S^{2,l}_{TM}=-\frac{1}{\mu}<\Delta n &[\mu\Delta h^T+\Delta (n^T-
\dot{h}^{TL})]+\\ 
&+\mu\Delta (n^T-\dot{h}^{TL})\Delta (n^T-\dot{h}^{TL})+\\ 
&-2\Delta \dot{n}^L\Delta (n^T-\dot{h}^{TL})+\Delta^2h^{TL}\Delta (n^T-
\dot{h}^{TL})+\\ 
&-2\mu\Delta \dot{n}^L\Delta h^T-\Delta \ddot{h}^T\Delta (n^T-\dot{h}^{TL})+\\ 
&+\Delta \ddot{h}^L\Delta (n^T-\dot{h}^{TL})-\mu\Delta \dot{h}^{TL}\Delta 
\dot{h}^L>,\tag 2.17
\endalign
$$
donde hemos destacado la combinaci\'on invariante de ca\-li\-bre $n^T-\dot{h}^{TL}$. 
Ha\-ce\-mos la sustituci\'on

$$
N=n^T-\dot{h}^{TL},\tag 2.18
$$
que es equivalente a hacer una fijaci\'on, parcial, de calibre, ya que 
corresponder\'{\i}a a tomar $h^{TL}=0$. De esta forma $n$ aparece como un 
multiplicador de Lagrange asociado al v\'{\i}nculo

$$
\mu \Delta h^T+\Delta N=0,\tag 2.19
$$

que permite despejar $N$ en funci\'on de $h^T$. El rol de $n$, como 
multiplicador, esta acorde con la forma como transforma bajo cambios de 
calibre.
Luego de sustituir (2.18) y (2.19) en (2.17) llegamos a la forma reducida 
final de $S^{2,l}_{TM}$

$$
S^{2,l(red)}_{TM}=<\Delta \dot{h}^T\Delta \dot{h}^T+\Delta h^T
(\Delta -\mu^2)\Delta h^T>,\tag 2.20
$$
donde no aparecen $h^L$ y $n^L$. Estos pueden fijarse con el calibre 
residual que tenemos. En (2.20) observamos que esta acci\'on co\-rres\-pon\-de a la de 
un s\'olo grado din\'amico de libertad con masa $\mu$ y energ\'{\i}a definida positiva. 
Este resultado no depende del signo de $\mu$ en (2.1) y s\'{\i} del signo de la 
acci\'on de Einstein, ya que cambiarle el signo a $S_E$ equivale a multiplicar 
$S^{2,l}_{TM}$ por -1 y cambiar $\mu$ por $-\mu$.

\vskip 5mm
\noi
{\bf 3. La acci\'on intermedia, o la gravedad masiva Vectorial de 
Chern-Simons linealizada}
\vskip 3mm
\noi
{\bf 3.1 An\'alisis covariante}
\vskip 3mm
Una manera alterna de describir una teor\'{\i}a de spin 2 masivo, se logra 
con la acci\'on intermedia [17,18]

$$
S^l_{VCS}=\frac{1}{2}<2h_{pa}\ep^{pmn}\partial_m\omega_n{}^a-\ep^a{}_{bc}
\ep^{pmn}\eta_{pa}\omega_m{}^b\omega_n{}^c-\mu h_{pa}\ep^{prs}\partial_r
h_s{}^a>,\tag 3.1
$$
donde el sub\'{\i}ndice $VCS$ proviene del hecho que es la linealizaci\'on de 
la teor\'{\i}a curva llamada gravedad masiva vectorial de Chern-Simons, 
introducida posteriormante [43] a la presentaci\'on de la acci\'on intermedia 
[17,18]. Esta acci\'on de spin 2 masivo fu\'e con\-si\-de\-ra\-da inicialmente como una 
acci\'on ``intermedia" entre la acci\'on ``maestra" de tercer orden que es 
equivalente a $S^2_{AD}$ y $S^{2,l}_{TM}$ [18].

$S^l_{VCS}$ es invariante bajo las transformaciones de calibre

$$
\delta h_{mn}=\partial_m \xi_n,\tag 3.2
$$
al igual que la acci\'on $TM$ linealizada. 

Las ecuaciones de movimiento de $S^l_{VCS}$ son 

$$\align
\ep^{pmn}\partial_m\omega_n{}^a &-\mu\ep^{prs}\partial_rh_s{}^a=0,
\tag 3.3,a\\ 
\ep^{pmn}\partial_mh_n{}^a &-\omega^{ap}+\eta^{ap}\omega_l{}^l=0,\tag 3.3,b
\endalign
$$
donde (3.3,b) al igual que (1.3,a) permite despejar $\omega_p{}^a$ en 
funci\'on de $h_p{}^a$. Con esto conseguimos el sistema de segundo orden, que 
satisface $h_p{}^a$

$$
(\frac{1}{2}\ep^{pma}\ep^{srb}-\ep^{pmb}\ep^{sra})\partial_m\partial_rh_{sb}-
\mu \ep^{pmn}\partial_mh_n{}^a=0.\tag 3.4
$$
Podemos ver cual es el espectro f\'{\i}sico de la teor\'{\i}a si escribimos 
(3.4) en funci\'on de los proyectores de las distintas partes de $h_n^b$

$$\align
[\sq (P^2_S-P^0_S) &-\mu \sq^{1/2}(P^2_{+S}-P^2_{-S}-P^0_{BS}-P^0_{SB})+\\ 
-\frac{\mu}{2}\sq^{1/2}&(P^1_{+S}-P^1_{-S}+P^1_{+SE}-P^1_{-SE}+P^1_{+E}+\\ 
&\ \ \ \ \ \ \ \ \ \ \ \ \ \ \ -P^1_{-E}+P^1_{+ES}-P^1_{-ES})]h=0, \tag 3.5
\endalign
$$
\noi
y escogemos el calibre arm\'onico $\partial_mh^{mn}=0$ que es equivalente 
a\footnote"*"{\ninerm{Ver ap\'endice B}}

$$\align
h^T & = Th\\
&=[P^2_S+P^0_S+P_B^0+\frac{1}{2}(P^1_{SE}+P^1_{ES}+P^1_S+P^1_E)]h,\tag 3.6,a
\endalign
$$
o

$$
[P^0_W+\frac{1}{2}(P^1_S+P^1_E-P^1_{SE}-P^1_{ES})]h=0.\tag 3.6,b
$$
Es inmediato notar que $P^0_S\wt{h}=P^0_B\wt{h}=P^1_S\wt{h}=P^1_E\wt{h}=0$, 
donde $\wt{h}=\sq^{1/2}h$, y que 

$$
(\sq^{1/2}\mp \mu )P^2_{\pm S}\wt{h}=0,\tag 3.7
$$
donde se muestra que $\wt{h}$ representa una excitaci\'on masiva de masa $\mu$ 
y helicidad +2. Si cambiamos $\mu$ por $-\mu$ la helicidad ser\'a -2. As\'{\i} 
$S^l_{VCS}$ describe una excitaci\'on masiva de helicidad $2\mu /|\mu |$.

\vskip 5mm
\noi
{\bf 3.2 Descomposici\'on 2+1 y la forma invariante de calibre $S^l_{VCS}$}
\vskip 3mm

Verifiquemos lo heur\'{\i}sticamente encontrado haciendo la des\-com\-po\-si\-ci\'on 
2+1 de $S^l_{VCS}$. 
En un primer paso tendremos

$$\align
S^l_{VCS}=\frac{1}{2}< &-2h_{00}[\ep_{ij}\partial_i\omega_{j0}-\mu \ep_{ij}
\partial_ih_{j0}]+\\
&-2\omega_{00}[\omega_{ii}+\ep_{ij}\partial_ih_{j0}]\\ 
&+h_{0k}[\ep_{ij}\partial_i\omega_{jk}-\mu\ep_{ij}\partial_ih_{jk}]+\\ 
&+2\omega_{0k}[\omega_{k0}+\ep_{ij}\partial_ih_{jk}]+2h_{i0}\ep_{ij}
\dot{\omega}_{j0}+\\ 
&-\mu h_{i0}\ep_{ij}\dot{h}_{j0}-2h_{ik}\ep_{ij}\dot{\omega}_{jk}+
\mu h_{ik}\ep_{ij}h_{jk}+\\
&+\omega_{ii}\omega_{jj}-\omega_{ij}\omega_{ji}>,\tag 3.8
\endalign
$$
donde hemos partido de la expresi\'on de primer orden, pues, facilita el 
proceso de reducci\'on. Observamos que $h_{00}, h_{0k}$ son multiplicadores 
de Lagrange (lo que es de esperarse, pues bajo cambios de calibre transforman 
como $\delta h_{00}=\dot{\xi}_0$, $\delta h_{0k}=\dot{\xi}_k$); as\'{\i} como 
$\omega_{00}$ y $\omega_{0k}$.

Hacemos la descomposici\'on para 

$$\align
h_{00} &=n,\tag 3.9,a\\ 
h_{i0} &=\ep_{ij}\partial_j(n^T+v^L)+\partial_i(n^L-v^T),\tag 3.9,b\\
h_{0i} &=\ep_{ij}\partial_j(n^T-v^L)+\partial_i(n^L+v^T),\tag 3.9,c\\ 
h_{ij} &=(\delta_{ij}\Delta -\partial_i\partial_j)h^T+\partial_i\partial_j
h^L+\\
&\quad\quad\quad\quad\quad+(\ep_{ik}\partial_k\partial_j+\ep_{jk}\partial_k\partial_i)h^{TL}+\ep_{ij}
V,\tag 3.9,d\\ 
\omega_{00} &=\gamma ,\tag 3.9,e\\  
\omega_{i0} &=\ep_{ij}\partial_j(\gamma^T+\lambda^L)+
\partial_i(\gamma^L-\lambda^T),\tag 3.9,f\\ 
\omega_{0i} &=\ep_{ij}\partial_j(\gamma^T-\lambda^L)+
\partial_i(\gamma^L+\lambda^T),\tag 3.9,g\\  
\omega_{ij} &=(\delta_{ij}\Delta -\partial_i\partial_j)
\omega^T+\partial_i\partial_j\omega^L+\\ 
&\quad \quad \quad \quad \quad +(\ep_{ik}\partial_k\partial_j+\ep_{jk}\partial_k\partial_i)
\omega^{TL}+\ep_{ij}\lambda ,\tag 3.9,h
\endalign
$$ 
Con estas definiciones resolvemos los v\'{\i}nculos asociados a $h_{00}$, 
$h_{0k}$, $\omega_{00}$ y $\omega_{0k}$, obteniendo

$$\align 
\Delta\omega^T &=\mu\Delta h^T,\tag 3.10,a\\ 
\Delta\omega^L &=-\frac{\Delta (\Delta -\mu^2)}{\mu}h^T,\tag 3.10,b\\ 
\Delta\omega^{TL}+\lambda &=\mu (\Delta h^{TL}+V),\tag 3.10,c\\
\Delta (n^T+v^L) &=\frac{\Delta^2}{\mu} h^T,\tag 3.10,d\\ 
\omega_{i0} &=\ep_{ij}\partial_j\Delta h^T+\partial_i(\Delta h^{TL}+V).
\tag 3.10,e
\endalign
$$
Al sustituir la soluci\'on de los v\'{\i}nculos en (3.8), llegamos, casi, a 
la forma final de la acci\'on re\-du\-ci\-da

$$\align
S^l_{VCS} &=<2\Delta \dot{h}^T\Delta\omega^{TL}+\Delta h^T(\Delta -\mu^2)
\Delta h^T+\mu^2\Delta h^{TL}\Delta h^{TL}+\\ 
&\ \ \ \ \ +\mu^2VV+2\mu^2V\Delta h^{TL}-2\mu \Delta h^{TL}\Delta \omega^{TL}-
2\mu\Delta \omega^{TL}V>\\ 
&=<2\Delta\dot{h}^T\Delta\omega^{TL}+\Delta h^T(\Delta -\mu^2)\Delta h^T-
2\mu\Delta\omega^{TL}(V+\Delta h^{TL})+\\ 
&\ \ \ \ \ +\mu^2(\Delta h^{TL}+V)(\Delta h^{TL}+V)>,\tag 3.11
\endalign
$$
donde observamos que todav\'{\i}a hay un v\'{\i}nculo cuadr\'atico asociado a 
$\Delta h^{TL}+V$, cuya soluci\'on es

$$
\Delta h^{TL}+V=\frac{\Delta}{\mu}\omega^{TL}.\tag 3.12
$$
sustituyendo (3.12) en (3.11) llegamos a la forma final de $S^l_{VCS}$

$$
S^{l\ (Red)}_{VCS}=<2\Delta \dot{h}^T\Delta\omega^{TL}+\Delta h^T
(\Delta -\mu^2)\Delta h^T-\Delta \omega^{TL}\Delta\omega^{TL}>.\tag 3.13
$$

Definimos

$$\align
Q &=\sqrt{2}(-\Delta )h^T,\tag 3.14,a\\ 
\pi & =\sqrt{2}(-\Delta )\omega^{TL},\tag 3.14,b
\endalign
$$
y $S^{l(red)}_{VCS}$ se escribe como

$$
S^{l\ (red)}_{VCS}=<\pi \dot{Q}-\frac{1}{2}Q(-\Delta +\mu^2)Q-\frac{1}{2}
\pi\pi >,\tag 3.15
$$
que, justamente, muestra que $S^l_{VCS}$ describe una s\'ola excitaci\'on masiva, 
con masa $\mu$, y la teor\'{\i}a tiene energ\'{\i}a definida positiva, independiente del 
signo de $\mu$.

Observemos que luego de resolver los v\'{\i}nculos asociados a $\omega_{00}$, 
$\omega_{0k}$, $h_{00}$ y $h_{0k}$, quedan indeterminados entre otras 
variables $n^L-v^T$ y $h^L$ y que, luego de sustituir la soluci\'on de ellos, 
estas dos variables ya no aparecen. Miremos como cambian las distintas 
componentes de $h_{mn}$ 
$(\xi_m=(\xi ,\xi_i=\ep_{ij}\partial_j\xi^T+\partial_i\xi^L))$

$$\align
\delta h_{00}=\dot{\xi}\ \ &,\ \ \delta h_{0i}=\dot{\xi}_i,\tag 3.16,a\\ 
\delta (n^T+v^L)=0\ \ &,\ \ \delta (n^L-v^T)=\xi ,\tag 3.16,b\\
\delta h^T=0\ \ &,\ \ \delta h^L=\xi^L, \tag 3.16,c\\
\delta h^{TL}=\frac{1}{2}\xi^T\ \ &,\ \ \delta V=-\frac{1}{2}\delta\xi^T.
\tag 3.16,d
\endalign
$$
Observamos que $h_{00}$, $h_{0i}$ transforman acorde a su rol. Luego de 
sustituir (3.10), la acci\'on (3.11) todav\'{\i}a es invariante de calibre y 
se expresa s\'olo en funci\'on de cantidades invariantes de calibre (ver que 
$\delta_\xi (\Delta h^{TL}$+V)=0), ya que no fu\'e necesario explotar la 
invariancia de calibre para resolver los v\'{\i}nculos. As\'{\i}, estas 
variables que ``desaparecen" son fijadas libremente y $S^{l(red)}_{VCS}$ no 
depende del calibre que fijemos en su reducci\'on. Esto puede verse si 
partimos de la acci\'on de segundo orden de $S^l_{VCS}$, que escribimos como

$$
S^l_{VCS}=-\frac{1}{2}<h_{pa}(G^{pa}+\mu \ep^{prs}\partial_rh_s{}^a)>, 
\tag 3.17
$$
donde observamos que el factor que multiplica a $h_{pa}$ es trans\-ver\-so en $p$ 
e in\-va\-rian\-te de ca\-li\-bre, lo cual asegura que $S^l_{VCS}$ es invariante 
de calibre pues cambia como una derivada total. Si de\-fi\-ni\-mos 
$W^{pa}\equiv \ep^{prs}\partial_rh_s{}^a$, tenemos que

$$\align
W_{00} &=\Delta (n^T+v^L),\tag 3.18,a\\ 
W_{0i} &=\ep_{ik}\partial_k\Delta h^T+\partial_i(\Delta h^{TL}+V),
\tag 3.18,b\\ 
W_{i0} &=\ep_{ik}\partial_k(n-\dot{n}^L+\dot{v}^T)+\partial_i
(\dot{n}^T+\dot{v}^L),\tag 3.18,c\\ 
W_{ij} &=(\delta_{ij}\Delta -\partial_i\partial_j)(n^T-v^L-\dot{h}^{TL}-
\frac{1}{(-\Delta )}\dot{V})+\\ 
&\ \ \ \ \ +\partial_i\partial_j(\dot{h}^{TL}-\frac{1}{(-\Delta )}\dot{V})+\\
&\ \ \ \ \ +\frac{1}{2}(\ep_{ij}\partial_k\partial_j+\ep_{jk}\partial_k
\partial_i)(n^L+V^T-\dot{h}^L+\dot{h}^T)+\\ 
&\ \ \ \ \ +\frac{1}{2}\ep_{ij}(n^L+v^T-\dot{h}^L-\dot{h}^T),\tag 3.18,d\\
G_{00} &=-\Delta^2h^T,\tag 3.18,e\\ 
G_{0i} &=G_{i0}=\ep_{ij}\partial_j(\Delta \dot{h}^{TL}-\Delta n^T)-\partial_i
\Delta \dot{h}^T,\tag 3.18,f\\
G_{ij} &=(\delta_{ij}\Delta -\partial_i\partial_j)(2\dot{n}^L-n-\ddot{h}^L)-
\partial_i\partial_j\ddot{h}^T+\\ 
&\ \ \ \ \ +(\ep_{ik}\partial_k\partial_j+\ep_{jk}\partial_k\partial_i)
(\ddot{h}^{TL}-\dot{n}^T).\tag 3.18,g
\endalign
$$
As\'{\i}, expresamos $S^l_{VCS}$ como

$$\align
S^l_{VCS} = < &\Delta h^T\Delta (n-2\dot{n}^L+\ddot{h}^L)+\Delta (n^T-
\dot{h}^{TL})\Delta (n^T-\dot{h}^{TL})+\\ 
&+\mu\Delta (n^T+v^L)(\dot{n}^L-\dot{v}^T-n)+\\
&+\mu\Delta (\Delta h^{TL}+V)(\dot{h}^L-v^T-n^L)+\\
&+\mu\Delta h^T(\Delta h^{TL}+\dot{V}-\Delta (n^T-v^T))>\tag 3.19
\endalign
$$
que podemos escribir de forma mas sugestiva como

$$\align
S^l_{VCS}=< &\rho (\Delta h^T-\mu (n^T+v^L))+\sigma (\mu (\Delta h^{TL}+V)-
\Delta \dot{h}^T)+\theta^2+\\ 
&-\mu \Delta h^T[(\Delta \dot{h}^{TL}+\dot{V})+\mu \Delta h^T-\Delta 
(n^T+v^L)]>,\tag 3.20
\endalign
$$
donde $\rho ,\sigma$ y $\theta$ son las combinaciones invariantes de calibre

$$\align
\rho &=\Delta (n-(\dot{n}^L-\dot{v}^T)),\tag 3.21,a\\ 
\sigma &=\Delta (\dot{h}^L-(n^L+v^T)),\tag 3.21,b\\ 
\theta &=\frac{\Delta}{2}((n^T-v^L)-2\dot{h}^{TL}+(n^T+v^L)-2\mu h^T)\\ 
&=\Delta (n^T-\dot{h}^{TL}-\mu h^T),\tag 3.21,c
\endalign
$$
Haciendo variaciones respecto a $\rho ,\sigma$ y $\theta$, llegamos a 

$$
S^l_{VCS}=<\Delta h^T(\sq -\mu^2)\Delta h^T>,\tag 3.22
$$ 
que corresponde a un s\'olo grado de libertad con energ\'{\i}a definida 
positiva.

\vskip 5mm
\noi
{\bf 3.3 An\'alisis can\'onico de $S^l_{VCS}$}
\vskip 3mm
Hacemos, ahora el an\'alisis can\'onico de la acci\'on linealizada de la gravedad 
$VCS$. Esto ser\'a utilizado para estudiar la equivalencia can\'onica entre 
$S^l_{VCS}$ y $S^2_{AD}$ [44]. Partimos de la forma de $S^l_{VCS}$ a 
segundo orden

$$\align
S^l_{VCS}=\frac{1}{2}<\ep^{pmb}\partial_ph_{ma}\ep^{rsa}\partial_rh_{sb}&-
\frac{1}{2}\ep^{pmn}\partial_ph_{mn}\ep^{lrs}\partial_lh_{rs}+\\
&-\mu h_{pa}\ep^{prs}\partial_rh_s{}^a>.\tag 3.23
\endalign
$$

Hacemos la descomposici\'on 2+1 y llamamos $n,N_i,M_i,H_{ij}+\ep_{ij}V$ 
respectivamente a $h_{00}$, $h_{i0}$, $h_{0i}$, $h_{ij}$, como en (1.25). 
Llegamos as\'{\i} a

$$\align
S^l_{VCS}= < &\dot{N}_i(\frac{\mu}{2}\ep^{ij}N_j+\partial_iH_{jj}-\partial_j
H_{ij})+\\ 
&+\dot{H}_{ij}[\delta_{ij}(\mu V+\partial_kM_k-\frac{1}{2}\dot{H}_{kk})+
\frac{1}{2}\dot{H}_{ij}+\\ 
&\ \ \ \ \ \ \ \ -\frac{1}{2}(\partial_iM_j+\partial_jM_i)-\frac{\mu}{4}
(\ep_{ik}H_{kj}+\ep_{jk}H_{ki})]+\\ 
&+n[(\Delta \delta_{ij}-\partial_i\partial_j)H_{ij}+\mu \ep_{ij}\partial_i
N_j]-\mu \ep_{ij}\partial_iH_{jk}M_k+\\ 
&-\frac{1}{4}(N_i+M_i)\Delta (N_i+M_i)-\frac{1}{4}\partial_i(N_i+M_i)
\partial_j(N_j+M_j)+\\
&-\mu V\partial_iM_i>.\tag 3.24
\endalign
$$
Si definimos los momentos conjugados a las variables din\'amicas $N_i,H_{ij}$

$$
\pi_i\equiv \frac{\partial \Cal{L}}{\partial \dot{N}_i}\ \ ,\ \ 
\pi_{ij}\equiv \frac{\partial \Cal{L}}{\partial \dot{H}_{ij}},\tag 3.25
$$
sucede que podemos despejar $\dot{H}_{ij}$

$$
\dot{H}_{ij}=\pi_{ij}-\delta_{ij}(\pi_{ll}-\mu V)+\frac{1}{2}
(\partial_iM_j+\partial_jM_i)+\frac{\mu}{4}(\ep_{ik}H_{kj}+\ep_{jk}H_{ki}),
\tag 3.26
$$
y tenemos el v\'{\i}nculo primario

$$
\varphi_i=\pi_i+\partial_jH_{ij}-\partial_iH_{jj}-\frac{\mu}{2}\ep_{ij}N_j.
\tag 3.27
$$

$n$ es un multiplicador de Lagrange. Sin embargo, escribimos primero el 
hamiltoniano, cuya densidad es

$$\align
\Cal{H}_c &=\pi_{ij}\dot{H}_{ij}+\pi_i\dot{N}_i-\Cal{L}\\ 
&=\frac{1}{4}(N_i\Delta N_i)+\frac{1}{4}\partial_iN_i\partial_jN_j+
\frac{1}{2}\pi_{ij}\pi_{ij}-\frac{1}{2}\pi_{ii}\pi_{jj}+\\ 
&\ \ \ \ \ \ +\frac{\mu^2}{8}H_{ij}H_{ij}-\frac{\mu^2}{16}H_{ii}H_{jj}-
\frac{\mu^2}{2}\ep_{ij}H_{ik}\pi_{kj}+\\ 
&\ \ \ \ \ \ +n[\partial_i\partial_jH_{ij}-\Delta H_{ii}-\mu \ep_{ij}
\partial_iN_j]+\\ 
&\ \ \ \ \ \ +M_i[\frac{1}{2}\Delta N_i-\frac{1}{2}\partial_i\partial_jN_j+
\frac{3}{4}\mu \ep_{kl}\partial_kH_{li}-\frac{\mu}{4}\ep_{ij}\partial_kH_{jk}-
\partial_j\pi_{ij}]+\\ 
&\ \ \ \ \ \ +\mu V[\pi_{ii}-\mu V],\tag 3.28
\endalign
$$
donde observamos que adem\'as de $n$, $M_i$ y $V$ tambi\'en son 
multiplicadores. Los v\'{\i}nculos primarios adicionales son

$$\align
\varphi &\equiv V-\frac{1}{2\mu}\pi_{ii}\tag 3.28,a\\ 
\psi &\equiv -(\delta_{ij}\Delta -\partial_i\partial_j)H_{ij}-\mu \ep_{ij}
\partial_iN_j\tag 3.28,b\\ 
\psi_i &\equiv \frac{1}{2}(\delta_{ij}\Delta -\partial_i\partial_j)N_j+
\frac{3}{4}\mu\ep_{kl}\partial_kH_{li}-\frac{\mu}{4}\ep_{ij}\partial_kH_{jk}-
\partial_j\pi_{ij}.\tag 3.28,c
\endalign
$$

Sustituimos $\varphi$ en $H_c$ y comenzamos el procedimiento can\'onico con

$$
\Cal{H}_c=\Cal{H}_0+n\psi +M_i\psi_i +\lambda_i\varphi_i,\tag 3.29
$$
con $\psi ,\psi_i,\varphi$ como en (3.28) y

$$\align
\Cal{H}_0=& \frac{1}{4}N_i\Delta N_i+\frac{1}{4}\partial_iN_i\partial_jN_j+
\frac{1}{2}\pi_{ij}\pi_{ij}-\frac{1}{4}\pi_{ii}\pi_{jj}+\\ 
&+\frac{\mu^2}{8}H_{ij}H_{ij}-\frac{\mu^2}{16}H_{ii}H_{jj}-\frac{\mu}{2}
\ep_{ij}H_{ik}\pi_{kj}.\tag 3.30
\endalign
$$
El \'algebra entre los v\'{\i}nculos primarios es

$$
\{\Omega_A(x),\Omega_B(y)\}=M_{AB}(x)\delta^{(2)}(\ovr{x}-\ovr{y}),
\tag 3.31
$$
con $\Omega_A=\{ \psi ,\psi_i,\varphi_i \}$, 
$\Omega_B=\{ \psi ,\psi_k,\varphi_k \}$ y 

$$
M_{AB}(x)=\left( \matrix
0 & 0 & \mu \ep_{kl}\partial_l\\ 
0 & 0 & 0\\ 
\mu \ep_{il}\partial_l & 0 & -\ep_{ik}
\endmatrix \right) \tag 3.32
$$

La conservaci\'on de los v\'{\i}nculos s\'olo aporta relaciones entre los 
multiplicadores, por lo que el procedimiento termina. Podemos tomar la 
combinaci\'on $\theta \equiv \psi -\partial_i\varphi_i$ por $\psi$, y entonces 
el \'algebra de los v\'{\i}nculos tendr\'{\i}a $(\Omega_A=(\theta ,\psi_i,\varphi_i))$

$$
M_{AB}(x)=\left( \matrix
0 & 0 & 0\\ 
0 & 0 & 0\\ 
0 & 0 & -\mu \ep_{ik}
\endmatrix \right), \tag 3.33
$$
y $\Cal{H}_c$ ser\'{\i}a 
$(\theta =-\partial_i\pi_i-\frac{\mu}{2}\ep_{ij}\partial_iN_j)$

$$
\Cal{H}_c=\Cal{H}_0+n\theta +M_i\psi_i +(\lambda_i+\partial_in)\varphi_i.\tag 3.34
$$
$M_{AB}(x)$ muestra que $\theta$ y $\psi_i$ son de primera clase. Las 
transformaciones de calibre son:

$$
\delta (*)=\int \{(*),\xi_a(x')\Psi_a(x')\}d^2x'\tag 3.35
$$
donde $\Psi_a$ son los v\'{\i}nculos de primera clase y $\xi$ es el 
par\'ametro de la transformaci\'on. As\'{\i}

$$\align
\delta H_{ij} & =\int d^2x'\{H_{ij}(x),\xi (x')\theta (x')+
\xi_k(x')\psi_k(x')\}\\ 
& =\int d^2x'\{ H_{ij}(x),-\xi^k(x')(\partial_l\pi_{lk}(x')\}\\ 
&=\frac{1}{2}(\partial_i\xi_j+\partial_j\xi_i),\tag 3.36,a
\endalign
$$
y an\'alogamente

$$
\delta N_i=\partial_i\xi ,\tag 3.36,b
$$
como era de esperarse. Para los momentos

$$\align
\delta \pi_i &=\frac{1}{2}\ep_{ij}\partial_j\xi -\frac{1}{2}(\delta_{ij}
\Delta -\partial_i\partial_j)\xi_j, \tag 3.37,a\\
\delta \pi_{ij} &=\frac{\mu}{4}(\delta_{ij}\ep_{kl}\partial_k\xi_l-
(\ep_{ik}\partial_k\xi_j+\ep_{jk}\partial_k\xi_i)),\tag 3.37,a
\endalign
$$
y para $V$

$$\align
\delta V &=\frac{1}{2\mu}\delta \pi_{ii}=\frac{1}{2}\ep_{ij}\partial_i\xi_j\\ 
&=\frac{1}{2}\ep_{ij}\delta h_{ij}.\tag 3.37,b
\endalign
$$

Por \'ultimo el hamiltoniano es invariante de calibre, trivialmente, por la 
conservaci\'on de los v\'{\i}nculos de primera clase

$$
\delta \Cal{H}_c=0\tag 3.38
$$

\vskip 5mm
\noi
{\bf 4. Equivalencia can\'onica entre $S^2_{AD}$ y $S^l_{VCS}$}
\vskip 3mm
\noi
{\bf 4.1. El conjunto com\'un de los v\'{\i}nculos}
\vskip 3mm
Hab\'{\i}amos llegado en el an\'alisis de la teor\'{\i}a autodual a que el 
problema din\'amico correspond\'{\i}a a tomar

$$
\Cal{H}_0^{(AD)}=\frac{m^2}{2}(H_{ij}H_{ij}-H_{ii}H_{jj}),\tag 4.1
$$
sometido a los v\'{\i}nculos

$$\align
\theta & =-\frac{1}{m}\partial_i\pi_i+mH_{ii}+\frac{1}{2}\ep_{ij}\partial_iN_j
,\tag 4.2,a\\
\varphi_i &=\pi_i+\frac{m}{2}\ep_{ij}N_j,\tag 4.2,b\\
\psi_i &=\ep_{kl}\partial_kH_{li}-mN_i,\tag 4.2,c\\
\varphi_{ij} &= \pi_{ij}-\frac{m}{4}(\ep_{ik}H_{kj}+\ep_{jk}H_{ki}).\tag 4.2,d
\endalign
$$
En este problema hab\'{\i}an sido eliminados $V$ y su momento conjugado $\pi$ 
ya que $V=\pi =0$.

An\'alogamente para la acci\'on linealizada de la gravedad $VCS$ tenemos

$$\align
\Cal{H}_0^{(VCS)}= &\frac{\mu^2}{8}H_{ij}H_{ij}-\frac{\mu^2}{16}H_{ii}H_{jj}+
\frac{1}{2}\pi_{ij}\pi_{ij}-\frac{1}{4}\pi_{ii}\pi_{jj}+\\
&-\frac{\mu}{2}\ep_{ij}H_{ik}\pi_{kj}+\frac{1}{4}N_i\Delta N_i+
\frac{1}{4}\partial_iN_i\partial_jN_j, \tag 4.3
\endalign
$$
sometido a los v\'{\i}nculos 

$$\align
\theta &=-\partial_i\pi_i-\frac{\mu}{2}\ep_{ij}\partial_iN_j,\tag 4.4,a\\
\psi_i &= \frac{1}{2}(\delta_{ij}\Delta -\partial_i\partial_j)N_j +
\frac{3}{4}\mu \ep_{kl}\partial_kH_{li}-\frac{\mu}{4}\ep_{ij}\partial_kH_{jk}
-\partial_j\pi_{ij},\tag 4.4,b\\
\varphi_i &= \pi_i+\partial_jH_{ij}-\partial_iH_{jj}-\frac{\mu}{2}\ep_{ij}N_j,
\tag 4.4,c
\endalign
$$
donde $V$ fu\'e eliminado a trav\'es del v\'{\i}nculo $V=(1/2\mu )\pi_{ii}$. 
Si miramos nuevamente a $\varphi_{ij}$ de la autodual antes de sustituir 
$V=0$ (ec. (1.28,c)) observamos que la relaci\'on entre $V$ y $\pi_{ii}$ es 
la misma que aqu\'{\i}. Tomaremos de aqu\'{\i} en adelante $m=\mu$.

Observamos que si utilizo los v\'{\i}nculos de la teor\'{\i}a autodual (4.2) 
en la densidad $\Cal{H}_0^{(VCS)}$ obtengo $\Cal{H}_0^{(AD)}$ (ver la ec. 
(4.20), mas adelante). As\'{\i}, los tres v\'{\i}nculos adicionales de la 
teor\'{\i}a autodual podr\'{\i}an corresponder a una fijaci\'on de calibre en 
la teor\'{\i}a $VCS$ linealizada. De hecho

$$\align
\theta (\text{de la $VCS$}) &=-\partial_i\varphi_i (\text{de la $AD$}),
\tag 4.5,a\\
\psi_i(\text{de la $VCS$}) &=-\partial_j\varphi_{ij}-\frac{1}{2\mu}\ep_{ik}
\partial_k(\partial_l\varphi_l+\mu \theta )(\text{de la $AD$}),\tag 4.5,b\\
\varphi_i(\text{de la $VCS$}) &=\varphi_i-\ep_{ij}\psi_j (\text{de la $AD$}),
\tag 4.5,c
\endalign
$$
Por lo tanto, lo que ``sobra" en (4.2) deber\'{\i}a corresponder a una 
fijaci\'on de calibre. Esto es 

$$\align
\chi &=\ep_{ij}\partial_i\varphi_j,\tag 4.6,a\\
\chi_1 &=\theta +\frac{1}{\mu}\partial_i\varphi_i,\tag 4.6,b\\
\chi_2 &=\delta_{ij}\Delta \varphi_{ij}. \tag 4.6,c
\endalign
$$
Es f\'acil ver que estas \'ultimas definici\'ones son el ``faltante" de 
(4.2). Del conjunto inicial $\theta ,\varphi_i,\varphi_{ij},\psi_i$ la correspondencia 
de sus distintas componentes con los v\'{\i}nculos de la teor\'{\i}a $VCS$ linealizada 
es: $\theta (VCS)\sim \partial_i\varphi_i(AD)$, 
$\psi_i(VCS)\sim \partial_j\varphi_{ij}(AD)$, $\varphi_i(VCS)\sim \psi_i(AD)$. 
Luego $\chi \sim \ep_{ij}\partial_i\varphi_j(AD)$, 
$\chi_1\sim \theta (AD)$, $\chi_2\sim \varphi_{ii}(AD)$.

El conjunto com\'un de v\'{\i}nculos de las dos teor\'{\i}as es entonces

$$\align
\zeta &\equiv \theta (VCS,\ (4.4,a))\tag 4.7,a\\
\zeta_i &\equiv \psi_i(VCS,\ (4.4,b))\tag 4.7,b\\ 
\chi &=\ep_{ij}\partial_i\pi_j-\frac{\mu}{2}\partial_iN_i\tag 4.7,c\\
\chi_1 &=\mu H_{ii}+\ep_{ij}\partial_iN_j \tag 4.7,d\\
\chi_2 &=\Delta \pi_{ii}\tag 4.7,e\\
\varphi_i &= (VCS,\ (4.4,c))\tag 4.7,f
\endalign
$$

\vskip 5mm
\noi
{\bf 4.2 El hamiltoniano invariante de calibre}
\vskip 3mm
El conjunto de v\'{\i}culos (4.7) nos muestra, que para la teor\'{\i}a $AD$, 
hay un sector de los v\'{\i}nculos que se comporta como v\'{\i}nculos de 
primera clase. A\'un mas, al calcular la integral funcional de esta 
teor\'{\i}a, en la medida aparece [45] $(det\{\Omega_A,\Omega_B\})^{1/2}$, 
donde $\Omega_A$ representa a todos los v\'{\i}nculos. Es f\'acil ver que

$$
(det\{\Omega_A,\Omega_B\})^{1/2}=det\{\zeta_a,\chi_b\}
(det\{\varphi_i,\varphi_j\})^{1/2}, \tag 4.8
$$
donde, nuevamente, vemos que la teor\'{\i}a corresponde a una teor\'{\i}a de 
calibre, con trans\-for\-ma\-cio\-nes de calibre generadas por los $\zeta_a$, con 
fijaciones de calibre $\chi_a$, y sujeta a los v\'{\i}nculos adicionales de 
segunda clase $\varphi_i$. Nos proponemos, entonces, hallar el 
correspondiente hamiltoniano, invariante de calibre $\wt{H}_0^{(AD)}$.

La forma mas general que debe tener $\wt{H}_0^{(AD)}$ es [36,44]

$$\align
\wt{H}_0^{(AD)}=H_0^{(AD)}&+<\alpha_a\zeta_a>+<\beta_a(x,z)\chi_a(z_1)>+\\
&+<\beta_{ab}(x,z_1,z_2)\chi_a(z_1)\chi_b(z_2)>,\tag 4.9
\endalign
$$
donde no se agregan t\'erminos con productos de mas de 2 v\'{\i}nculos, 
pues tratamos con una teor\'{\i}a cuadr\'atica. Pedimos que

$$
\{\wt{H}_0^{(AD)},\zeta_a(y)\}=0\ \ (=V_a{}^b\zeta_b).\tag 4.10
$$
Si

$$
\{\beta_{ab}(x,z_1,z_2),\zeta_a(y)\}=0,\tag 4.11
$$
una soluci\'on de (4.10) es

$$\align
\{ & H^{(AD)}_0,\zeta_c(y)\}+<\beta_a(x,z_1)\{\chi_a(z_1),\zeta_c(y)\}>=0,
\tag 4.12,a\\
< \chi_a(z_1)&\{\beta_a(x,z_1),\zeta_c(y)\}>+\\
&+2<\beta_{ab}(x,z_1,z_2)\chi_a(z_1)
\{\chi_b(z_2),\zeta_c(y)\}>=0,\tag 4.12,b
\endalign
$$
y las $\alpha_a$ quedan indeterminadas. Veremos que para una combinaci\'on 
particular de los v\'{\i}nculos de primera clase 
$\wt{H}_0^{(AD)}=H_0^{(VCS)}$.

Para mayor simplicidad de c\'alculos descomponemos

$$\align
\zeta_i &=\ep_{ij}\partial_j\zeta^T+\partial_i\zeta^L\\
& \equiv \ep_{ij}\partial_j\wt{\zeta}_1+\partial_i\wt{\zeta}_2,\tag 4.13
\endalign
$$
y tomamos como convenci\'on $a,b=0,1,2$ donde ning\'un sub\'{\i}ndice se 
sobreentiende como 0 (i.e. $\chi =\chi_0,\wt{\zeta}=\wt{\zeta}_0=\zeta$). 
El \'algebra entre los $\chi_a$'s y los $\wt{\zeta}_a$'s es

$$\align
&\{\chi_a(x),\wt{\zeta}_b(y)\}=\\ 
&\quad \quad \quad \left( \matrix
-\mu \Delta^x\delta^{(2)}(\ovr{x}-\ovr{y}) & 0 & \frac{1}{2}\Delta^x
\delta^{(2)}(\ovr{x}-\ovr{y})\\
0 & -\mu \delta^{(2)}(\ovr{x}-\ovr{y}) & 0\\
0 & 0 & \mu \Delta^x\delta^{(2)}(\ovr{x}-\ovr{y})\endmatrix 
\right)
\endalign$$
$$
\tag 4.14
$$
y adem\'as

$$\align
\{H_0^{(AD)},\wt{\zeta}\}&=0,\tag 4.15,a\\
\{H_0^{(AD)},\wt{\zeta}_1\}&=-\mu^2(-\Delta )^{-1}(\delta_{ij}\Delta -
\partial_i\partial_j)H_{ij},\tag 4.15,b\\
\{H_0^{(AD)},\wt{\zeta}_2\}&=-\mu^2(-\Delta )^{-1}\ep_{ij}\partial_i
\partial_kH_{jk}.\tag 4.15,c
\endalign
$$

Vamos a (4.12,a) y obtenemos

$$\align
\beta (x,z) &=0,\tag 4.16,a\\
\beta_1(x,z) &=\mu K(x-z)(\delta_{ij}\Delta -\partial_i\partial_j)H_{ij},
\tag 4.16,b\\
\beta_2(x,z) &=<-\mu K(x-z)K(z-y)\ep_{ij}\partial_i\partial_kH_{jk}(y)>_y,
\tag 4.16,c
\endalign
$$
donde $(-\Delta )K(x-y)=-\delta^{(2)}(\ovr{x}-\ovr{y})$. Los corchetes entre 
los $\beta_a$'s y los $\wt{\zeta}_a$'s dan todos, salvo uno, nulos. \'Este es

$$
\{\beta_2(x,z_1),\wt{\zeta}_2(y)\}=\frac{\mu}{2}\delta^{(2)}(x-z_1)
\delta^{(2)}(z_1-z_2).\tag 4.17
$$

Vamos a (4.12,b) y encontramos que el \'unico $\beta_{ab}$ no nulo es

$$
\beta_{22}(x,z_1,z_2)=-\frac{1}{4}K(x-z_1)K(z_1-y).\tag 4.18
$$
Finalmente tenemos que

$$\align
\wt{H}^{(AD}_0=H^{(AD)}_0 +&<\alpha_a\wt{\zeta}_a>+<-\mu (-\Delta )^{-1}\chi_1
(\delta_{ij}\Delta -\partial_i\partial_j)H_{ij}+\\ 
-& \mu ((-\Delta )^{-1}\chi_2)((-\Delta )^{-1}\ep_{ij}\partial_i\partial_k
H_{jk})+\\ 
-& \frac{1}{4}((-\Delta )^{-1}\chi_2)((-\Delta )^{-1}\chi_2)>.\tag 4.19
\endalign
$$
donde se entiende que $(-\Delta )^{-1}(*)(x)=-\int d^2yK(x-y)(*)(y)$.

La b\'usqueda de cual es la combinaci\'on de v\'{\i}nculos de primera clase que 
nos lleva a $H_0^{(VCS)}$ no es trivial. Sin embargo, podemos inducir 
$H_0^{(AD)}$ a partir de $H_0^{(VCS)}$ con el conjunto de v\'{\i}nculos (4.7). 
Encontramos, entonces que 

$$\align
H_0^{(VCS)}=H_0^{(AD)}+<&\wt{\zeta}_1(\chi_1+\wt{\zeta}_1+2\mu (-\Delta )^{-1}
\partial_i\partial_jH_{ij}+\mu H_{jj})+\\
&+\wt{\zeta}_2(2\wt{\zeta}_2+2(-\Delta )^{-1}\ep_{ij}\partial_i\partial_k
H_{kj}-\frac{1}{2}(-\Delta )\chi_2)+\\
&-\mu ((-\Delta )^{-1}\chi_1)(\delta_{ij}\Delta -\partial_i\partial_j)H_{ij}+\\
&-\mu ((-\Delta )^{-1}\chi_2)((-\Delta )^{-1}\ep_{ij}\partial_i\partial_k
H_{jk})+\\ 
&-\frac{1}{4}((-\Delta )^{-1}\chi_2)((-\Delta )^{-1}\chi_2)>.\tag 4.20
\endalign
$$

Al comparar con (4.19) encontramos que para la combinaci\'on particular de 
los $\wt{\zeta}_a$'s con

$$\align
\alpha &=0,\tag 4.21,a\\
\alpha_1 &=\wt{\zeta}_1+\chi_1+2\mu (-\Delta )^{-1}\partial_i\partial_j
H_{ij}+\mu H_{jj},\tag 4.21,b\\
\alpha_2 &=2\wt{\zeta}_2+2(-\Delta )^{-1}\ep_{ij}\partial_i\partial_kH_{kj}-
\frac{1}{2}(-\Delta )^{-1}\chi_2,\tag 4.21,c
\endalign
$$

Resulta que $\wt{H}_0^{(AD)}=H_0^{(VCS)}$. Tenemos as\'{\i}, una equivalencia 
can\'onica entre la teor\'{\i}a autodual y la $VCS$ linealizada.

\vskip 5mm
\noi
{\bf 4.3 La extensi\'on ``invariante de calibre" de $H_0^{(VCS)}$}
\vskip 3mm

Vimos que es posible obtener una extensi\'on invariante de ca\-li\-bre de 
$H_0^{(AD)}$ que es igual a $H_0^{(VCS)}$. En el proceso todav\'{\i}a nos 
quedaron unos v\'{\i}nculos de segunda clase. As\'{\i}, podr\'{\i}amos 
to\-mar\-los a ellos como punto de partida en la teor\'{\i}a $VCS$ y obtener una 
extensi\'on de $H_0^{(VCS)}$, $\wt{H}_0^{(VCS)}$, que corresponda a una 
teor\'{\i}a con s\'olo v\'{\i}nculos de primera clase.

Por razones que se aclarar\'an en breve, vamos a considerar el conjunto de 
v\'{\i}nculos (4.7) y le agregamos los dos v\'{\i}nculos, de segunda clase, 
que ten\'{\i}amos antes de eliminar a $V$ y su momento conjugado $\pi$. El 
conjunto de v\'{\i}nculos ser\'a: $\varphi_i$, $\varphi$ (ecuaciones (3.28,a) 
y (4.4,c)) y $\wh{\varphi}\equiv \pi$, que corresponden al sector de 
v\'{\i}nculos remanente, mas los v\'{\i}nculos de primera clase $\zeta_a$ y sus 
fijaciones de calibre $\chi_a$.

Mirando el \'algebra entre el conjunto de v\'{\i}nculos $\varphi_i$, 
$\varphi$ y $\wh{\varphi}$ notamos que podemos tomar a uno de los 
$\varphi_i$', y a $\varphi$ \'o $\wh{\varphi}$ como v\'{\i}nculos de ``primera 
clase'', y a los que quedan como sus respectivas fijaciones. Apoyados en esto, 
escogemos la siguiente designaci\'on

$$\align
\zeta_3 &\equiv \wt{\varphi}=\pi ,\tag 4.22,a\\ 
\zeta_4 &\equiv -\partial_i \varphi_i=-\partial_i\pi_i+\frac{\mu}{2}\ep_{ij}
\partial_iN_j+(\delta_{ij}\Delta-\partial_i\partial_j)H_{ij}, \tag 4.22,b\\ 
\chi_3 &\equiv \mu \varphi -\frac{1}{\mu}\ep_{ij}\partial_i\varphi_j=
\mu V-\frac{1}{2}\pi_{ii}-\frac{1}{2}\ep_{ij}\partial_i\pi_j -\frac{1}{2}
\partial_iN_i+\\ 
&\quad\quad\quad\quad\quad\quad\quad\quad\quad\quad\quad\quad\quad\quad\quad\quad\ -\frac{1}{\mu}\ep_{ij}\partial_i\partial_k
H_{kj},\tag 4.22,c\\    
\chi_4 &\equiv
-\ep_{ij}\partial_i\varphi_j=-\ep_{ij}\partial_i\pi_j-
\frac{\mu}{2}\partial_iN_i-\ep_{ij}\partial_i\partial_kH_{kj},\tag 4.22,d 
\endalign
$$ 
donde estamos sugiriendo a
$-\frac{1}{\mu}\ep_{ij}\partial_i\varphi_j$ y  $-\ep_{ij}\partial_i\varphi_j$
como fijaciones de calibre. Los \'unicos  corchetes no nulos entre v\'{\i}nculos

son 

$$\align
\{\chi_3(x),\zeta_3(y)\} &=\mu \delta^{(2)}(\ovr{x}-\ovr{y}), \tag 4.23,a\\
\{\chi_4(x),\zeta_4(y)\} &=-\mu \delta^{(2)}(\ovr{x}-\ovr{y}), \tag 4.23,b
\endalign
$$

Suponemos

$$\align
\wt{H}_0^{(VCS)}=\wt{H}_0^{(VCS)}&+<\alpha_{a'}\zeta_{a'}>+<\beta_{a'}(x,z)
\chi_{a'}(z)>+\\
&+<\beta_{a'b'}(x_1z_1,z_2)\chi_{a'}(z_1)\chi_{b'}(z_2)>, \tag 4.24
\endalign
$$
donde $a'=3,4$. Siguiendo el m\'etodo ya expuesto en la subsecci\'on 
anterior obtenemos que $\beta_3(x,z_1)$ y 

$$
\beta_4(x,z)=\frac{\mu}{2}K(x-z)(\mu \ep_{ij}\partial_i\partial_kH_{kj}(x)-
\partial_i\partial_j\pi_{ij}(x)). \tag 4.25
$$
Luego observamos que $\{\beta_{a'}(x,z_1),\zeta_{b'}(y)\}=0$ por lo que 
$\beta_{a'b'}(x,z_1,z_2)$ es una soluci\'on. As\'{\i}

$$
\wt{H}_0^{(VCS)}=H_0^{(VCS)}+<\alpha_{a'}\zeta_{a'}>+\frac{\mu}{2}
<(-\Delta )^{-1}\chi_4(\mu \ep_{ij}\partial_i\partial_kH_{kj}-
\partial_i\partial_j\pi_{ij})>.\tag 4.26
$$
Los $\alpha_{a'}$'s en (4.26) son arbitrarios. Para alguna escogencia, el 
modelo ser\'a equivalente can\'onicamente al de una teor\'{\i}a con s\'olo 
v\'{\i}nculos de primera clase.

Usando los v\'{\i}nculos, convenientemente escogidos, podemos mirar como 
transforman $H_{ij}$, $N_i$ y $V$ $(\delta (*)=\int\{(*),\lambda_{a'}(x^1), 
\zeta_{a'}(x^1)\}d^2x^1)$

$$\align
\delta H_{ij} &=\frac{1}{2}(h_{ij}+h_{ji})=0,\tag 4.27,a\\
\delta N_j &=\delta h_{j0}=\partial_j\lambda_4,\tag 4.27,b\\
\delta V &= \frac{1}{2}\ep_{ij}\delta h_{ij}=\lambda_3. \tag 4.27,c
\endalign
$$
Esta transformaci\'on es reminiscente de las transformaciones de Lorentz 
li\-nea\-li\-za\-das para los dreibeins

$$
\delta_Lh_{mn}=-\ep_{mnl}l^l.\tag 4,28,c
$$
si

$$
l^0=\lambda_3\ \ ,\ \ l^i=-\ep_{ij}\partial_j\lambda_4.\tag 4.28,b
$$
Si pedimos consistencia al linearizar con el hecho de que 
$\delta g_{mn}=\delta (e_m{}^ae_n{}^b$ $\eta_{ab})$, entonces $\delta h_{00}=0$ 
y $\delta h_{0i}=-\delta h_{i0}$. Luego para 
$\omega_n{}^a=-(1/2)\ep^{prs}\partial_p h_{rs}+\ep^{ars}\partial_rh_{sn}$, 
es inmediato probar que

$$
\delta \omega_n{}^a=-\partial_n l^a, \tag 4.29
$$
que corresponde a las transformaciones de Lorentz (linealizadas) para la 
co\-ne\-xi\'on. La forma particular de los $l^l$ es de tal forma que no choca con
las  transformaciones de calibre originales $(\delta h_{mn}=\partial_m\xi_n)$. 
Por ejemplo, la parte transversa de $h_{j0}$ es invariante bajo ambas 
transformaciones.

Todo lo antes expuesto nos hace suponer que para alguna escogencia de los 
$\alpha_{a'}$'s en $\wt{H}_0^{VCS}$ debe suceder que 
$\wt{H}_0^{(VCS)}=H_0^{(TM)}$, la cual es trivialmente invariante bajo estas 
transformaciones li\-ne\-a\-li\-za\-das de Lorentz.

\newpage

$\ $

\pageno=47
\headline={\ifnum\pageno=47\hfil\else\hss\tenrm \folio\ \fi}

\vskip 1cm

\centerline{\catorce Cap\'{\i}tulo {\catorcebf IV}}

\vskip 1cm

\centerline{\dseisbf LA GRAVEDAD MASIVA}

\vskip 3mm

\centerline{\dseisbf VECTORIAL DE}

\vskip 3mm

\centerline{\dseisbf CHERN-SIMONS}

\vskip 2cm

En este cap\'{\i}tulo presentamos la acci\'on curva cuya linealizaci\'on
corresponde a la acci\'on $S^l_{VCS}$ que fu\'e analizada en el cap\'{\i}tulo
anterior, y que constituye una alternativa como teor\'{\i}a de gravedad masiva
curva en D = 2+1. Esta acci\'on se obtiene sum\'andole a la acci\'on de
Einstein un t\'ermino, construido con los dreibeins, que es topol\'ogico en los
\'{\i}ndices de universo. Este t\'ermino es an\'alogo al de $CS$ vectorial, de
ah\'{\i} la denominaci\'on de t\'erminos de $CS$ tri\'adico. Mostraremos
algunas propiedades de esta teor\'{\i}a en contraposici\'on a la otra
teor\'{\i}a masiva existente.

\noi
{\bf 1.- La acci\'on dentro de un marco jer\'arquico de simetr\'{\i}as}
\vskip 3mm
Ya hemos se\~nalado la particularidad de que no es posible, en dimensi\'on 2+1,
tener part\'{\i}culas sin masa y con spin distinto de cero. As\'{\i}, si 
desearamos hablar de gravitones, estos no tendr\'{\i}an spin 2, tal como 
sucede en dimensiones mayores. En otro orden de ideas, la acci\'on de 
Einstein en 2+1 dimensiones

$$
S_E=-\frac{1}{2\kappa^2}\int d^3x\sqrt{-g}R,\tag 1.1
$$
no posee din\'amica local [40]. Ya en el cap\'{\i}tulo anterior 
se\~nalamos la existencia de dos posibilidades como modelos de spin 2 
masivos: la gravedad $TM$ linealizada y la gravedad $VCS$ linealizada. 
Estas corresponden a la linealizaci\'on de las respectivas teor\'{\i}as 
curvas. La segunda de estas es la que introduciremos en este cap\'{\i}tulo.

Como teor\'{\i}a curva la gravedad $TM$ corresponde a la acci\'on [14]

$$
S_{TM}=-\frac{1}{\mu}S_{CS}-S_E , \tag 1.2
$$
donde

$$
S_{CS}=\frac{1}{2\kappa^2}<\omega_{pa}\ep^{pmn}\partial_m\omega_n{}^a-
\frac{1}{3}\ep^{pmn}\ep_{abc}\omega_p{}^a\omega_m{}^b\omega_n{}^c>,\tag 1.3
$$
es la clase caracter\'{\i}stica de Chern-Simons. Esta se obtiene de la 
densidad de Hir\-ze\-bru\-ch-Pontryagin en dimensi\'on 3+1

$$
{}^*RR\equiv \frac{1}{2}\ep^{mnls}R_{mnrt}R_{ls}{}^{rt}=\partial_mX^m,
\tag 1.4
$$
luego de integrar $X^3$ omitiendo toda dependencia de $x^3$. En (1.3) 
$\omega_p{}^a=\omega_p{}^a(e)$ de tal forma que la torsi\'on es nula

$$
T_{mn}{}^a=D_me_n{}^a-D_ne_m{}^a=0.\tag 1.5
$$
$S_E$ viene dada en su forma, equivalente, en tr\'{\i}adas
\footnote"*"{\ninerm{Ver ap\'endice A}}

$$
S_E=\frac{1}{2\kappa^2}<e_{pa}\ep^{pmn}R_{mn}^{*a}>.\tag 1.6
$$

Resaltamos el hecho de que en (1.2) $S_E$ participa con signo distinto al que 
aparece en otras dimensiones. Sin embargo, a nivel linealizado el signo de 
$S_E$ es importante para que la teor\'{\i}a tenga energ\'{\i}a definida 
positiva.

$S_{CS}$ es invariante bajo transformaciones locales conformes

$$
\delta e_p{}^a=\frac{1}{2}\rho (x)e_p{}^a,\tag 1.7,a
$$
bajo transformaciones locales de Lorentz

$$
\delta e_p{}^a=e_p{}^b l_b{}^a(x), \tag 1.7,b
$$
y bajo difeomorfismos

$$
\delta e_p{}^a=\xi^l(x)\partial_le_p{}^a+\partial_p\xi^l(x)e_l{}^a.
\tag 1.7,c.
$$
Esto se hace claro si notamos que en general

$$
\delta S_{CS}=\frac{1}{\kappa^2}<\frac{1}{e}\ep^{rls}(\frac{1}{2}e_{pa}e_{rb}-
e_{ra}e_{pb})D_l\delta e_s{}^bR^{**pa}>, \tag 1.8
$$
donde hemos usado el hecho de que, como $T_{mn}{}^a=0$,

$$
e\delta \omega_m{}^a=\ep^{rls}(\frac{1}{2}e_m{}^ae_{rb}-e_{mb}e_r{}^a)
D_l\delta e_s{}^b.\tag 1.9
$$

$S_E$ es invariante s\'olo bajo (1.7,b) y (1.7,c). En $S_{TM}$ se ha perdido, 
entonces, la invariancia conforme. As\'{\i}, aunque $S_{CS}$ y $S_E$ no 
tienen din\'amica local, al combinarlos tenemos una teor\'{\i}a con una 
excitaci\'on masiva de spin 2 a expensas de que perdimos la invariancia 
conforme de $S_{CS}$.

Existe otra posibilidad de dar din\'amica local a $S_E$ si la combinamos con 
el t\'ermino de $CS$ tri\'adico [43]

$$
S_{TCS}=\frac{1}{2\kappa^2}<e_{pa}\ep^{prs}\partial_re_s{}^a>.\tag 1.10
$$
Este t\'ermino es invariante s\'olo bajo (1.7,c), por lo que al combinarlo 
con la acci\'on de Einstein tendr\'{\i}amos una teor\'{\i}a curva que es 
invariante bajo difeomorfismos y no bajo transformaciones de Lorentz. Estas 
\'ultimas cobran importancia cuando se quiere interpretar la teor\'{\i}a, sin 
embargo, no nos ocuparemos de eso. La p\'erdida de invariancia de Lorentz ha 
sido tambi\'en observada en teor\'{\i}as vectoriales de calibre abelianas con 
un t\'ermino de Chern-Simons cuando se genera din\'amicamente un campo 
magn\'etico que no se anula [48]. La acci\'on propuesta es [43,47]

$$
S_{VCS}=S_E-\mu S_{TCS},\tag 1.11
$$
cuya linealizaci\'on es la acci\'on intermedia [17] $S_{VCS}^l$ y como vimos 
describe excitaci\'on masiva de helicidad $2\mu /|\mu |$, dependiendo del 
signo de $\mu$ en (1.11). Hay una formulaci\'on curva con dreibeins [50], 
donde se introduce un t\'ermino llamado de $CS$ traslacional el cual 
b\'asicamente constituye un acoplamiento del dreiben con la torsi\'on 
(i.e. $S_{CS}^{(tras.)}\sim <e_{pa}\ep^{pmn}T_{mn}^a>$). Este tiene 
``lo que le falta'' a $S_{TCS}$ para ser invariante Lorentz pero la 
teor\'{\i}a tiene torsi\'on no nula por lo cual no la tratamos aqu\'{\i}.

Tenemos, entonces, un marco jer\'arquico de simetr\'{\i}as. Empezamos con el 
t\'ermino $S_{CS}$, de tercer orden (ya que $\omega_p{}^a\sim \partial e_m^b$), 
invariante bajo transformaciones locales conformes, de Lorentz y de 
difeomorfismos. Sigue la acci\'on de Einstein, $S_E$, de segundo orden 
(ya que $R^{**pa}\sim \partial \omega_m{}^b$), que no es invariante 
conforme. Por \'ultimo tenemos el t\'ermino $S_{TCS}$, de primer orden e 
invariante  s\'olo bajo difeomorfismos.

Si hacemos variaciones en $S_{CS}$, obtenemos

$$
\delta S_{CS}=0\sim -C^{pl}=0,\tag 1.12
$$
donde\footnote"${}^\dagger$"{\ninerm{Ver ap\'endice A}}

$$
C^{pl}\equiv \frac{1}{\sqrt{-g}}\ep^{pmn}{\Cal{D}}_m\wt{R}_n{}^l,\tag 1.13
$$
es el tensor de Cotton. Este tensor es sim\'etrico y de divergencia 
covariante nula debido a las identidades de Bianchi que satisface el tensor 
de Einstein $G_p{}^m$, adem\'as tiene, expl\'{\i}citamente, traza nula. 
As\'{\i}, s\'olo puede acoplarse a fuentes sin masa y con tensor de 
energ\'{\i}a-momentum sin traza.

Siguiendo con $S_E$, las ecuaciones de movimiento provenientes de hacer 
variaciones en ella, son

$$
\delta S_E=0\sim G^{pl}=0,\tag 1.14
$$
donde $G^{pl}=R^{pl}-\frac{1}{2}g^{pl}R$ es el tensor de Einstein. Ya dijimos 
que $S_E$ no tiene grados din\'amicos locales. Al aclopar $S_E$ con materia 
tendremos que el espacio tiempo es localmente plano fuera de las fuentes, ya 
que en dimensi\'on 2+1 el tensor de Riemann es equivalente al de Einstein

$$
R_{mn}{}^{ls}=-\ep_{mnr}\ep^{lst}G_t{}^r.\tag 1.15
$$
Sin embargo, este espacio, localmente plano, tiene estructura to\-po\-l\'o\-gi\-ca y 
geo\-m\'e\-tri\-ca no trivial [14,46,49].

En este mismo nivel de simetr\'{\i}a est\'a la acci\'on topol\'ogica masiva, 
$S_{TM}$. Sus ecuaciones de movimiento son

$$
\delta S_{TM}=0\sim \frac{1}{\mu}C^{pl}-G^{pl}=0.\tag 1.16
$$
Esta acci\'on, por la presencia de $S_E$, si puede acoplarse con fuentes 
masivas. Ya que $C_p{}^p=0$, tenemos que en ausencia de fuentes externas la 
curvatura $R$ es nula.

En el nivel de menor simetr\'{\i}a est\'a $S_{TCS}$. Si hacemos variaciones 
en ella, e imponemos que la torsi\'on sea nula, tendremos

$$
\delta S_{TCS}=0\sim\omega_m{}^p-\delta_m{}^p\omega_r{}^r=0,\tag 1.17
$$
donde

$$
\omega_m{}^p\equiv \omega_m{}^a e_a{}^p, \tag 1.18
$$
La ecuaci\'on (1.17) adem\'as dice que $\omega_m{}^p=0$; as\'{\i}, el 
t\'ermino $S_{TCS}$ es trivial por s\'{\i} solo.

El objeto $\omega_m{}^p$ no es necesariamente sim\'etrico, ni tiene traza 
nula. Su relaci\'on con el tensor de Einstein $G_s{}^p$ es

$$
G_s{}^p=\frac{1}{\sqrt{-g}}\ep^{pmn}{\Cal{D}}_m \omega_{ns}+\frac{1}{2}
\ep^{pmn}\ep_{str}\omega_m^t\omega_m{}^r,\tag 1.19
$$
y la simetr\'{\i}a de $G_s{}^p$ induce que

$$
{\Cal{D}}_t\omega_s{}^s-{\Cal{D}}_s\omega_t{}^s=\sqrt{-g}\ep_{pst}\omega^{sp}
\omega_r{}^r.\tag 1.20
$$
El primer miembro de (1.20) es la divergencia covariante de (1.17), y su 
segundo miembro es nulo si $\omega_{sp}=\omega_{ps}$, o si $\omega_r{}^r=0$. 
Luego, al acoplar con materia la simetr\'{\i}a del $T^{mn}$ implicar\'a su 
conservaci\'on. Igualmente si $T^m{}_m=0$ tambi\'en ${\Cal{D}}_mT^{mn}=0$ por 
(1.20). 

Tomando traza en (1.19), obtenemos

$$
R=-\frac{2}{\sqrt{-g}}\ep^{pmn}{\Cal{D}}_p\omega_{mn}-
(\omega_m{}^r\omega_r{}^m-\omega_m{}^m\omega_r{}^r).\tag 1.21
$$

Finalmente, veamos las ecuaciones de la gravedad $VCS$. Tenemos que las 
ecuaciones son
$$
E^{pa}\equiv R^{**pa}-\mu \ep^{prs}\partial_re_s{}^a=0,\tag 1.22,a
$$

$$
\ep^{pmn}D_me_n{}^a=0,\tag 1.22,b
$$
Cuando la torsi\'on es nula, la transformaci\'on de difeomorfismos puede 
escribirse como

$$
\delta e_p{}^a=D_p\xi^a+\ep^a{}_{bc}e_p{}^b l_\xi{}^c,\tag 1.23,a
$$
con

$$
l_\xi{}^c\equiv -\xi^n\omega_n{}^c.\tag 1.23,b
$$
As\'{\i}

$$\align
\delta_\xi S_{VCS}&=\frac{1}{k^2}<\delta e_{pa}E^{pa}>\\
&=<-\xi_aD_pE^{pa}+\mu \ep^{rst}e_{sb}e_{tc}\omega_r{}^b\omega_n{}^c
\xi^ae_a{}^n>,
\endalign
$$
de donde obtenemos la identidad de Bianchi

$$
D_pE^{pa}-\mu \ep^{rst}\omega_r{}^be_{sb}\omega_n^ce_{tc}e^{an}=0.
\tag 1.24,a
$$
Las identidades (1.24,a) se reescriben como

$$
{\Cal{D}}_p(G_n{}^p-\mu (\omega_n{}^p-\delta_n{}^p\omega_r{}^r))-
\mu \ep^{rst}\omega_{rs}\omega_{nt}=0,\tag 1.24,b
$$
por lo que al aclopar con una fuente sim\'etrica, esta deber\'a ser 
covariantemente conservada.

De (1.22) llegamos a las ecuaciones de movimiento

$$
\wt{R}^{mp}-\mu \omega^{mp}=0,\tag 1.25
$$
donde observamos que $\omega^{mp}=\omega^{pm}$. Adem\'as en ausencia de 
fuentes externas la curvatura

$$
R=4\mu \omega_l{}^l,\tag 1.26
$$
no es necesariamente nula.

De (1.25) podemos relacionar las ecuaciones de la gravedad $VCS$ y la 
gravedad $TM$. De hecho, si aplicamos $(-g)^{-1/2}\ep^{srm}{\Cal{D}}_r$ a 
(1.25), usando (1.19), obtendremos

$$
\frac{1}{\mu}C^{sp}-G^{sp}=-\frac{1}{2}\ep^{pmn}\ep^s{}_{rt}\omega_m{}^r
\omega_n{}^t,\tag 1.27
$$
que corresponde a (1.16), pero con un t\'ermino no homog\'eneo. Al linealizar, 
este t\'ermino no aparece y por tanto afirmamos que las ecuaciones de 
$S_{TM}^{2,l}$ son como el ``rotor" de las de la acci\'on $S_{VCS}^l$.

\newpage

$\ $

\pageno=54

\headline={\ifnum\pageno=54\hfil\else\hss\tenrm \folio\ \fi}

\vskip 1cm

\centerline{\catorce Cap\'{\i}tulo {\catorcebf V}}

\vskip 1cm

\centerline{\dseisbf ROTURA DE SIMETR\'IA}

\vskip 2cm

Hemos visto que tenemos teor\'{\i}as masivas de spin 1 y 2 donde todav\'{\i}a
quedan invariancia respecto a determinadas transformaciones locales. En este
cap\'{\i}tulo estudiamos la posibilidad de ''rom\-per'' estas simetr\'{\i}as,
apareciendo un cuadro de estrecha similitud entre las teor\'{\i}as de spines
distintos. Veremos que la teor\'{\i}a con mas simetr\'{\i}a (la $S_{TM}^{2,l}$)
no permite este tipo de proceso.

\noi
{\bf 1.- Teor\'{\i}a de Proca-Chern-Simons}

\vskip 3mm

\noi
{\bf 1.1.- La acci\'on como producto de un proceso de rotura espont\'anea de 
simetr\'{\i}a. }
\vskip 3mm

Partimos de la acci\'on a primer orden [51,52]
$$\align
S=<P_rP^{r\dagger}&-P^rD_r\varphi -P^{r\dagger}(D_r\varphi )^\dagger
-(h|\varphi |^2+U^2)(|\varphi |^2-V^2)^2+\\
&-\frac{1}{2}f^rf_r+f_r\ep^{rmn}\partial_na_n-\frac{\mu}{2}a_r\ep^{rmn}
\partial_ma_n>,\tag 1.1
\endalign
$$
donde $P_r,P^\dagger_r,\varphi ,\varphi^\dagger$ son variables 
independientes. El acoplamiento entre el campo escalar y el campo de calibre 
$a_r$ es minimal, as\'{\i}, $D=\partial_r-iea_r$. La acci\'on es invariante 
bajo las transformaciones infinitesimales 
$$\align
\delta a_r &=e^{-1}\partial_r\xi ,\tag 1.2,a\\
\delta \varphi &= i\xi \varphi ,\tag 1.2,b\\
\delta P_r &=-i\xi P_r ,\tag 1.2,c\\
\delta f_r &=0 ,\tag 1.2,d.
\endalign
$$

El procedimiento de rotura espont\'anea de simetr\'{\i}a es el usual: 
$<\varphi >_{vacio}=V\neq 0$; as\'{\i}, cambiamos 
$\varphi \to \varphi '=\varphi -V$ y por conveniencia cambiamos 
$P_r,P_r^\dagger \to P^r{}'+ieVa_r,P_r^{\dagger '}-ieVa_r$. En t\'erminos de 
estas nuevas variables, $S$ pasa a ser 
$$\align
S=<P^{r\dagger}P_r&-P^rD_r\varphi -P^{r\dagger}(D_r\varphi )^\dagger +ieVa_r
[(D_r\varphi )^\dagger-D_r\varphi ]+\\
&-[h|\varphi |^2_+hV(\varphi +\varphi^\dagger)+hV^2+U^2]
[|\varphi |^2+V(\varphi +\varphi^\dagger)]^2>+\\
&\ \ \ \ +S_{PCS},\tag 1.3
\endalign
$$
donde hemos omitido las ``$'$''. $S_{PCS}$ es 
$$
S_{PCS}=\frac{1}{2}<-f^rf_r+2f_r\ep^{rmn}\partial_ma_n-\mu a_r\ep^{rmn}
\partial_ma_n-m^2a_ra^r>,\tag 1.4
$$
con
$$
m^2\equiv 2e^2V^2.\tag 1.5
$$
Llamamos a $S_{PCS}$ la acci\'on de Proca-Chern-Simons. En el sector escalar 
(1.3) puede verse que s\'olo se propaga la excitaci\'on escalar real 
$\sim \varphi +\varphi^\dagger$ y que la otra excitaci\'on escalar posible  
$\sim \varphi -\varphi^\dagger$ ya no est\'a [53]. Nosotros centraremos 
nuestra atenci\'on en la parte vectorial de (1.3). Esta parte ten\'{\i}a, 
antes del procedimiento de rotura espont\'anea de simetr\'{\i}a, una sola 
escitaci\'on de masa $\mu$ y helicidad +1. Luego del proceso, el t\'ermino 
$m^2a_ra^r$ rompe la invariancia de calibre y como veremos $S_{PCS}$ propaga 
dos excitaciones masivas con masas distintas, a diferencia de la acci\'on de 
Proca.
 
\newpage
\vskip 5mm
\noi
{\bf 1.2.- Analisis Covariante}
\vskip 3mm

Las ecuaciones de movimiento provenientes de (1.4) son
$$\align
& f^r =\ep^{rmn}\partial_ma_n,\tag 1.6,a\\
& \ep^{rmn}(\partial_rf_m-\mu\partial_ra_m)-m^2a^n=0.\tag 1.6,b
\endalign
$$
Si sustituimos (1.6,a) en (1.6,b) obtenemos la ecuaci\'on de segundo orden
$$
\sq a_n-\partial_n\partial^la_l-\mu \ep^{rm}{}_n\partial_ra_m-m^2a_n=0,
\tag 1.7
$$
que trae como consecuencia que $a_n$ es transverso, asegur\'andonos la no 
propagaci\'on de la componente de spin 0 de $a_n$.

En el lenguaje de proyectores (1.7) se ``ve" como 
$$
\sq P_n{}^la_l-\mu \sq^{1/2}\xi_n{}^la_l-m^2a_n=0,\tag 1.8
$$
de donde
$$
(\sq^{1/2}+m_\mp )(\sq^{1/2}-m_\pm )P_\pm a=0, \tag 1.9
$$
con $(\ep \equiv \mu /m)$
$$
m_\pm =\frac{m\ep}{2}(\sqrt{1+\frac{4}{\ep^2}}\pm 1).\tag 1.10
$$

Si acoplamos $a_m$ con una fuente externa conservada, i.e. su\-ma\-mos un 
t\'er\-mi\-no de la forma $a_mJ^m$ a la acci\'on, obtenemos que
$$
a_m=-\bold{\Delta}_m{}^lJ_l,\tag 1.11,a
$$
con 
$$\align
\Delta_m{}^l=\frac{1}{m_+ +m_-}\Bigl[\frac{m_+}{\sq -m^2_+}& (P_m{}^l+
\frac{\sq^{1/2}}{m_+}\xi_m{}^l)+\\
&+\frac{m_-}{\sq -m^2_-} (P_m{}^l-\frac{\sq^{1/2}}{m_-}\xi_m{}^l)\Bigr]
\tag 1.11,b
\endalign
$$
En (1.11,b) observamos que el propagador est\'a constituido por 2 t\'erminos, 
los cuales son proporcionales al propagador de la acci\'on autodual. Esto 
puede verse r\'apidamente si consideramos la acci\'on $S_{AD}$ (Cap\'{\i}tulo 
II, ecuaci\'on (1.6)) con un t\'ermino adicional de acoplamiento $a_mJ^m$ con 
$\partial_mJ^m=0$. Las ecuaciones de movimiento en funci\'on de proyectores 
es
$$
[\sq^{1/2}(P_+-P_-)-mP]a=-\frac{1}{m}PJ.\tag 1.12
$$
Invertimos (1.12), llegando a que
$$
a_m=-\Delta_{(AD)m}{}^lJ_l,\tag 1.13,a
$$
con
$$
\Delta_{(AD)m}{}^l=\frac{1}{\sq -m}\Bigl(P_m{}^l+\frac{\sq^{1/2}}{m}
\xi_m{}^l\Bigr).\tag 1.13,b
$$

Observamos que los t\'erminos en (1.11,b) son efectivamente proporcionales 
a los de la teor\'{\i}a autodual pero con masa $m_+$ o $m_-$. Adem\'as (1.8) 
puede factorizarse como [34]
$$
-(\sq^{1/2}(P_+-P_-)-m_+(P_++P_-))(-\sq^{1/2}(P_+-P_-)-m_-(P_++P_-))a=0,
\tag 1.14
$$
que es el producto de las condiciones de autodualidad co\-rres\-pon\-dien\-tes 
(Cap\'{\i}tulo II, ecuaci\'on (1.4)). Cuando $\mu \to 0$, $m_\pm \to m$ y la 
e\-cua\-ci\'on (1.14) es el producto de las ``ra\'{\i}ces" de la acci\'on de 
Proca.

\noi
\vskip 5mm
{\bf 1.3.- Descomposici\'on 2+1 y la energ\'{\i}a}
\vskip 3mm

Si partimos de (1.4), haciendo la descomposici\'on 2+1, llegamos inicialmente 
a 
$$\align
S_{PCS}=\frac{1}{2}< & f(f-2\Delta a^T)+a(m^2a+2\mu\Delta a^T-2\Delta f^T)+\\
&+2\mu a^L\Delta \dot{a}^T+2\dot{a}^L\Delta f^T-2\dot{a}^T\Delta f^T-
f^T\Delta f^T+\\
&-f^L\Delta f^L+m^2a^T\Delta a^T+m^2a^L\Delta a^L>, \tag 1.15\endalign
$$
donde hemos tomado $a_0=a$, $a_i=\ep_{ij}\partial_ja^T+\partial_ia^L$, como 
en el Cap\'{\i}tulo II (a\-n\'a\-logamente para $f_m$). Observamos que $f$ y 
$a$ son multiplicadores a\-so\-cia\-dos a v\'{\i}nculos cuadr\'aticos que 
permiten despejarlos
$$
f=\Delta a^T\ \ ,\ \ a=\frac{1}{m^2}(\Delta f^T-\mu\Delta a^T).\tag 1.16
$$
Al sustituir (1.16) en la acci\'on llegamos a la forma desvinculada de 
$S_{PCS}$. Luego de definir
$$\align
Q &\equiv (-\Delta )^{1/2}a^T\ \ ,\ \ \pi \equiv (-\Delta )^{1/2}f^L,
\tag 1.17,a\\
q &\equiv \frac{1}{m}(-\Delta )^{1/2}(f^T-\mu a^T)\ \ ,\ \ 
p=m(-\Delta )^{1/2}a^L,\tag 1.17,b
\endalign
$$
la acci\'on reducida toma su forma final
$$\align
S^{(red)}_{PCS}=<\pi \dot{Q}&+p\dot{q} -\frac{1}{2}pp-\frac{1}{2}\pi\pi +
-\frac{1}{2}q(-\Delta )q\\
&-\frac{1}{2}Q(-\Delta +m^2)Q+\frac{1}{2}(mQ+\mu q)(mQ+\mu q)>,\tag 1.18
\endalign
$$
que muestra que el hamiltoniano es definido positivo.

En (1.18) observamos que pareciera existir alg\'un acoplamiento entre los dos 
grados de libertad, ya que tenemos un t\'ermino en $S_{PCS}^{(red)}$ de la 
forma $m\mu qQ$. Sin embargo vimos que el propagador es la suma de dos 
propagadores ``autoduales''. As\'{\i}, el sistema debe estar desacoplado. De 
hecho, si hacemos la transformaci\'on
$$\align
q_\pm &\equiv Q+\alpha_\pm q, \tag 1.19,a\\
p_\pm &\equiv \pm \frac{(\alpha_\pm \pi-p)}{\alpha_+-\alpha_-},\tag 1.19,b
\endalign
$$
con
$$
\alpha_\pm =\pm \frac{m}{m_\pm}, \tag 1.19,c
$$
el sistema se desacopla. Si adicionalmente redefinimos
$$\align
Q_\pm &= (\frac{m_\pm}{m_++m_-})^{1/2}q_\pm ,\tag 1.20,a\\
P_\pm &= (\frac{m_++m_-}{m_\pm})^{1/2}p_\pm ,\tag 1.20,b
\endalign
$$
llegamos a la expresi\'on final, desacoplada, de $S_{PCS}$
$$\align
S^{(red)}_{PCS}=<P_+\dot{Q}_+&+P_-\dot{Q}_--\frac{1}{2}P_+P_+-\frac{1}{2}
P_-P_--\frac{1}{2}Q_+(-\Delta +m^2_+)Q_++\\
&-\frac{1}{2}Q_-(-\Delta +m^2_-)Q_->. \tag 1.21
\endalign
$$

\vskip 5mm
\noi
{\bf 2. Teor\'{\i}a de Einstein autodual}

\vskip 3mm
\noi
{\bf 2.1 La acci\'on, an\'alisis covariante}

\vskip 3mm

A nivel linealizado la acci\'on de Einstein
$$
S_E^l=\frac{1}{2}<2h_{pa}\ep^{pmn}\partial_m\omega_n{}^a-(\omega_p{}^a
\omega_a{}^p-\omega_p{}^p\omega_a{}^a)>, \tag 2.1
$$
es invariante bajo las transformaciones locales
$$\align
\delta h_{pa} &=\partial_p\xi_a\ \ ,\ \ \delta \omega_{pa}=0 \tag 2.2,a\\
\delta h_{pa} &= -\ep_{pac}l^c\ \ ,\ \ \delta \omega_{pa}=-\partial_pl_a ,
\tag 2.2,b
\endalign
$$
donde (2.2,a) son las transformaciones de calibre y corresponden, a nivel 
curvo, a las transformaciones de difeomorfismos. (2.2,b) son las 
correspondientes transformaciones de Lorentz locales.
\footnote"*"{\ninerm{Ver ap\'endice A}}

El t\'ermino $CS$ tri\'adico linealizado
$$
S_{TCS}^l=\frac{1}{2}<h_{pa}\ep^{prs}\partial_rh_s{}^a>, \tag 2.3
$$
es invariante s\'olo bajo transformaciones de calibre. Por \'ultimo, el 
t\'ermino de Fierz-Pauli
$$
S_{FP}=\frac{1}{2}<h_p{}^ah_a{}^p-h_p{}^ph_a{}^a>, \tag 2.4
$$
no goza de ninguna de estas invariancias.

El contexto presentado se encuentra dentro del mismo esp\'{\i}ritu con que 
analizamos las teor\'{\i}as curvas, cuando presentamos los distintos 
t\'erminos $S^2_{CS}$, $S_E$ y $S_{TCS}$ como pertenecientes a un esquema 
jer\'arquico de simetr\'{\i}as. Aqu\'{\i}, partimos con la acci\'on de 
Einstein y terminamos con el t\'ermino de Fierz-Pauli. Cuando consideramos 
$S_E^l-\mu S^l_{TCS}$, vimos que corresponde a una excitaci\'on masiva de 
helicidad 2. Esta es s\'olo invariante bajo difeomorfismos. Decimos que la 
invariancia Lorentz ha sido rota y producto de esto 
tenemos el espectro f\'{\i}sico descrito. Cuando 
consideramos $-\mu S^l_{TCS}- m^2S_{FP}$, tenemos la teor\'{\i}a 
autodual. Decimos que la invariancia de calibre ha sido rota y producto de 
esto tenemos una excitaci\'on masiva de masa $m^2/|\mu |$ y helicidad $-2$
\footnote"${}^\dagger$"{\ninerm{Ver sub-secci\'on 1.2 del Cap\'{\i}tulo III}}. 
Este proceso es an\'alogo al caso que tomamos el t\'ermino de $CS$ vectorial 
y le sumamos el t\'ermino de Proca, dando como resultado la acci\'on $AD$ 
vectorial. Esta aparece si tomamos una teor\'{\i}a con un campo escalar 
acoplado minimalmente con un campo electromagn\'etico, donde tenemos el 
t\'ermino de $CS$ vectorial, en vez del t\'ermino de Maxwell y aplicamos el 
formalismo de Higgs para romper la simetr\'{\i}a de calibre. Por \'ultimo 
$S_E-m^2S_{FP}$ es la acci\'on de Fierz-Pauli, que corresponde por 
analog\'{\i}a a la acci\'on de Proca.

Vista la analog\'{\i}a con el caso vectorial, cabe, entonces, preguntar que 
sucede si consideramos la acci\'on [52,54,55,56]
$$
S^{(AD)}_E=S_E-\mu S^l_{TCS}-m^2S_{FP}, \tag 2.5
$$
la cual apodamos acci\'on de Einstein autodual. Miremos primero cual ser\'a 
su espectro f\'{\i}sico. Al hacer variaciones respecto a $\omega_m{}^a$, 
podemos obtener la expresi\'on de $\omega_m{}^a$ en funci\'on de $h_m{}^a$. 
Al sustituirla en la variaciones respecto a $h_{pa}$, llegamos a
$$\align
(\frac{1}{2}\ep^{pma}\ep^{srb}-\ep^{pmb}\ep^{sra})\partial_m\partial_rh_{sb}
-&\mu \ep^{prs}\partial_rh_s{}^a+\\
&-m^2(h^{ap}-\eta^{pa}h_s{}^s)=0. \tag 2.6
\endalign
$$

En funci\'on de proyectores y operadores de transferencia, (2.6) queda como
$$\align
[(\sq^{1/2}-&m)P^2_{S}-\mu \sq^{1/2}(P^2_{+S}-P^2_{-S})
+m^2(P^1_E-P^1_S)-(\sq -m^2)P^0_S\\
-&\frac{\mu}{2}\sq^{1/2}(P^1_{+S}+P^1_{+E}+P^1_{+SE}+P^1_{+ES}-P^1_{-S}
-P^1_{-E}-P^1_{-SE}-P^1_{-SE})+\\
&+m^2P^0_B+\mu \sq^{1/2}(P^0_{BS}+P^0_{SB})+\sqrt{2}m^2(P^0_{WS}+P^0_{SW})]
h=0.\tag 2.7
\endalign
$$
Aplicamos $P^0_{SW}$, $P^0_B$, $P^0_{WS}$ sobre (2.7) y conseguimos que
$$\align
&\sqrt{2}m^2P^0_Sh = 0,\tag 2.8,a\\ 
&(m^2P^0_B+\mu \sq^{1/2}P^0_{BS})h =0, \tag 2.8,b\\ 
&((\sq -m^2)P^0_{WS}-\mu \sq^{1/2}P^0_{WB}-\sqrt{2}m^2P^0_W)h=0.\tag 2.8,c
\endalign
$$
de donde obtenemos que $P^0_Sh=P^0_Bh=P^0_Wh=0$. Aplicando $P^1_{\pm S}$ y 
$P^1_{\pm E}$ sobre (2.7)
$$\align
&\pm \frac{\mu}{2}\sq^{1/2}(P^1_{\pm S}+P^1_{\pm SE})h +m^2P^1_{\pm S}h=0, 
\tag 2.9,a\\
&\pm \frac{\mu}{2}\sq^{1/2}(P^1_{\pm E}+P^1_{\pm ES})h +m^2P^1_{\pm E}h=0, 
\tag 2.9,b
\endalign
$$
de donde $P^1_Sh=P^1_Eh=0$. As\'{\i}, nos queda solo el sector de spin 2
$$
(\sq \mp \mu \sq^{1/2}-m^2)P^2_{\pm S}h=0, \tag 2.10
$$
o
$$
(\sq^{1/2}-m_\pm )(\sq^{1/2}+m_\mp )P^2_{\pm S}h=0, \tag 2.10$^{'}$
$$
Observamos que los \'unicos polos corresponder\'an a una excitaci\'on de masa 
$m_+$ con helicidad 2 y otra de masa $m_-$ con helicidad $-2$. $m_\pm$ tienen 
la misma forma que en el caso vectorial $(\ep =\mu /m)$
$$
m_\pm =\frac{m\ep}{2}(\sqrt{1+\frac{4}{\ep^2}}\pm 1). \tag 2.11
$$
A semejanza del caso vectorial, la ecuaci\'on (2.10), 
verificada por la parte trans\-ver\-sa y sin traza, tambi\'en se factoriza en un 
producto de condiciones de auto dualidad (Cap\'{\i}tulo III, ecuaci\'on 
(1.12))
$$\align
-(\sq^{1/2}(P^2_{+S}&-P^2_{-S})-m_+P^2_S)(-\sq^{1/2}(P^2_{+S}-P^2_{-S})-
m_-P^2_S)h^{Tt}=\\ 
&=[(\sq-m_+m_-)P^2_S-(m_+-m_-)\sq^{1/2}(P^2_{+S}-P^2_{-S})]h^{Tt}\\
&=[(\sq-m_+m_-)P^2_S-\mu \sq^{1/2}(P^2_{+S}-P^2_{-S})]h^{Tt},\tag 2.12
\endalign
$$
de obtenemos (2.10) aplicando $P^2_{+S}$ o $P^2_{-S}$. Cabe resaltar que si 
$K_{AD}(m)h=0$, son las ecuaciones de movimiento de la 
acci\'on autodual y $K^{(AD)}_E(m_+,m_-)h=0$ las de acci\'on de Einstein 
autodual, resulta que
$$
-K_{AD}(m_+)K_{AD}(m_-)h \neq K^{(AD)}_E(m_+,m_-)h, \tag 2.13
$$
a menos que reduzcamos $h_{ma}$ a sus grados f\'{\i}sicos de libertad 
(i.e. $h_{ma}^{Tt}$).

Miremos el propagador. Si acoplamos $h_{ma}$ con un corriente $J^{ma}$ las 
ecuaciones son ahora
$$
K^{(AD)}_Eh=-\kappa J.\tag 2.14
$$
Aplicando las reglas ortogonalidad entre operadores y pro\-yec\-to\-res
\footnote"${}^\star$"{\ninerm{Ver ap\'endice B}}, de igual manera a como 
hemos hecho en cap\'{\i}tulos anteriores, es posible invertir (2.14)
$$
h=-\Delta^{(AD)}_EJ, \tag 2.15,a
$$
donde
$$\align
\Delta_E^{(AD)}=&\frac{\kappa P^2_{+S}}{(\sq^{1/2}-m_+)(\sq^{1/2}+m_-)}+
\frac{\kappa P^2_{-S}}{(\sq^{1/2}+m_+)(\sq^{1/2}-m_-)}+\\ 
&+\kappa \frac{(m_+-m_-)\sq^{1/2}}{2m_+^2m_-^2}
(P^1_{+S}+P^1_{-SE}+P^1_{+E}+P^1_{-ES})+\\
&-\kappa \frac{(m_+-m_-)\sq^{1/2}}{2m_+^2m_-^2}
(P^1_{-S}+P^1_{+SE}+P^1_{-E}+P^1_{+ES})+\\
&+\frac{\kappa }{m_+m_-}\Bigl[(P^1_E-P^1_S)+P^0_B+\frac{1}{\sqrt{2}}(P^0_{SW}+
P^0_{WS})\Bigr]+\\
&+\frac{\kappa }{m_+^3m_-^3}{\Bigl[(m_+^2+m_-^2-m_+m_-)\sq -m_+^2m^2_-)}P^0_W+\\
&\ \ \ \ \ \ \ \ \ -\sqrt{2}m_+m_-(m_+-m_-)\sq^{1/2}(P^0_{BW}+
P^0_{WB})\Bigr].\tag 2.15,b
\endalign
$$
Para ver la presencia de los propagadores autoduales de cada ex\-ci\-ta\-ci\'on, 
debemos hallar en principio el propagador de $S^2_{AD}$ (Ca\-p\'{\i}\-tu\-lo III, 
ecuaci\'on (1.13)). Si sumamos a $S_{AD}^2$ un t\'ermino de la forma 
$<\kappa h_{mn}J^{mn}>$, las ecuaciones de movimiento ser\'an
$$\align
[(\sq^{1/2}&-m)P^2_{+S}-(\sq^{1/2}+m)P^2_{-S}+\frac{1}{2}(\sq^{1/2}-2m)
(P^1_{+S}-P^1_{-E})+\\
&+\frac{\sq^{1/2}}{2}(P^1_{+SE}+P^1_{+ES}-P^1_{-SE}-P^1_{-ES})+\\
&+\frac{1}{2}(\sq^{1/2}+2m)(P^1_{+S}-P^1_{-S})-\sq^{1/2}(P^0_{BS}+P^0_{SB})+\\ 
&+m(P^0_S+P^0_B+\sqrt{2}(P^0_{WS}+P^0_{SW})]h=-\frac{\kappa }{m}J.\tag 2.16
\endalign
$$

Despu\'es de cierto trabajo, usando las reglas de ortogonalidad entre los 
distintos operadores obtenemos que
$$
h=-\Delta^{2(+)}_{(AD)}(m)J\tag 2.17
$$
donde el super\'{\i}ndice (+) nos indica que la excitaci\'on tiene 
helicidad $+2$, $m$ es la masa de la misma. El propagador para cuando se 
propaga una excitaci\'on de masa $m$ y helicidad $-2$ se obtiene, 
simplemente, sustituyendo $m$ por $-m$ en $\Delta^{2(+)}_{(AD)}(m)$. 
Lo llamamos $\Delta^{2(-)}_{(AD)}(m)$. Expl\'{\i}citamente
$$\align
\Delta^{2(+)}_{(AD)}(m) &=\frac{\kappa }{m(\sq^{1/2}-m)}P^2_{+S}-
\frac{\kappa }{m(\sq^{1/2}+m)}P^2_{-S}+\frac{\kappa }{m^2}(P^1_E-P^1_S)+\\ 
&-\frac{\kappa \sq^{1/2}}{2m^3}(P^1_{+S}+P^1_{+E}-P^1_{-S}-P^1_{-E})+\\
&+\frac{\kappa \sq^{1/2}}{2m^3}(P^1_{+SE}+P^1_{+ES}-P^1_{-SE}-P^1_{-ES})+\\
&+\frac{\kappa }{2m^4}[(\sq -m^2)P^0_W+\sqrt{2}m\sq^{1/2}(P^0_{WB}+P^0_{BW})]+\\
&+\frac{\kappa }{m^2}[P^0_B+\frac{1}{\sqrt{2}}(P^0_{SW}+P^0_{WS})].\tag 2.18
\endalign
$$
Entonces es inmediato ver que
$$
\Delta^{(AD)}_E=\frac{1}{m_++m_-}[m_+\Delta^{(+)}_{(AD)}(m_+)+
m_-\Delta^{(-)}_{(AD)}(m_-)], \tag 2.19
$$
donde, al igual que en el caso vectorial, el propagador es la suma de 
propagadores ``autoduales''. Las restricciones que por razones f\'{\i}sicas 
deba cumplir $J$ ser\'an las mismas tanto en la teor\'{\i}a de Einstein 
autodual como en la teor\'{\i}a autodual.

\newpage
\vskip 5mm
\noi
{\bf 2.2 Descomposici\'on 2+1}
\vskip 3mm

Partimos de (2.5) y llegamos inicialmente a 
$$\align
S^{(AD)}_E=< &h_{00}[\mu \ep_{ij}\partial_ih_{j0}-\ep_{ij}\partial_i 
\omega_{j0}-m^2h_{ii}]-\omega_{00}[\omega_{ii}-\ep_{ij}\partial_ih_{j0}]+\\
&+h_{0k }[-\mu \ep_{ij}\partial_ih_{jk }+\ep_{ij}\partial_i\omega_{jk }+m^2
h_{k 0}]+\\
&+\omega_{0k }[\ep_{ij}\partial_ih_{jk }+\omega_{k 0}]+\frac{\mu}{2}h_{ik }
\ep_{ij}\dot{h}_{jk }+\\
&-\frac{\mu}{2}h_{i0}\ep_{ij}\dot{h}_{j0}+h_{i0}\ep_{ij}\dot{\omega}_{j0}-
h_{ik }\ep_{ij}\dot{\omega}_{jk }+\\
&-\frac{m^2}{2}(h_{ij}h_{ji}-h_{ii}h_{jj})-\frac{1}{2}(\omega_{ij}
\omega_{ji}-\frac{1}{2}\omega_{ii}\omega_{jj})>,\tag 2.20
\endalign
$$
donde observamos que $h_{00}$, $h_{0k }$, $\omega_{00}$ y $\omega_{0k }$ son 
multiplicadores, cuyos v\'{\i}nculos asociados permiten despejar 
$\omega_{k 0}$ y $h_{k 0}$, y 
$$\align
\omega_{ii} &= \frac{1}{m^2}(\delta_{ij}\Delta -\partial_i\partial_j)
(\omega_{ij}-\mu h_{ij}), \tag 2.21,a\\
h_{ii} &= \frac{1}{m^4}(\delta_{ij}\Delta -\partial_i\partial_j)
(\mu^2+m^2)(h_{ij}-\mu \omega_{ij}), \tag 2.21,b
\endalign
$$
Hacemos la descomposici\'on usual
$$
h_{ij}=(\delta_{ij}\Delta -\partial_i\partial_j)h^T+\partial_i\partial_j
h^L+(\ep_{ik }\partial_k \partial_j+\ep_{jk }\partial_k \partial_i)h^{TL}+
\ep_{ij}V, \tag 2.22,a
$$
$$
\omega_{ij}=(\delta_{ij}\Delta -\partial_i\partial_j)\omega^T+\partial_i
\partial_j\omega^L+(\ep_{ik }\partial_k \partial_j+\ep_{jk }\partial_k \partial_i)
\omega^{TL}+\ep_{ij}\lambda, \tag 2.22,b
$$
que sustituimos en (2.20) junto con la soluci\'on de los v\'{\i}nculos 
(2.21) (donde despejamos $\Delta h^L$ y $\Delta \omega^L$), 
adem\'as de $\omega_{k 0}$ y $h_{k 0}$. Llegamos as\'{\i} a
$$\align
S_E^{(AD)}=<&2\Delta \dot{h}^T[\Delta\omega^{TL}-\mu\Delta h^{TL}]+
2\Delta\dot{\omega}^T\Delta h^{TL}+\lambda \lambda +m^2VV+\\
&-\Delta \omega^{TL}\Delta \omega^{TL}-m^2\Delta h^{TL}\Delta h^{TL}+\Delta 
h^T(\Delta -m^2)\Delta h^T+\\
&+\Delta (\omega^T-\mu h^T)\frac{\Delta}{m^2}\Delta (\omega^T-\mu h^T)-
\Delta \omega^T\Delta \omega^T>.\tag 2.22
\endalign
$$
Haciendo las definiciones
$$\align
Q &\equiv \sqrt{2}\Delta h^T\ \ ,\ \ q\equiv \sqrt{2}
\frac{\Delta (\omega^T-\mu h^T)}{m}\tag 2.23,a\\
\pi &\equiv \sqrt{2}\Delta \omega^{TL}\ \ , \ \ p\equiv \sqrt{2}m
\Delta h^{TL}\tag 2.23,b
\endalign
$$
la acci\'on toma su forma final
$$\align
S^{(AD)(red)}_E=<\pi \dot{Q}&+p\dot{q} -\frac{1}{2}pp-\frac{1}{2}\pi\pi 
-\frac{1}{2}q(-\Delta )q\\
&-\frac{1}{2}Q(-\Delta +m^2)Q-\frac{1}{2}(mQ+\mu q)(mQ+\mu q)>,\tag 2.24
\endalign
$$
que corresponde a 2 excitaciones, con energ\'{\i}a definida positiva. El 
aspecto de (2.24) es exactamente el mismo que (1.18) del caso vectorial. 
As\'{\i}, sabemos que las dos excitaciones estan desacopladas y tienen masas 
$m_+$ y $m_-$. Este resultado reafirma lo que ha venido ocurriendo a lo largo 
del trabajo respecto a la analog\'{\i}a entre los casos vectoriales y 
tensoriales.

Este esquema se repite con las teor\'{\i}as de spin 3 y suponemos que pueda 
generalizarse para cualquier spin entero [20].

\vskip 5mm
\noi
{\bf 3.- La no viabilidad de romper la simetr\'{\i}a para la teor\'{\i}a $TM$}

\vskip 3mm
\noi
{\bf 3.1.- La teor\'{\i}a con los dos t\'erminos de $CS$}
\vskip 3mm

Hemos visto que podemos romper, consistentemente, la si\-me\-tr\'{\i}\-a de la 
teor\'{\i}a $S_{VCS}$ linealizada. La teor\'{\i}a $TM$ es invariante bajo 
transformaciones locales de calibre y de Lorentz. Esto nos presenta un marco 
de simetr\'{\i}as mas amplio para ``romper''. A\-na\-li\-za\-mos primero la 
posibilidad de que se pierda s\'olo la invariancia Lorentz.

La acci\'on de la teor\'{\i}a $TM$ es
$$
S^{2,l}_{TM}=\frac{1}{4\mu}\Bigl<-\ep^{lmr}\partial_lh^{(s)}_{mp}G^{pa}
(h^{(s)})\eta_{ra} + \mu h^{(s)}_{pa}G^{pa}(h^{(s)})\Bigr>,\tag 3.1  
$$
donde, como siempre
$$
h^{(s)}_{ma}=h_{ma}+h_{am}\tag 3.2
$$
$S^{2,l}_{TM}$ es claramente invariante bajo las transformaciones men\-cio\-na\-das. 
En particular la invariancia Lorentz en (3.1) es trivial pues la acci\'on 
depende s\'olo de $h^{(s)}_{ma}$. Podemos reescribir (3.1) si notamos que la 
expresi\'on de $\omega_m{}^a$, en funci\'on de $h_m{}^a$, cuando la torsi\'on 
es nula
$$
\omega_a{}^m=-\frac{1}{2}\delta_a{}^m\ep^{rse}\partial_rh_{sl}+\ep^{mrs}
\partial_rh_s{}^a, \tag 3.3
$$
nos permite obtener
$$\align
\ep^{pmn}\partial_m\omega_n{}^a &=-G^{pa}(h^{(s)}), \tag 3.4,a\\
\frac{1}{2}\Bigl<\omega_{pa}\omega^{ap}-\omega_p{}^p\omega_a{}^a\Bigr>&=
-\frac{1}{4}\Bigl<h^{(s)}_{mn}G^{mn}(h^{(s)})\Bigr>=S_E. \tag 3.4,b
\endalign
$$
As\'{\i}
$$
S^{2,l}_{TM}=\frac{1}{2\mu}\Bigl<\omega_{pa}\ep^{pmn}\partial_m\omega_n{}^a - 
\mu(\omega_{pa}\omega^{ap}-\omega_p{}^p\omega_a{}^a)\Bigr>, \tag 3.5
$$
con $\omega_{pa}$ dado por (3.3).

Una expresi\'on a primer orden de $S^{2,l}_{TM}$ es 
$$\align
S^{2,l}_{TM_1}=\frac{1}{2\mu}\Bigl<\omega_{pa}&\ep^{pmn}\partial_m
\omega_n{}^a-\mu(\omega_{pa}\omega^{ap}-\omega_p{}^p\omega_a{}^a)+\\
&+2\mu\lambda_p{}^a\ep^{prs}(\partial^rh_{sa}-\omega_r{}^b\ep_{bsa})\Bigr>,
\tag 3.6
\endalign
$$
donde hemos agregado un ``multiplicador" $\lambda_p{}^a$ de forma que se 
forza el ``v\'{\i}nculo" (3.3). (3.6) es invariante bajo las transformaciones 
de calibre: $\delta h_{pa}=\partial_p\xi_a$, $\delta\omega_{pa}=0$ y las de 
Lorentz: $\delta h_{pa}=-\ep_{pab}l^b$, $\delta\omega_{pa}=-\partial_pl_a$. 
En ambos casos $\delta\lambda_p{}^a=0$.

Rompemos la invariancia de Lorentz sumando, a la acci\'on, el t\'ermino 
$S_{TCS}^l$. As\'{\i}, partimos de la acci\'on [55,56]
$$\align
S_{(CS+CS)} &= S^{2,l}_{TM_1} - m S^{l}_{TCS}\\ 
&=\frac{1}{2\mu}\Bigl<\omega_{pa}\ep^{prs}\partial_r\omega_s^a-\mu 
(\omega_{pa}\omega^{ap}-\omega_p{}^p\omega_a{}^a)+\\
&\ \ \ \ \ \ \ \ + 2\mu \lambda_p{}^a\ep^{prs}(\partial_rh_{sa}-\omega_r{}^b
\ep_{bsa})+\\
&\ \ \ \ \ \ \ \ -m\mu h_{pa}\ep^{prs}\partial_rh_s{}^a\Bigr>. \tag 3.7
\endalign
$$
Las ecuaciones de movimiento provenientes de hacer variaciones respecto a 
$\omega_{pa}$, $\lambda_{pa}$ y $h_{pa}$ son
$$\align
\ep^{prs}\partial_r\omega_s{}^a &-\mu(\omega^{ap}-\eta^{pa}\omega_r{}^r)
-\mu(\lambda^{ap}-\eta^{pa}\lambda_r{}^r)=0, \tag 3.8,a\\
\ep^{prs}\partial_rh_s{}^a &- (\omega^{ap} - \eta^{pa}\omega_r{}^r)=0 
\tag 3.8,b\\
\ep^{prs}\partial_r&(\lambda_s{}^a-mh_s{}^a)=0, \tag 3.8,c
\endalign
$$
donde (3.8,b) es la ecuaci\'on equivalente a (3.3).

Si tomamos la divergencia respecto a $p$ en (3.8) y sacamos su parte 
antisim\'etrica obtenemos que $\omega_{pa}$ y $\lambda_{pa}$ son 
sim\'etricos; y para $\omega_{pa}$, $\lambda_{pa}$ y $h_{pa}$ se cumple
$$
\partial_a(*)^{pa}-\partial^p(*)_a{}^a=0. \tag 3.10
$$
Tomando la traza de (3.8), tendremos en consecuencia que
$$\align
m\ep^{prs}\partial_rh_{sp}&=\omega_r{}^r=\lambda_r{}^r=0, \tag 3.11,a\\
\partial_p\omega^{pa}&=\partial_p\lambda^{pa}=0. \tag 3.11,b
\endalign
$$
As\'{\i} que de $\omega_{pa}$ y $\lambda_{pa}$ nos quedar\'an s\'olo sus 
partes transversas y sin traza. Tomando divergencia respecto a ``$a$" en 
(3.8) y con (3.11,a) tendremos que si escogemos el calibre 
$\partial_ph^{pa}=0$, entonces $h_{pa}$ es sim\'etrico y como resultado, al 
igual que para $\lambda_{pa}$ y $\omega_{pa}$ nos quedar\'a solo su parte 
transversa y sin traza. Tenemos, por tanto, que el sistema (3.8) describe 
excitaciones de spin 2 puro.

Al tomar en cuenta que s\'olo importan las partes $\omega_{pa}^{Tt}$, 
$\lambda_{pa}^{Tt}$ y $h_{pa}^{Tt}$, obtenemos que para $h_{pa}^{Tt}$ se 
cumple
$$
\ep^{prs}\partial_r\sq h^{Tt_a}_s - \mu\sq h^{Tt_{pa}} + m\mu\ep^{prs}
\partial_r h^{Tt_a}_s=0, \tag 3.12
$$
que al proyector en sus partes $h_{pa}^{Tt_{(+)}}$ y $h_{pa}^{Tt_{(-)}}$, nos 
da
$$
(\sq^{1/2}(\sq^{1/2}\mp\mu )- m\mu )\sq^{1/2}h^{Tt_{(\pm )}}_{pa}=0, 
\tag 3.13
$$
o
$$
(\sq^{1/2}\mp M_+)(\sq^{1/2}\pm M_-)\sq^{1/2}h^{Tt_{(\pm )}}_{pa}=0, 
\tag 3.13
$$
con $(\ep =\mu /m)$
$$
M_\pm =\frac{m\ep}{2}\Bigl(\sqrt{1+\frac{4}{\ep}}\pm 1\Bigr) .\tag 3.14
$$
La posible excitaci\'on sin masa no aparece. Esto se ve mas claro al hacer la
descomposici\'on $2+1$ de $S_{(CS+CS)}$. Sin embargo, notaremos que el sistema 
no es viable pues carece de energ\'{\i}a definida positiva.

Al hacer la descomposici\'on 2+1 obtenemos inicialmente
$$\align
S_{(CS+CS)}=\frac{1}{2\mu}\Bigl<2\omega_{0a}&C^a+2h_{0a}D^a+2\lambda_{0a}
E^a+\\
&+\omega_{i0}\ep_{ij}\dot{\omega}_{j0}-\omega_{ik }\ep_{ij}\dot{\omega}_{jk }
-m\mu h_{i0}\ep_{ij}\dot{h}_{j0}+\\ 
&+m\mu h_{ik }\ep_{ij}\dot{h}_{jk }+2\mu\lambda_{i0}\ep_{ij}\dot{h}_{j0}+\\  
&-2\mu\lambda_{ik }\ep_{ij}\dot{h}_{jk }+\mu (\omega_{ii}\omega_{jj}-
\omega_{ij}\omega_{ji})+\\
&+2\mu(\lambda_{ii}\omega_{jj}-\lambda_{ij}\omega_{ji})\Bigr>, \tag 3.15
\endalign
$$
donde los v\'{\i}nculos asociados a los multiplicadores $\omega_{0a}$, 
$h_{0a}$ y $\lambda_{0a}$, son
$$\align
C^0 &\equiv -\ep_{ij}\partial_i\omega_{j0}-\mu\omega_{ii}-
\mu\lambda_{ii}=0, \tag 3.16,a\\
C^k &\equiv \ep_{ij}\partial_i\omega_{jk }+\mu\omega_{k 0}+\mu\lambda_{k 0}=0, 
\tag 3.16,b\\
D^0 &\equiv m\mu\ep_{ij}\partial_ih_{j0}-\mu\ep_{ij}\partial_i\lambda_{j0}=0, 
\tag 3.16,c\\
D^k  &\equiv -m\mu\ep_{ij}\partial_ih_{jk }+\mu\ep_{ij}\partial_i\lambda_{jk }=0,
\tag 3.16,d\\
E^0 &\equiv -\mu\omega_{ii}-\mu\ep_{ij}\partial_ih_{j0}=0, \tag 3.16,e\\
E^k  &\equiv \mu\omega_{k 0}+\mu\ep_{ij}\partial_ih_{jk }=0. 
\tag 3.16,f
\endalign
$$
Descomponemos $h_{ij}$ y $\omega_{ij}$ como en (2.22). A $\lambda_{ij}$ lo 
descomponemos igualmente
$$
\lambda_{ij}=(\delta_{ij}\Delta -\partial_i\partial_j)\rho^T+
\partial_i\partial_j\rho^L+(\ep_{ik }\partial_k \partial_j+\ep_{jk }\partial_k 
\partial_i)\rho^{TL}+\ep_{ij}\rho \tag 3.17.
$$
Del conjunto de v\'{\i}nculos (3.16) despejamos $\omega_{k 0}$, $\lambda_{k 0}$, 
$\Delta\omega^L$, $\Delta\rho^{TL}$, $\Delta\rho^T$, $\Delta\rho^L$ y la 
parte transversa de $h_{k 0}$, $\Delta N^T$, obteniendo
$$\align
\omega_{k 0}&=\ep_{k j}\partial_j\Delta h^T + \partial_k (\Delta h^{TL}+V),
\tag 3.18,a\\
\lambda_{k 0}&=\frac{1}{\mu}[\ep_{k j}\partial_j
(\Delta\omega^T-\mu\Delta h^T)+\\ 
&\ \ \ \ \ \ \ \ \ \ \ \ \ \ \partial_k (\Delta\omega^{TL}
+\lambda -\mu\Delta h^{TL}-\mu v)], \tag 3.18,b\\ 
\Delta\omega^L&=\frac{(\Delta -m\mu )\Delta\omega^T}{m\mu}-\frac{\Delta^2}{m}
h^T, \tag 3.18,c\\
\Delta \rho^{TL}+\rho &=m(\Delta h^{TL}+V),\tag 3.18,d\\
\Delta \rho^T &=m\Delta h^T,\tag 3.18,e\\
\Delta \rho^L&=-\frac{\Delta^2}{m\mu}\omega^T+\frac{1}{m\mu}((m+\mu )\Delta 
-m^2\mu )\Delta h^T, \tag 3.18,f\\
\Delta N^T &=\frac{1}{m\mu}\Delta^2(\omega^T-\mu h^T). \tag 3.18,g
\endalign
$$
Al sustituir (3.18) en (3.15), llegamos a
$$\align
S_{(CS+CS)}=\frac{1}{\mu}\Bigl<2\mu&\Delta\dot{h}^{T}H^{TL}+2\Delta
\dot{\omega}^T\Delta\omega^{TL}+\\ 
&+2\Delta\omega^T(\Delta -m\mu )\Delta h^T-\mu\Delta\omega^T\Delta\omega^T +\\
&-\mu\Delta h^T\Delta\Delta h^T-2\mu\Delta\omega^{TL}(H^{TL} + \rho ) +\\
&+2\mu\rho(\Delta\omega^{TL}+\lambda )-\mu\Delta\omega^{TL}\Delta\omega^{TL}+
\mu\lambda\lambda\Bigr>, \tag 3.19
\endalign
$$
donde
$$
H^{TL} \equiv m(\Delta h^{TL} + V) - \rho .\tag 3.20
$$
Observamos que la parte longitudinal de $h_{k 0}$, $\Delta N^L$, no aparece 
en la acci\'on. Esta constituye la parte sensible a transformaciones de 
calibre de esta variable. As\'{\i}, (3.20) es expl\'{\i}citamente invariante 
de calibre $(\delta H^{TL}=0)$. $\lambda$ y $\rho$ aparecen como 
multiplicadores de Lagrange asociados a v\'{\i}nculos cuadr\'aticos que 
permiten despejarlos
$$
\lambda = - \rho = \Delta\omega^{TL}. \tag 3.21
$$

Al sustituir $\lambda$ y $\rho$, y hacer las redefiniciones
$$\align
Q_1 &=\sqrt{2}\Delta h^T\ \ ,\ \ P_1=\sqrt{2}H^{TL},\tag 3.22,a\\
Q_2 &=\sqrt{2}\frac{\Delta}{\mu}\omega^{TL}\ \ ,\ \ 
P_2=\sqrt{2}\Delta\omega^{TL}, \tag 3.22,b
\endalign
$$
llegamos a la forma final, reducida, de $S_{(CS+CS)}$
$$\align
S^{(red)}_{(CS+CS)}=\Bigl<\dot{Q}_1P_1 &+\dot{Q}_2P_2+\frac{1}{2}P_1P_1-
\frac{1}{2}P_2P_2+\\
&-\frac{1}{2}(P_1+P_2)^2-\frac{1}{2}Q_2(-\Delta )Q_2+\\
&-\frac{1}{2}(mQ_1+\mu Q_2)^2+\frac{1}{2}(Q_1-Q_2)(-\Delta )(Q_1-Q_2)+\\
&+\frac{1}{2}m^2Q_1Q_1 \Bigr>, \tag 3.23
\endalign
$$
con energ\'{\i}a no definida positiva. As\'{\i}, la teor\'{\i}a $TM$ 
linealizada no permite una rotura espont\'anea, consistente, de su 
simetr\'{\i}a bajo transformaciones de Lorentz.

\noi
{\bf 3.2.- La teor\'{\i}a TM con todas sus simetr\'{\i}as rotas}
\vskip 3mm

Miremos, ahora, que sucede si consideramos la posibilidad, mas fuerte, de 
poder romper todas las simetr\'{\i}as de la teor\'{\i}a $TM$, linealizada. 
Esto se consigue si le sumamos a $S^{2,l}_{TM}$ el t\'ermino de Fierz-Pauli. 
Partimos entonces de la acci\'on [55,56]
$$\align
S_{(TM+FP)}&=S^{2,l}_{TM_1}+m^2S_{FP}\\
&=\frac{1}{2\mu}\Bigl<\omega_{pa}\ep^{prs}\partial_r\omega_s{}^a-\mu 
(\omega_{pa}\omega^{pa}-\omega_p{}^p\omega_a{}^a)+\\
&\ \ \ \ \ \ \ \ \ \ +2\mu \lambda_p{}^a\ep^{prs}(\partial_rh_{sa}-
\omega_r{}^b\ep_{bsa})+\\
&\ \ \ \ \ \ \ \ \ \ +m^2\mu(h_{pa}h^{ap}-h_p{}^ph_a{}^a)\Bigr>.\tag 3.24
\endalign
$$
Las ecuaciones de movimiento son
$$\align
&\ep^{prs}\partial_r\omega_s{}^a-\mu(\omega^{ap}-\eta^{pa}\omega_r{}^r)-
\mu (\lambda^{ap}-\eta^{pa}\lambda_r{}^r)=0,\tag 3.25,a\\
&\ep^{prs}\partial_r\lambda_s{}^a+m^2(h^{ap}-\eta^{pa}h_r{}^r)=0,\tag 3.25,b\\
&\ep^{prs}\partial_rh_s{}^a-(\omega^{ap}-\eta^{pa}h_r{}^r)=0,\tag 3.25,c
\endalign
$$
Si tomamos divergencia respecto a $p$, sacamos la parte an\-ti\-si\-m\'e\-tri\-ca y la 
traza de (3.25), obtenemos que tanto para $\omega_{pa}$, $\lambda_{pa}$ y 
$h_{pa}$, se cumple
$$\align
&\partial_a(*)^{pa}-\partial^p(*)_a{}^a=0,\tag 3.26,a\\
&\ep_{pal}(*)^{pa}=0,\tag 3.26,b\\
&(*)_a{}^a=0,\tag 3.26,c
\endalign
$$
qued\'andonos solo la parte se spin 2 de cada uno. Las ecuaciones para 
$h_{pa}^{Tt(\pm)}$ son
$$
[\sq (\sq^{1/2}\mp \mu)\pm m^2\mu ]h_{pa}^{Tt(\pm)}=0. \tag 3.27
$$

Las posibles excitaciones, masivas tendr\'an masas que constituyen las 
ra\'{\i}ces positivas de $x^3\mp \mu x^2\pm m^2\mu$. Para la helicidad $+2$ 
observamos que $x^3$ crece mas lento que $\mu x^2-m^2\mu$, si $x<\mu$. Luego, 
si $\mu$ es grande comparado con $m^2$, tenemos la posibilidad de que existan 
2 ra\'{\i}ces positivas de la correspondiente ecuaci\'on para las masas. Una 
ser\'a menor que $\mu$ y la otra mayor que $\mu$ (estamos suponiendo que 
$\mu >0$). Las llamamos $m_{+1}$ y $m_{+2}$. La tercera ra\'{\i}z es negativa, 
pero constituye el opuesto de la \'unica ra\'{\i}z positiva de la ecuaci\'on 
de las masas para la helicidad $-2$. La llamamos $m_-$. Las otras dos 
ra\'{\i}ces de la ecuaci\'on para esta helicidad $(-2)$ son $-m_{+1}$ y 
$-m_{+2}$. Si cambiamos $\mu$ por $-\mu$ se invierten las helicidades 
del espectro. Si cambiamos $m^2$ por $-m^2$ aparecen ra\'{\i}ces complejas. 
Por lo tanto, para $\mu >0$, (3.27) se factoriza como
$$
(\sq^{1/2}\mp m_{+1})(\sq^{1/2}\mp m_{+2})(\sq^{1/2}\pm m_-)h_{pa}^{Tt(\pm)}
=0. \tag 3.28
$$
Veamos, sin embargo, que la energ\'{\i}a del sistema no es definida positiva.

Al hacer la descomposici\'on $2+1$, nos encontramos, al igual que en el caso 
que tratamos anteriormente, con que $\omega_{0a}$, $\lambda$ y $h_{0a}$ son 
multiplicadores de Lagrange asociados a los v\'{\i}nculos
$$\align
\ep_{ij}\partial_i\omega_{j0}+\mu\omega_{ii}+\mu\lambda_{ii}&=0, \tag 3.29,a\\
\ep_{ij}\partial_i\omega_{jk }+\mu\omega_{k 0}+\lambda_{k 0}&=0, \tag 3.29,b\\
-\ep_{ij}\partial_i\lambda_{j0}+m^2 h_{ii}&=0,\tag 3.29,c\\
\ep_{ij}\partial_i\lambda_{jk }-m^2h_{k 0}&=0, \tag 3.29,d\\
\omega_{ii}+\ep_{ij}\partial_i\lambda_{j0}&=0, \tag 3.29,e\\
\omega_{k 0}+\ep_{ij}\partial_ih_{jk }&=0, \tag 3.29,f
\endalign
$$
de donde (con las descomposiciones (2.22) y (3.17))

$$\align
\omega_{k 0}&=\ep_{k j}\partial_j\Delta h^T+\partial_k (\Delta h^{TL}+V), 
\tag 3.30,a\\
\lambda_{k 0}&=\frac{1}{\mu}[\ep_{k j}\partial_j(\Delta\omega^T-\mu\Delta h^T)+
\partial_k (\Delta\omega^{TL}+\\
&\ \ \ \ \ \ \ \ \ \ \ \ \ \ \ +\partial_k (\Delta\omega^{TL}
+\lambda -\mu\Delta h^{TL}-\mu V)], \tag 3.30,b\\
h_{k 0}&=-\frac{1}{m^2}[\ep_{k j}\partial_j\Delta\rho^T+\partial_k 
(\Delta\rho^{TL}+\rho )], \tag 3.30,c\\
\Delta\omega^L&=-\frac{\Delta^2}{m^2}\rho^T-\Delta\omega^T, \tag 3.30,d\\
\Delta\rho^L&=\frac{\Delta(\Delta -m^2)}{m^2}\rho^T+\frac{\Delta^2}{\mu}
h^T, \tag 3.30,e\\
\Delta h^L&=\frac{-\Delta^2}{m^2\mu}\omega^T+(\frac{\Delta-m^2}{m^2})
\Delta h^T .\tag 3.30,f
\endalign
$$
Al sustituir (3.29) en la acci\'on, tendremos
$$\align
S_{(TM+FP)}\Bigl|_{(3.29)}\Bigr.&=\frac{1}{2\mu}\Bigl<\omega_{i0}\ep_{ij}
\dot{\omega}_{j0}-\omega_{ik }\ep_{ij}\dot{\omega}_{jk }+
2\mu\lambda_{i0}\ep_{ij}\dot{h}_{j0}+\\
&\ \ \ \ \ \ \ \ \ -2\mu\lambda_{ik }\ep_{ij}
\dot{h}_{jk }+\mu(\omega_{ii}\omega{jj}-\omega_{ij}\omega_{ji})+\\
&\ \ \ \ \ \ \ \ \ +2\mu (\lambda_{ii}\omega{jj}-\lambda_{ij}\omega_{ji})
 -m^2\mu (h_{ii}h_{jj}-h_{ij}h_{ji})\Bigr>\\
&=\frac{1}{\mu}\Bigl<2\mu\Delta \dot{h}^T\Delta \rho^{TL}+2\Delta 
\dot{\omega}^T\Delta \omega^{TL}+2\mu \Delta \dot{\rho}^T\Delta h^{TL}+\\
&\ \ \ \ \ \ \ \ \ -\mu \Delta \omega^T\Delta \omega^T+\mu\lambda\lambda +
\mu \Delta \omega^{TL}\Delta \omega^{TL}+\\
&\ \ \ \ \ \ \ \ \ +2\Delta h^T\Delta\Delta \omega^T-\mu \Delta h^T
(\Delta -m^2)\Delta h^T+\\
&\ \ \ \ \ \ \ \ \ +m^2\mu \Delta h^{TL}\Delta h^{TL}-m^2\mu VV-
\frac{\mu}{m^2}\Delta \rho^T\Delta\Delta \rho^T+\\
&\ \ \ \ \ \ \ \ \ -2\mu \Delta \rho^T\Delta \omega^T+2\mu\lambda\rho -
2\mu\Delta\omega^{TL}\Delta \rho^{TL}\Bigr>,\tag 3.31
\endalign
$$
donde, entonces, $\lambda$, $\rho$ y $V$ aparecen como multiplicadores 
asociados a v\'{\i}nculos cuadr\'aticos.

Su soluci\'on, $\lambda =V=\rho =0$, la sustituimos y hacemos las 
redefiniciones
$$\align
\sqrt{2}\Delta h^T&=Q_1\ \ ,\ \ \sqrt{2}\Delta\rho^{TL}=P_1, \tag 3.32,a\\
\sqrt{2}\frac{\Delta\omega^T}{\mu}&=Q_2\ \ ,\ \ \sqrt{2}\Delta
\omega^{TL}=P_2, \tag 3.32,b\\
\sqrt{2}\frac{\Delta\rho^T}{m}&=Q_3\ \ ,\ \ \sqrt{2}\Delta h^{TL}
m=P_3. \tag 3.32,c
\endalign
$$
Llegamos, as\'{\i}, finalmente a la forma reducida de $S_{(TM+FP)}$
$$\align
S^{(red)}_{(TM+FP)}=\Bigl<P_1\dot{Q}_1&+P_2\dot{Q}_2+P_3\dot{Q}_3+\frac{1}{2}
P_1P_1+\\
&-\frac{1}{2}(P_1+P_2)^2+\frac{1}{2}P_3P_3-\frac{1}{2}Q_2(-\Delta )Q_2+\\
&+\frac{1}{2}Q_3(-\Delta )Q_3+\frac{1}{2}(Q_1-Q_2)(-\Delta )(Q_1-Q_2)+\\
&+\frac{1}{2}m^2Q_1Q_1+\frac{1}{2}m^2Q_3Q_3+\\
&-\frac{1}{2}(\mu Q_2+mQ_3)^2\Bigr>,\tag 3.33
\endalign
$$
la cual muestra que la energ\'{\i}a es no definida positiva.

Concluimos que tampoco podemos romper, consistentemente, las simetr\'{\i}as 
de $S^{2,l}_{TM}$, agregando un t\'ermino de Fierz-Pauli.

\newpage

$\ $

\pageno=75
\headline={\ifnum\pageno=75\hfil\else\hss\tenrm \folio\ \fi}

\vskip 1cm

\centerline{\biggbf Cap\'{\i}tulo {\catbf VI}}

\vskip 1cm

\centerline{\dseis COMPORTAMIENTO ANY\'ONICO}

\vskip 2mm

\centerline{\dseis EN TEOR\'IAS VECTORIALES}

\vskip 2mm

\centerline{\dseis Y DE GRAVEDAD LINEALIZADA}

\vskip 2cm

PLanteamos en este cap\'{\i}tulo la posibilidad de implementar spin y
es\-ta\-d\'{\i}s\-ti\-ca fraccionaria en teor\'{\i}as acopladas con un campo
electromagn\'etico o con un campo gravitacional d\'ebil. Nuevamente aparece una
estrecha analog\'{\i}a entre las teor\'{\i}as vectoriales y spin 2.

Desde otro punto de vista esta cap\'{\i}tulo sirve tambi\'en como un ejemplo de
las teor\'{\i}as ya analizadas en presencia de fuentes externas.

\vskip 3mm
\noi
{\bf 1.- Spin y estad\'{\i}stica en dimensi\'on 2+1}
\vskip 3mm

En 3 dimensiones espaciales el grupo
de rotaciones SO(3) es  
no abeliano. As\'{\i}, cuando cuantizamos puede demostrarse que la componente 
del momento angular en la direcci\'on de un eje fijo tiene autovalores que son 
m\'ultiplos semienteros de $\hbar$. Para el spin, por ser un momento angular 
intr\'{\i}nseco, la situaci\'on es an\'aloga. Decimos, cuando tomamos 
$\hbar =1$, que el spin en 3 dimensiones espaciales, tiene valores 
semienteros o enteros. Adem\'as, las funciones de onda que 
representan estados de dos o mas part\'{\i}culas id\'enticas son sim\'etricas 
o antisim\'etricas, respecto al ``intercambio" de dos de estas si, 
respectivamente, su spin es entero o semientero. 

En 2 dimensiones espaciales la situaci\'on es distinta. Las rotaciones 
planares se realizan alrededor de un eje com\'un, por lo que no habr\'a 
reglas de conmutaci\'on no abelianas. As\'{\i}, el momento angular, y por 
ende el spin, no est\'a restringido a tomar valores que sean m\'ultiplos de 
$1/2$. Esta posibilidad real de tener part\'{\i}culas con cualquier spin nos 
lleva a introducir el concepto de ``anyon" (del ingles $any$ = cualquiera) 
con el que identificamos a las part\'{\i}culas que posean spin que no sea 
entero o semientero. Respecto a la estad\'{\i}stica, adem\'as de las usuales 
(fermi\'onica y bos\'onica) en 2 dimensiones espaciales tenemos la 
posibilidad de tener otro tipo de estad\'{\i}sticas. Esto aparece de manera
natural si tomamos en cuenta la topolog\'{\i}a del espacio de configuraciones 
de $n$ part\'{\i}culas id\'enticas [57,58,59].

Cuando un sistema cl\'asico es cuantizado, es descrito por funciones 
de onda $\psi (q)$ sobre el espacio de configuraciones $Q$. La 
correspondencia entre los estados cu\'anticos del sistema y estas funciones 
de onda no es uno a uno, ya que la predictibilidad no cambia si multiplicamos 
la funci\'on de onda que representa un estado por un factor de fase. Si $Q$ 
es simplemente conexo siempre es posible definir esta fase globalmente [58]. 
Cuando $Q$ no es simplemente conexo esto no es posible hacerlo siempre.

La evoluci\'on del sistema viene dada por la ecuaci\'on de Schr\"o\-din\-ger
$$
i\hbar \partial_t \psi =H_{op}\psi \tag 1.1
$$
donde $H_{op}$ es el hamiltaniano del sistema. $H_{op}$ es un operador local, 
as\'{\i}, cuando $Q$ no es simplemente conexo podemos ``levantarlo" 
un\'{\i}vocamente al espacio de recubrimiento universal $\wt{Q}$, de $Q$ y 
hacer mec\'anica cu\'antica con funciones de onda univaluadas 
$\wt{\psi}(\wt{q})$ sobre $\wt{Q}$, quedando por resolver bajo que 
condiciones volvemos a $Q$.

Cuando tenemos a $\wt{Q}$, \'este viene dado junto a un mapa de recubrimiento 
$\pi :\wt{Q}\to Q$. Adem\'as, el grupo fundamental de $Q,\pi_1(Q)$, act\'ua 
libremente sobre $\wt{Q}$, y $Q$ es el cociente de $\wt{Q}$ bajo esta acci\'on
$$
\frac{\wt{Q}}{\pi_1(Q)}=Q. \tag 1.2
$$
Se dice que $\wt{Q}$ es un fibrado principal sobre $Q$ con grupo de 
estructura $\pi_1(Q)$. Esto puede aclararse si construimos 
expl\'{\i}citamente a $\wt{Q}$. Una manera de hacerlo es la siguiente [58,59]: 
Tomamos un punto $q_0\in Q$ fijo y formamos el conjunto, $\Cal{P}Q$, de todos 
los caminos que empiezan en $q_0$ y terminan en algun $q\in Q$. Llamemos a 
los elementos de $\Cal{P}Q$, $\Gamma_q$. Sea $[\Gamma_q]$ la clase de 
equivalencia de todos los caminos homot\'opicos a $\Gamma_q$. Para cada punto 
$q$ habr\'an varias clases $[\Gamma_q]$ asignadas. Puede probarse que
$$
\wt{Q}=\{[\Gamma_q],q\in Q\}.\tag 1.3
$$

El grupo fundamental $\pi_1(Q)$ es el conjunto de las clases de equivalencia 
asociadas a $q_0$
$$
\pi_1(Q)=\{[\Gamma_{q_0}],q_0\in Q\ {\text{fijo}}\}. \tag 1.4
$$
La estructura de grupo emerge si definimos el producto entre clases
$$
[\Gamma_{q_0}][\Gamma_{q_0}^{'}]=[\Gamma_{q_0}\cup \Gamma_{q_0}^{'}], 
\tag 1.5,a
$$
donde $\Gamma_{q_0}\cup \Gamma_{q_0}^{'}$ se entiende como el camino, 
cerrado, que resulta de recorrer $\Gamma_{q_0}$ y luego $\Gamma_{q_0}^{'}$. 
El inverso de cada elemento es la clase de equivalencia
$$
[\Gamma_{q_0}]^{-1}\equiv [\Gamma_{q_0}^{-1}], \tag 1.5,b
$$
donde $\Gamma_{q_0}^{-1}$ es $\Gamma_{q_0}$ recorrido en sentido inverso. La 
identidad son todos los caminos homot\'opicos a $q_0$
$$
e=[\Gamma_{q_0}][\Gamma_{q_0}]^{-1}=``q_0". \tag 1.5,c
$$
La acci\'on de $\pi_1(Q)$ sobre $\wt{Q}$ viene dada por
$$
[\Gamma_{q_0}]\to [\Gamma_{q_0}][\Gamma_q]\equiv [\Gamma_{q_0}\cup 
\Gamma_q]. \tag 1.6
$$

El mapa de recubrimiento es
$$
\pi :\wt{Q}\to Q;\ \pi ([\Gamma_{q}])=q. \tag 1.7
$$
Observamos que $\pi^{-1}(q)$ son puntos que difieren por una acci\'on de 
$\pi_1(Q)$. As\'{\i}, todo lazo $\gamma$ sobre alg\'un punto $q\in Q$ se 
levanta como un camino, $\wt{\gamma}$, en $\wt{Q}$ que empieza y termina en 
la fibra $\pi^{-1}(q)$.

$\wt{Q}$ puede descomponerse en la uni\'on de ``dominios 
fundamentales" [57,59], cada uno de los cuales es isomorfo a $Q$. Estos 
contienen una y s\'olo una pre\-ima\-gen de cada punto $q\in Q$ y podemos 
pasar de un dominio fundamental a otro por la acci\'on de $\pi_1(Q)$ sobre 
$\wt{Q}$. Definimos,ahora, una funci\'on de onda multivaluada 
$\psi (q)$ que toma los valores $\wt{\psi}([\gamma ]\wt{q})$, donde 
$[\gamma ]$ recorre todo $\pi_1(Q)$. Si queremos que para dos puntos $\wt{q}$ 
y $\wt{q}'$ en la fibra de $q$, $\wt{\psi}(\wt{q})$ y $\wt{\psi}(\wt{q}')$ 
tengan la misma predictibilidad f\'{\i}sica, entonces
$$
\wt{\psi}([\gamma ]\wt{q})=a([\gamma ])\wt{\psi}(\wt{q}), 
\ \ \ \ \ \ \forall \wt{q}\in \wt{Q} \tag 1.8
$$
donde $|a([\gamma ])|=1$, para todas las $[\gamma ]\in \pi_1(Q)$. Puede 
probarse que $a([\gamma ])$ es una representaci\'on unitaria 
unidimensional de $\pi_1(Q)$. Tenemos entonces que se induce una acci\'on de 
$\pi_1(Q)$ sobre $\psi (q)$ en el sentido que si movemos a $q$ sobre un lazo 
$\gamma$ en $Q$ las funciones de onda difieren por la fase $a([\gamma ])$.

Finalmente, debido a que el principio de superposici\'on se cum\-ple entre 
funciones de onda que adquieren el mismo $a([\gamma ])$ por la acci\'on de 
$\pi_1(Q)$, tendremos distintos sectores en el espacio de Hilbert de estados 
caracterizados por el factor $a([\gamma ])$ que adquieren las funciones de 
onda $\psi (q)$ al mover $q$ sobre un lazo $\gamma$ en $Q$ [57]. Cuando $Q$ 
es el espacio de configuraciones de n part\'{\i}culas id\'enticas $M_n$, la 
acci\'on de $\pi_1(M_n)$ sobre las funciones de onda corresponde justamente a 
hacer un intercambio. Veamos entonces como es $M_n$ [60-63].

Consideramos $n$ part\'{\i}culas id\'enticas en $\Cal{R}^d$. En principio 
tendremos que 
$$
M_n \subset \undersetbrace{n-veces}\to{\Cal{R}^d\times \Cal{R}^d\times \cdots 
\times \Cal{R}^d}\equiv \Cal{R}^{nd}.\tag 1.9
$$
La indistinguibilidad se hace presente si tomamos el cociente de 
$\Cal{R}^{nd}$ por la acci\'on del grupo sim\'erico $S_n$. Adem\'as pediremos 
que las part\'{\i}culas no colisionen. Esto corresponde a eliminar todos los 
puntos diagonales de $\Cal{R}^{nd}$, (i.e.)
$$
\Delta \equiv \cases 
\ovr{r}_1,\cdots ,\ovr{r}_n \in \Cal{R}^d \ \ 
t.q. & \ovr{r}_i=\ovr{r}_j\ \text{para al}\\ 
 & \ \text{ menos un par {\it (i,j)} con}\ i\neq j.
\endcases \Biggr\} \tag 1.10
$$
As\'{\i}, tomamos [60,61]
$$
M_n =\frac{(\Cal{R}^{nd}-\Delta)}{S_n}. \tag 1.11
$$

Nos interesa como es $\pi_1(M_n)$. Para $d\geq 3$ resulta que [60,61] 
$\pi_1(M_n)=S_n$. 

Si definimos $\sigma_i$ como el operador abstracto que efect\'ua el 
intercambio del objeto $i$ con el $i+1$, en un arreglo de $n$ objetos, 
resulta que todo elemento de $S_n$ es generado por $1$, 
$\sigma_1\cdots \sigma_{n-1}$, donde $1$ representa la identidad de $S_n$. 
Estos generadores verifican
$$\align
\sigma_i^2 &= 1, \tag 1.12,a\\
\sigma_i\sigma_j&=\sigma_j\sigma_i\ \ \ si\ \ \ |i-j|\geq 2, \tag 1.12,b\\
\sigma_i\sigma_{i+1}\sigma_i&=\sigma_{i+1}\sigma_i\sigma_{i+1}.\tag 1.12,c
\endalign
$$

Nos interesa las representaciones unitarias de $S_n$. Le asignamos una fase 
$e^{-i\theta_j}$ a cada $\sigma_j$. La propiedad (1.12,c) no dice que 
$\theta_i=\theta_{i+1}=\theta$. Luego (1.12,a) implica que $2\theta=2k\pi$, 
con $k$ entero. As\'{\i} $\theta =0$ \'o $\pi$ (mod $2\pi$). Para $g\in S_n$
$$
a([g])=1 \ \ \ \ \ \ \forall g,\tag 1.13,a
$$
o
$$
a([g])=\pm 1 \ \ \text{dependiendo si la permutaci\'on es par o impar}. 
\tag 1.13,b
$$

Para dimensi\'on espacial 2, sucede que [10,57,58] $\pi_1(M_n)=B_n$, donde $B_n$ 
se le conoce como el grupo de trenzas (braid group). Este puede definirse, 
al igual que $S_n$ en t\'erminos de la identidad y las $\sigma_i$'s, salvo 
que no verifican (1.12,a). Por lo tanto para $B_n$
$$
a([g])=e^{(i\theta \sum_{k}\ep_k)}, \tag 1.14
$$
donde $\ep_k$ es +1 \'o $-1$ dependiendo de la ocurrencia, respectivamente, 
de $\sigma_i$ o $\sigma_i^{-1}$.

Tenemos, por tanto, que la acci\'on de $\pi_1(M_n)$ sobre las funciones de 
onda es $d\geq 3$
$$
\psi \to \cases
\psi &, \text{estad\'{\i}stica de Bose-Einstein}\\
\pm \psi &, \text{estad\'{\i}stica de Fermi-Dirac},
\endcases
\tag 1.15
$$
y en $d=2$ 
$$
\psi \to a([g])\psi , \text{con $a(g)$}\ \ \ \text{como en (1.14)}.\tag 1.16
$$

Observamos en $d\geq 3$ la ocurrencia natural de la estad\'{\i}sticas 
usuales. En $d=2$ si $\theta =0$ o $\pi$ tendremos la estad\'{\i}stica 
bos\'onica o fermi\'onica y en otro caso tenemos una estad\'{\i}stica no 
usual que suele llamarse ``fraccionaria''.

En 2 dimensiones espaciales, si miramos el intercambio como un suceso que 
ocurre en determinado intervalo de tiempo, el \'angulo entre la posici\'on 
relativa de 2 part\'{\i}culas barre, en caso de haber intercambio, un 
\'angulo de $\pi$ o $-\pi$. As\'{\i}, podemos reescribir $a(g)$ como [61]
$$
a([g])=e^{\frac{-i\theta}{\pi}\sum_{ij}\Delta \phi_{ij}}\tag 1.17
$$
donde $\Delta \phi_{ij}$ es el cambio angular del vector relativo entre las 
par\-t\'{\i}\-cu\-las $i$ y $j$.

\vskip 5mm
\noi
{\bf 2.- Implementaci\'on din\'amica de la estad\'{\i}stica fraccional}
\vskip 3mm

La amplitud de transici\'on entre dos configuraciones de un sistema, usando 
la formulaci\'on de Feynman para la mec\'anica cu\'an\-ti\-ca, es (cuando $Q$ es 
simplemente conexo)
$$
K(q',t';q,t)=\int \Cal{D}q e^{iS/\hbar}, \tag 2.1
$$
donde $S$ es la acci\'on cl\'asica y $\Cal{D}q$ denota que la suma se hace 
sobre todas las trayectorias posibles, $\alpha$, entre $q$ y $q'$. Si $Q$ no 
es simplemente conexo debemos ir a $\wt{Q}$ y al volver resulta que esta
amplitud ser\'a la suma ponderada [59,62-64] 
$$
K(q't';q,t)=\sum_{[\gamma ]\in \pi_1(Q)}a([\gamma ])K_\gamma (q',t';q,t), 
\tag 2.2
$$
donde $K_\gamma$ es el propagador parcial sobre todos los caminos $\alpha$ 
etiquetados 
por $[\gamma ]$. Para esto \'ultimo definimos un conjunto de ca\-mi\-nos 
standards, $\Gamma_q$, desde un punto fijo $q_0$ a todos los puntos de $Q$. 
Luego construimos el camino cerrado $\Gamma_q\alpha \Gamma_{q'}^{-1}$ el cual 
pertenece a alguna de las clases de equivalencia $[\gamma ]$ de $\pi_1(Q)$. 
La libertad para escoger $q_0$ y el hecho de que $K(q'',t'';q',t')$ 
$K(q',t';$ $q,t)=K(q'',t'';q,t)$ nos lleva a la conclusi\'on que 
$a([\gamma ])$, al igual que en (1.8), es una representaci\'on de $\pi_1(Q)$.

En $\wt{Q}$ la ley de propagaci\'on de las funciones $\wt{\psi}(\wt{q})$ se 
realiza integrado sobre uno de los dominios fundamentales
$$
\wt{\psi}(\wt{q}')=\int_F\Bigl[\sum_{[\gamma ]\in \pi_1(Q)]}\wt{K}(\wt{q}',t';
[\gamma ]\wt{q},t)a[\gamma ]\Bigr]\wt{\psi}(\wt{q})d\wt{q}\tag 2.3
$$
lo cual es equivalente a integrar sobre todo $\wt{Q}$. Es preciso, por 
consistencia, que $\wt{\psi}([\gamma ]\wt{q})=a[\gamma ]\wt{\psi}(\wt{q})$ 
siempre, esto es, debe ser propagada por (2.3). As\'{\i} 
$\wt{K}([\gamma ]\wt{q}',$ $t';[\wt{\gamma}]q,t)=K(\wt{q}',t';q,t)$ y 
$\wt{\psi}([\gamma ']\wt{q}')=a[\gamma ']\psi (\wt{q}')$ en (2.3). Tenemos, por 
tanto, que las funciones de onda evolucionan dentro de un mismo sector del 
espacio de Hilbert.

Pensando en el espacio de configuraciones, de dos part\'{\i}culas id\'enticas 
en el plano, y teniendo en cuenta (1.17) el factor $a([\gamma ])$ en (2.2) 
puede reescribirse como 
$$
a([\gamma ])=e^{\Bigl(\frac{-i\theta}{\pi}\int_{[\gamma ]}d\Phi \Bigr)}
\tag 2.4
$$
donde $\Phi$ es el \'angulo del vector posici\'on relativo entre las 
part\'{\i}culas con respecto a una direcci\'on fija. Introduciendo (2.4) en 
(2.2), la parte correspondiente a los caminos standards $\Gamma_q$ y 
$\Gamma_{q'}$ contribuye como un factor de fase global [61], as\'{\i} la 
parte relevante que queda es
$$
K'(q',t';q,t)=\int \Cal{D}q e^{i\ov{S}/\hbar} \tag 2.5a
$$ 
con
$$
\ov{S}=\int^{t'}_t dt\Bigl[\Cal{L}-\frac{\theta}{\pi}\frac{d\Phi}{dt}\Bigr]
\tag 2.5,b
$$
Tenemos as\'{\i} que la adici\'on de este t\'ermino a la acci\'on determina la 
estad\'{\i}stica sin afectar las ecuaciones cl\'asicas de movimiento.

La \'ultima discusi\'on sugiere que consideremos la mec\'anica cu\'an\-ti\-ca de 
dos part\'{\i}culas id\'enticas con el t\'ermino de interacci\'on [65]
$$
L_i=\frac{\theta}{\pi}\int \dot{\Phi}dt, \tag 2.6
$$
donde $\theta$ es un par\'ametro num\'erico, llamado par\'ametro 
estad\'{\i}stico. En el sentido de (1.7): $\theta$ es 0 \'o $\pi$ si queremos 
considerar, respectivamente, 2 bosones o 2 fermiones, y otros valores si 
queremos considerar anyones.

El t\'ermino de interacci\'on puede reescribirse como
$$
L_I=\int q\ovr{A}(\ovr{r}_{rel})\cdot \frac{d\ovr{r}_{rel}}{dt}dt,\tag 2.7
$$
donde
$$
A_i(\ovr{r}_{rel})=-\frac{\theta}{\pi q}\ep_{ij}\partial_j\ln r_{rel},\tag 2.8
$$
que corresponde a un punto de flujo en el origen del sistema de referencia 
relativo. Esto pues $B=\ep_{ij}\partial_iA_j=\frac{2\theta}{q}\delta 
(\ovr{r}_{rel})$. El flujo del ``campo magn\'etico" $B$ es 
$\Phi =\frac{2\theta}{q}$. Desde este punto de vista cada part\'{\i}cula 
``ve" a la otra como un punto de flujo magn\'etico. En este cuadro podemos
decir que los anyones se comportan como part\'{\i}culas con carga ficticia $q$
y flujo $\Phi$, relacionados por el v\'{\i}nculo $2\theta = q\Phi$. Pensando en
la situaci\'on que se presenta cuando se estudia el efecto Aharanov-Bohm, la
fase que adquirir\'{\i}a la funci\'on de onda de una part\'{\i}cula que rodee
al punto de flujo ser\'{\i}a $q\Phi$. Definimos entonces
$$
\alpha \equiv q \oint dx^iA_i,\tag 2.9
$$
como el ``par\'ametro de comportamiento any\'onico", el cual usaremos en las 
si\-gui\-entes secciones. Su relaci\'on con el par\'ametro estad\'{\i}stico es 
$$
\theta =\frac{\alpha}{2}. \tag 2.10
$$

\vskip 5mm
\noi
{\bf 3.- Comportamiento any\'onico en teor\'{\i}as vectoriales}
\vskip 3mm

\noi
{\bf 3.1.- Teor\'{\i}a de CS pura y la TM vectorial}
\vskip 3mm

Partimos de la acci\'on 
$$
S=S_{part}+S_{CS}+<a_mJ^m>, \tag 3.1
$$
donde la acci\'on de la part\'{\i}cula es la usual
$$
S_{part}=-m\int ds ,\tag 3.2,a
$$
y 
$$
S_{CS}=-\frac{\mu}{2}<\ep^{mnl}a_m\partial_na_l>, \tag 3.2,b
$$
que corresponde al t\'ermino de CS. $J^m$ es una corriente conservada. En el 
l\'{\i}mite no relativista el lagrangiano $L_{part}+a_mJ^m$ puede verse como 
el usual $\frac{1}{2}mv^2+a_mv^mq$, si $J^m=<q\dot{x}^m\delta (x-x(\tau ))>$. 
La ecuaci\'on de movimiento para $a_m$ es 
$$
\mu \ep^{mnl}\partial_na_l=J^m .\tag 3.3
$$
La componente cero de (3.3) constituye el v\'{\i}nculo  
$$
\mu B-J^0=0 ,\tag 3.4
$$
donde $B\equiv \ep_{ij}\partial_ia_j$ es el "campo magn\'etico". Cuando $J^m$ 
corresponde a una carga $q$
$$
q=\mu \Phi ,\tag 3.5
$$
donde $\Phi$ es el flujo del campo magn\'etico. Tenemos as\'{\i} la 
t\'{\i}pica relaci\'on vincular del modelo any\'onico, vista en la secci\'on 
anterior. Adem\'as, como $J^0$ es nulo en todas partes salvo en la posici\'on 
de la part\'{\i}cula, $B=F_{12}=0$, lo que indica que $a_i$ es puro calibre, 
y que la part\'{\i}cula es efectivamente un punto de flujo. El par\'ametro de 
comportamiento any\'onico es
$$\align
\alpha_{CS}=q\oint dx^ia_i &=q\Phi \\
&=\frac{q^2}{\mu},\tag 3.6
\endalign
$$
que corresponder\'a al parametro estad\'{\i}stico
$$
\theta =\frac{q^2}{2\mu}.\tag 3.7
$$

Estos resultados se verifican cuando buscamos las soluciones est\'aticas para 
$a_i$ y calculamos $\alpha$ [29,30]. Si a la acci\'on (3.1) le sumamos la 
acci\'on de Maxwell, asint\'oticamente el t\'ermino dominante es el de CS. El 
v\'{\i}nculo (3.4) es ahora la ley de Gauss modificada
$$
\partial_iE^i-\mu B=-J^0. \tag 3.8
$$
Como el campo el\'ectrico cumple la ec. de Klein-Gordon con masa $\mu$, 
\'este decae con $r$ como $r^\alpha e^{-\mu r}$, donde $\alpha$ es alguna 
potencia. As\'{\i} $\int_{R^2}d^2x \partial_iE^i=0$, y entonces al 
integrar sobre todo el espacio se sigue verificando (3.5).

Las ecuaciones de movimiento para $a_m$ cuando el t\'ermino de Maxwell est\'a 
presente son 
$$
\partial_mF^{mn}-\mu \ep^{nml}\partial_ma_l=-J^n, \tag 3.9
$$
donde en el l\'{\i}mite en que $\mu$ es muy grande recuperamos la teor\'{\i}a 
de CS pura. Como deseamos comparar con el caso gravitacional, tomamos como 
corriente $J^m$ a 
$$
J^0=q\delta^{(2)}(\ovr{r}),\ J^i=g\ep_{ij}\partial_j\delta^{(2)}(\ovr{r}), 
\tag 3.10
$$
que co\-rres\-pon\-de a una carga q en el o\-ri\-gen, con momento mag\-n\'e\-ti\-co dipolar 
$g$\footnote"*"{\ninerm En 3 dimensiones espaciales podemos definir el 
momento mag\-n\'e\-ti\-co como $\ovr{m}=\frac{1}{2}I\int \ovr{r}\times d\ovr{l}$, 
haciendo la identificaci\'on $Id\ovr{l}\to \ovr{J}dV$ tendremos, entonces, 
que $\ovr{m}=\int \ovr{r}\times \ovr{J}dV$. ``Traduciendo" este 
concepto a 2 dimensiones espaciales 
$``\ovr{r}\times \ovr{J}"=\ep_{ij}x^iJ^j$. Obteniendo con (3.10) 
$\text{''$\ovr{m}''$}=g$.}. 
Tomamos el ansatz
$$
a_0=a(r),\ a_i=\ep_{ij}\partial_jV(r)+\partial_i\lambda (r). \tag 3.11
$$
Podemos escoger $\lambda (r)=0$. As\'{\i} el sistema (3.9), queda como
$$\align
\Delta a-\mu \Delta V &=q\delta^{(2)}(\ovr{r})\tag 3.12,a\\
\Delta V-\mu a &=-g\delta^{(2)}(\ovr{r})\tag 3.12,b 
\endalign
$$
que podemos desacoplar sin ning\'un problema
$$\align
(-\Delta )(-\Delta +\mu^2)V &=(\mu q+g(-\Delta ))\delta^{(2)}(\ovr{r}),
\tag 3.13,a\\
(-\Delta +\mu^2)a &=-(q-\mu g)\delta^{(2)}(\ovr{r}).\tag 3.13,b
\endalign
$$
Este sistema lo resolvemos usando las funciones de Green de Yu\-ka\-wa y Coulomb, 
$Y(\mu r)$ y $C(\mu r)$, que verifican
$$\align
(-\Delta +\mu^2)Y(\mu r) &=\delta^{(2)}(\ovr{r}), \tag 3.14,a\\
(-\Delta )C(\mu r) &=\delta^{(2)}(\ovr{r}),\tag 3.14,b
\endalign
$$
donde
$$\align
Y(\mu r) &=\frac{1}{2\pi}K_0(\mu r),\tag 3.15,a\\
\Cal{C}(\mu r) &=-\frac{1}{2\pi}\ln (\mu r).\tag 3.15,b
\endalign
$$
$K_0(\mu r)$ es la funci\'on de Bessel modificada de orden 0. En 
$C(\mu r)$, $\mu$ puede cambiarse por cualquier constante con 
dimensiones de masa sin alterar (3.14,b), la funci\'on de esta constante es 
que el argumento del logaritmo sea adimensionado. La soluci\'on al sistema es
$$\align
a(r) &=-(q-\mu g)Y(\mu r), \tag 3.16,a\\
V(r) &=-\frac{(q-\mu g)}{\mu}Y(\mu r)+\frac{q}{\mu}C(\mu r). \tag 3.16,b
\endalign
$$

La soluci\'on del sistema (3.13) en el limite de $\mu$ grande (que corresponde 
a la teor\'{\i}a de CS pura) es $a(r)=(g/\mu )\delta^{(2)}(\ovr{r})$, 
$V(\ovr{r})=(q/\mu )\Cal{C}(r)$, que cuando $g=0$ corresponde a un punto de 
flujo con el comportamiento any\'onico planteado. Cuando $g\neq 0$, 
$\alpha_{CS}$ no cambia y $F_{mn}=0$ fuera de la posici\'on de la 
part\'{\i}cula.

Cuando el t\'ermino de Maxwell est\'a presente, usando el hecho de que 
a\-sin\-t\'o\-ti\-ca\-men\-te $K_0(x)\sim x^{-1/2}e^{-x}$, tenemos que cuando 
$\mu r>>0$
$$
a_0\sim 0,\ a_i\sim \frac{q}{\mu}\ep_{ij}\partial_jC(\mu r), \tag 3.17
$$
al igual que la teor\'{\i}a de CS pura. El par\'ametro de comportamiento 
any\'onico es para un c\'{\i}rculo de radio $R$
$$
\alpha_{TM}(R)=\frac{q^2}{\mu}-q(q-\mu g)RK_1(\mu r) ,\tag 3.18
$$
el cual tiende a $\alpha_{CS}$ cuando $R\to \infty$, como es de esperarse. En 
el caso particular en que $q-\mu g=0$ resulta que $\alpha_{TM}=\alpha_{CS}$ 
para cualquier contorno y las soluciones coinciden en las regiones externas a 
la fuente.

\vskip 5mm
\noi
{\bf 3.2.- Las teor\'{\i}as autodual y TM, y el problema de los aclopamientos 
no minimales}
\vskip 3mm

Se ha dicho que cuando hay rotura espont\'anea de simetr\'{\i}a, se pierde el 
comportamiento any\'onico [32,33]. De hecho si consideramos la teor\'{\i}a de CS 
pura con un t\'ermino adicional de la forma $-m^2a_ra^r$, que podr\'{\i}a 
venir de alg\'un proceso de rotura espont\'anea de simetr\'{\i}a, la acci\'on 
del campo $a_m$ corresponde a la teor\'{\i}a autodual con masa 
$m_{{}_{AD}}=m^2/|\mu |$. La ecuaci\'on (3.3) ser\'{\i}a ahora
$$
\mu \ep^{mnl}\partial_na_l+m^2a^m=J^m.\tag 3.19
$$
Si tomamos $J^m$ como en (3.10) y el ansatz (3.11), es inmediato obtener 
$$\align
a_{AD}(r) &=-\frac{m_{{}_{AD}}}{\mu}(q+m_{{}_{AD}}g)Y(m_{{}_{AD}}r)+
\frac{q}{\mu}\delta^{(2)}(\ovr{r}), \tag 3.20,a\\
V_{AD}(r) &=(\frac{m_{{}_{AD}}g+q}{\mu})Y(m_{{}_{AD}}r), \tag 3.20,b\\
\lambda_{AD}(r) &= 0,\tag 3.20,c
\endalign
$$
donde el argumento de la funci\'on de Green de Yukawa se corresponde con la 
masa de las excitaciones. En el caso particular en que $q+m_{{}_{AD}}g=0$ 
tendremos que $F_{mn}=0$. El par\'ametro de comportamiento 
any\'onico para un c\'{\i}rculo de radio $R$ es
$$
\alpha_{AD}(R)=\frac{q(q+m_{{}_{AD}}g)}{\mu}m_{{}_{AD}}RK_1(m_{{}_{AD}}R),
\tag 3.21
$$
que tiende a cero cuando $R\to \infty$, y es id\'enticamente cero 
$q+m_{{}_AD}g=0$.

En vista del equivalente entre las teor\'{\i}as $S_{TM}$ y $S_{AD}$ nos
preguntamos si ser\'a posible alg\'un acoplamiento no minimal que reproduzca en
una teor\'{\i}a las soluciones de la otra. La respuesta a esta pregunta es
afirmativa. 

Tomemos primero la teor\'{\i}a de CS con el t\'ermino de Maxwell (la TM) con 
un acoplamiento no minimal que corresponda a tomar $J^m$ en (3.9) como
$$
\wt{J}^m=\frac{1}{\mu}\ep^{mnl}\partial_mJ_l \tag 3.22
$$
y $J_l$ como en (3.10). Las soluciones est\'aticas ser\'an
$$\align
\wt{a}_{TM}(r) &=-(q-\mu g)Y-\frac{g}{\mu}\delta^{(2)}(\ovr{r}), \tag 3.23,a\\
\wt{V}_{TM}(r) &=-\frac{1}{\mu}(q-\mu g)Y(\mu r),\tag 3.23,b
\endalign
$$
y tomamos $\wt{\lambda}_{TM}=0$ como elecci\'on de calibre (lo cual es 
posible tomando est\'atico al par\'ametro de la transformaci\'on de 
calibre). Observamos que (3,23) corresponde a tomar $\mu \to -\mu$ y 
$m^2\to \mu^2$ en (3.20). As\'{\i} cuando tengo la teor\'{\i}a TM acoplada no 
minimalmente $(\sim a_m\ep^{mnl}\partial_nJ_l)$, las 
soluciones est\'aticas corresponden a las de la teor\'{\i}a $AD$ con 
acoplamiento minimal.

Rec\'{\i}procamente tomemos la teor\'{\i}a de CS con el t\'ermino de rotura 
de simetr\'{\i}a (la $AD$) con el acoplamiento no minimal, que corresponde a 
resolver (3.19) con $J^m$ como
$$
\wt{J}^m=m_{{}_{AD}}\ep^{mnl}\partial_n\Cal{G}_l, \tag 3.24,a
$$
donde
$$
\Cal{G}^0=qC(m_{{}_{AD}}r),\ \Cal{G}^i=g\ep_{ij}\partial_jC(m_{{}_{AD}}r),
\tag 3.24,b
$$
que corresponde a $\Cal{G}^m=(-\Delta )^{-1}J^m$, con $J^m$ como en (3.10). 
Las soluciones est\'aticas correspondientes, resultan ser
$$\align
\wt{a}_{AD}(r) &=-\frac{m_{{}_{AD}}}{\mu}(q+m_{{}_{AD}}g)Y(m_{{}_{AD}}r) 
\tag 3.25,a\\
\wt{V}_{AD}(r) &=(\frac{q+m_{{}_{AD}}g}{\mu})Y(m_{{}_{AD}}r)-
\frac{q}{\mu}C(m_{{}_{AD}}r), \tag 3.25,b\\
\wt{\lambda}_{AD}(r)&=0 .\tag 3.25,c
\endalign
$$
Estas soluciones corresponden a las de la topol\'ogica masiva si tomamos 
$m_{{}_{AD}}\to -\mu$, $\mu \to -\mu$. El par\'ametro de comportamiento
any\'onico es, para  un c\'{\i}rculo de radio $R$
$$
\wt{\alpha}_{AD}(R)=-\frac{q^2}{\mu}+q\frac{(q+m_{{}_{AD}}g)}{\mu}
m_{{}_{AD}}RK_1(m_{{}_{AD}}R), \tag 3.26
$$
donde para el caso en que $q+m_{{}_{AD}}g=0$, es igual al de la teor\'{\i}a 
de CS pura con par\'ametro $-\mu$. Es importante recalcar que en este caso 
particular como asint\'oticamente el par\'ametro de comportamiento any\'onico no
depende de  $m$.

El punto que no queda claro es cu\'al ser\'a el significado f\'{\i}sico de 
(3.24).

\vskip 5mm
\noi
{\bf 3.3.- La teor\'{\i}a de Hagen}
\vskip 3mm

Para la teor\'{\i}a de Hagen (ver cap\'{\i}tulo II), las ecuaciones del campo 
$a_m$ ser\' an
$$
\partial_m F^{nm}-\wt{\mu}\ep^{mnl}\partial_na_l=-\frac{1}{(1-\lambda )^2}
((1-\lambda )J^m+\frac{\lambda}{\wt{\mu}}\ep^{mnl}\partial_nJ_l), \tag 3.27
$$
donde $\wt{\mu}=\mu /(1-\lambda )$. Tomamos $J^m$ como en (3.10), y las 
soluciones est\'aticas son
$$\align
a_H(r) &=-\frac{1}{(1-\lambda )^2}(q-\wt{\mu}g)Y(\wt{\mu}r), \tag 3.28,a\\
V_H(r) &=-\frac{1}{(1-\lambda )^2}\frac{(q-\wt{\mu}g)}{\wt{\mu}}Y(\wt{\mu}r)+
\frac{q}{(1-\lambda )\wt{\mu}}C(\wt{\mu}r). \tag 3.28,b
\endalign
$$
El par\'ametro de comportamiento any\'onico, para un c\'{\i}rculo de radio 
$R$ es
$$
\alpha_H(R)=\frac{q^2}{\mu}-\frac{q}{(1-\lambda )^2}(q-\wt{\mu}g)RK_1
(\wt{\mu}R).\tag 3.29
$$

Observamos que cuando $\lambda \to 0$ tienden uniformemente a las soluciones 
correspondientes de la teor\'{\i}a TM. Para el par\'ametro de comportamiento
any\'onico  sucede lo mismo. Sin embargo es de notar que tanto
asint\'oticamente, como en  el caso particular, $q-\wt{\mu}g=0$, el
par\'ametro no depende  de $\lambda$. Esto \'ultimo esta en
correspondencia con el hecho de que  asint\'oticamente el t\'ermino dominante es
el de CS.

\vskip 5mm
\noi
{\bf 4.- Comportamiento any\'onico en teor\'{\i}as de gravedad linealizada}
\vskip 3mm

\noi
{\bf 4.1.- La posibilidad de tener anyones gravitacionales}
\vskip 3mm

La acci\'on de una part\'{\i}cula libre en un campo gravitacional es
$$
S_p=-m\int d\tau (-g_{mn}\frac{dx^m}{d\tau}\frac{dx^n}{d\tau})^{1/2}.\tag 4.1
$$
Cuando tomamos la aproximaci\'on de campo d\'ebil (o linealizamos) 
$g_{mn}=\eta_{mn}+\kappa h_{\ov{mn}}$ obtenemos
$$\align
S_p^l =-m\int d\tau &(-\eta_{mn}\frac{dx^m}{d\tau}\frac{dx^n}{d\tau})^{1/2}+\\
&+\frac{\kappa }{2}\int d^3x h_{\ov{mn}}(x)T^{mn}(X),\tag 4.2
\endalign
$$
donde
$$
T^{mn}(X)=m\frac{dx^m}{dt}\frac{dx^n}{dt}\frac{dt}{d\tau}\delta^{(2)}
(\ovr{x}-\ovr{X}(t)).\tag 4.3
$$
Si pasamos al l\'{\i}mite no relativista de (4.2), tendremos
$$
S_p^{l(nr)}=\frac{1}{2}m\int dt \dot{X}^2+\frac{\kappa m}{2}\int
dth_{00}(X)+\kappa m \int dth_{0i}(X)\dot{X}^i, \tag 4.4
$$
que es igual a la acci\'on de una part\'{\i}cula en un campo 
electromagn\'etico, si hacemos la identificaci\'on [26,27]
$$
\kappa mh_{\ov{0i}}\to qa_i,\ \ \ \frac{1}{2}\kappa mh_{\ov{00}}\to q a_0. 
\tag 4.5
$$
Este resultado nos induce a pensar en la posibilidad de implementar 
di\-n\'a\-mi\-ca\-men\-te estad\'{\i}sticas any\'onicas en el contexto de 
teor\'{\i}as de gravedad linealizada. Esto fu\'e presentado originalmente por 
Deser [29,30], con la teor\'{\i}a de gravedad topol\'ogica masiva linealizada. 
Sin embargo, veremos en la pr\'oxima subsecci\'on que con la teor\'{\i}a TM 
no se logra una gene\-ralizaci\'on uniforme de los resultados para 
teor\'{\i}as vectoriales [66]. La raz\'on es que el t\'ermino topol\'ogico 
linealizado es de tercer orden y por lo tanto no domina sobre el de Einstein 
asint\'oticamente.

El efecto del t\'ermino de CS vectorial se generaliza para teor\'{\i}as de 
gravedad linealizada con el t\'ermino de CS tri\'adico (TCS) [66]. 
Consideremos, entonces, la acci\'on
$$
S=-m\int ds + S_{TCS}+\kappa <h_{mn}T^{mn}>,\tag 4.6
$$
donde al linealizar la acci\'on de la part\'{\i}cula (4.1) hemos partido con 
la m\'etrica en funci\'on de los dreibeins
$$
g_{mn}=e_m{}^ae_n{}^b\eta_{ab}, \tag 4.7
$$
y linealizamos los dreibeins, $e_p{}^a=\delta_p{}^a+\kappa h_p{}^a$, lo cual 
produce el acoplamiento $\kappa h_{mn}T^{mn}$ con $T^{mn}$ sim\'etrico. La 
relaci\'on entre $h_{\ov{mn}}$ y $h_{mn}$ es
$$
h_{\ov{mn}}=h_{mn}+h_{nm}. \tag 4.8
$$
En (4.6) $S_{TCS}$ es
$$
S_{TCS}=\frac{\mu}{2}<h_{pa}\ep^{prs}\partial_rh_s{}^a>. \tag 4.9
$$

Tomamos como fuente, $T^{mn}$
$$
T^{00}=m\delta^{(2)}(\ovr{r}),\ \ \ T^{0i}=\frac{1}{2}\sigma 
\ep_{ij}\partial_j\delta^{(2)}(\ovr{r}),\ \ \ T^{ij}=0, \tag 4.10
$$
que corresponde a una part\'{\i}cula con masa $m$ y momento angular 
intr\'{\i}nseco $\sigma$ 
$(\int d^2xT^{00}=m$, $\int d^2\ep_{ij}x^iT^{0j}=\sigma)$. Si hacemos 
la descomposici\'on $T+L$ de $h_{mn}$
$$\align
h_{00} &=n(r),\tag 4.11,a\\
h_{0i} &=\ep_{ij}\partial_j(n^T-v^L)+\partial_i(n^L+v^T),\tag 4.11,b\\
h_{i0} &=\ep_{ij}\partial_j(n^T+v^L)+\partial_i(n^L-v^T),\tag 4.11,c\\
h_{ij} &=(\delta_{ij}\Delta -\partial_i\partial_j)h^T+
(\ep_{i\kappa }\partial_\kappa \partial_j+\ep_{j\kappa }\partial_\kappa
\partial_i)h^{TL}+\partial_i \partial_jh^L+\ep_{ij}V, \tag 4.11,d
\endalign
$$
las ecuaciones de movimiento para las $h_{mn}$ son
$$
\mu \ep^{prs}\partial_rh_s{}^n=-\kappa T^{pn},\tag 4.12
$$
La componente $00$ de (4.12) es el v\'{\i}nculo
$$
\mu \ep_{ij}\partial_ih_{j0}=\kappa T^{00}, \tag 4.13
$$
que constituye la generalizaci\'on del v\'{\i}nculo (3.4), como veremos. En 
funci\'on de las componentes T+L, (4.12) queda como
$$\align
(-\Delta )(n^T+v^L)&=\frac{\kappa m}{\mu}\delta(\ovr{r}), \tag 4.14,a\\
\ep_{ij}\partial_j\Delta h^T+\partial_i(\Delta h^{TL}+V)&=
\frac{\kappa\sigma}{2\mu}\ep_{ij}\partial_j\delta (\ovr{r}), \tag 4.14,b\\
n&=\frac{\kappa\sigma}{2\mu}\delta (\ovr{r}), \tag 4.14,c
\endalign
$$
$$\align
(\delta_{ij}\Delta -\partial_i\partial_j)(n^T-v^L)&+\frac{1}{2}
(\ep_{ij}\partial_j\partial_\kappa +\ep_{\kappa j}\partial_j\partial_i)(n^L-v^T)+\\
&+\frac{1}{2}\ep_{i\kappa }\Delta (n^L-v^T)=0. \tag 4.14,d
\endalign
$$

Para resolver (4.14), notamos que si hacemos una transformaci\'on de calibre 
est\'atica , $n$, $n^T-v^L$, $n^L+v^T$, $n^T+v^L$ y $h^T$ son invariantes, y
$$
\delta h^L=\xi^L,\ \delta h^{TL}=\frac{1}{2}\xi^T,\ \delta V=-\frac{1}{2}
\Delta \xi^T,\ \delta (n^L-V^T)=\xi , \tag 4,15
$$
donde $\xi_0=\xi$, $\xi_i=\partial_i\xi^L+\ep_{ij}\partial_j\xi^T$ y 
$\delta h_{mn}=\partial_m\xi_n$. Fijamos el calibre $h^{TL}=0$, $h^L=h^T$ y 
tendremos 
$$\align
h_{00}(r) &=\frac{K\sigma}{2\mu}\delta(\ovr{r}), \tag 4.16,a\\
h_{0i}(r) &=0,\tag 4.16,b\\
h_{i0}(r) &=\frac{\kappa m}{\mu}\ep_{ij}\partial_jC(\mu r),\tag 4.16,c\\
h_{ij}(r) &=\frac{\kappa \sigma}{2\mu}\delta_{ij}\delta(\ovr{r}).\tag 4.16,d
\endalign
$$
La comparaci\'on con los casos vectoriales es a trav\'es de los 
$h_{\ov{mn}}$ de la linealizaci\'on de la m\'etrica. As\'{\i} con (4.8)
$$\align
h_{\ov{00}}(r) &=\frac{\kappa \sigma}{\mu}\delta(\ovr{r}), \tag 4.17,a\\
h_{\ov{0i}}(r) &=\frac{\kappa m}{\mu}\ep_{ij}\partial_j\Cal{C}(\mu r), 
\tag 4.17,b\\
h_{ij}(r) &= \frac{\kappa \sigma}{\mu} \delta_{ij}\delta (\ovr{r}). \tag 4.17,c
\endalign
$$
Luego de hacer la identificaci\'on $\kappa \sigma \to g$ y $\kappa m \to q$, 
notamos que $h_{\ov{00}}$ y $h_{\ov{0i}}$ corresponden a las soluciones para 
$a_0$ y $a_i$ en el sistema vectorial de CS puro (ecuaci\'on (3.12) en el 
l\'{\i}mite $\mu$ muy grande). El par\'ametro de comportamiento any\'onico 
ser\'a.
$$
\alpha_{TCS}\equiv \kappa m \oint dx^ih_{\ov{0i}}=\frac{(\kappa m)^2}{\mu}, 
\tag 4.18
$$
independientemente del contorno que escojamos. Este resultado es la 
ge\-ne\-ra\-li\-za\-ci\'on uniforme del obtenido para la teor\'{\i}a 
vectorial correspondiente.

\vskip 5mm
\noi
{\bf 4.2.- El par\'ametro de comportamiento any\'onico para la teor\'{\i}a VCS
linealizada y  la TM linealizada}
\vskip 3mm

Visto el resultado obtenido para la teor\'{\i}a TCS pura, miramos la 
teor\'{\i}a que resulta de agregar la acci\'on de Einstein linealizada a 
(4.6) [66]. Este ser\'a el an\'alogo a cuando tenemos el t\'ermino de Maxwell 
presente, para el caso vectorial. Las soluciones deben coincidir 
asint\'oticamente ya que en este l\'{\i}mite dominar\'a $S_{TCS}$.

Las ecuaciones de movimiento para $h_{mn}$ son ahora las correspondientes a 
la teor\'{\i}a VCS linealizada con la fuente (4.10)
$$
(\frac{1}{2}\ep^{pmn}\ep^{srl}-\ep^{pml}\ep^{srn})\partial_m\partial_rh_{sl}
+\mu \ep^{prs}\partial_sh_r{}^n=-\kappa T^{pn}.\tag 4.19
$$
La componente $00$ es el v\'{\i}nculo
$$
(\delta_{ij}\Delta -\partial_i\partial_j)h_{ij}-\mu \ep_{ij}\partial_ih_{j0}=
-\kappa T^{00}. \tag 4.20
$$
Al integrar (4.20) sobre todo el plano da la misma relaci\'on entre 
$\kappa m$ y el flujo de $\ep_{ij}\partial_ih_{j0}$ que se obtiene en (4.13). 
Esto se debe, al igual que en el caso vectorial, a que asint\'oticamente 
$\Delta h^T$ va como $r^\alpha e^{-\mu r}$ por lo que no contribuye.

Haciendo la misma descomposici\'on que (4.11), las ecuaciones se transforman 
en

$$\align
\Delta^2h^T+\mu \Delta (n^T+v^L) &=-\kappa m\delta^{(2)}(\ovr{r}), \tag 4.21,a\\
-\Delta n^T-\mu \Delta h^T &=-\frac{\kappa \sigma}{2}\delta^{(2)}(\ovr{r}),
\tag 4.21,b\\
\mu V+\mu \Delta h^{TL} &=0, \tag 4.21,c\\
-\Delta n^T-\mu n &=-\frac{\kappa \sigma}{2}\delta^{(2)}(\ovr{r}), \tag 4.21,d\\
n+\mu (n^T-v^L) &=0,\tag 4.21,e\\
\mu (n^L+v^T) &=0 \tag 4.21,f
\endalign
$$
que es invariante bajo (4.15). Fijamos $h^{TL}=0$, $(\Rightarrow V=0)$, 
$n^L=0(\Rightarrow V^T=0$ y $h^L=h^T$. Cada fijaci\'on la hacemos con una de 
las componentes de $\xi_m(\xi^T,\xi$ y $\xi^L$ respectivamente). Al 
desacoplar el sistema obtendremos
$$\align
n(r) &=\kappa \frac{(m+\mu\sigma )}{2}Y(\mu r)\tag 4.22,a\\
h^T(r)&=-\kappa \frac{(m+\mu\sigma )}
{2\mu^2}(C(\mu r)-Y(\mu r))\tag 4.22,b\\
n^T(r) &=-\kappa \frac{(m+\mu\sigma )}{2\mu}Y(\mu r)+\frac{\kappa m}{2\mu}
C(\mu r)\tag 4.22,c\\
v^L(r)&=\frac{\kappa m}{2\mu}C(\mu r).\tag 4.22,d
\endalign
$$
Resaltamos que $h_{mn}$ no es sim\'etrico

Las $h_{\ov{mn}}$ correspondientes son entonces
$$\align
h_{\ov{00}} &=\kappa (m+\mu\sigma )Y(\mu r), \tag 4.23,a\\
h_{\ov{0i}} &=\ep_{ij}\partial_j[-\frac{\kappa }{\mu}(m+\mu\sigma )Y(\mu r)+
\frac{\kappa m}{\mu}C(\mu r)],\tag 4.23,b\\
h_{\ov{ij}} &=\delta_{ij}\kappa (m+\mu\sigma )Y(\mu r),\tag 4.23,c
\endalign
$$
donde observamos que $h_{\ov{00}}\sim a_0$, $h_{\ov{0i}}\sim a_i$, en (3.16), 
luego de identificar $\kappa m\sim q$ $\kappa \sigma \sim g$.

El par\'ametro de comportamiento any\'onico es para un c\'{\i}rculo de radio 
$R$
$$
\alpha_{VCS}(R)=\frac{(\kappa m)^2}{\mu}-\kappa m\kappa (m+\mu\sigma )RK_1
(\mu R).\tag 4.24
$$

En (4.24) vemos que asint\'oticamente 
$\alpha_{VCS}(R\to \infty )=\alpha_{TCS}$. Tambi\'en sucede lo mismo para 
$m+\mu\sigma =0$ y no depender\'a del contorno. Asint\'oticamente
$$
h_{\ov{00}}\sim 0,\ h_{\ov{0i}}\sim \ep_{ij}\partial_j\frac{\kappa m}{\mu}C
(\mu r),\ h_{\ov{ij}}\sim 0, \tag 4.25
$$
que corresponde a la (4.17). Cuando $m+\mu\sigma =0$ las soluciones 
exteriores a las fuentes de ambos modelos ($TCS$ y $VCS$) coinciden.
Para la TM linealizada las ecuaciones de movimiento son [29]
$$
\frac{1}{2}(\ep^{lp}{}_r\partial_lG^{mr}+\ep^{lm}{}_r\partial_lG^{pr})-
\mu G^{mp}=\mu \kappa  T^{pm} \tag 4.26
$$
donde
$$
G^{pm}=-\frac{1}{2}\ep^{prs}\ep^{mlt}\partial_r\partial_lh_{\ov{st}}. 
\tag 4.27
$$
Con la descomposici\'on (4.11), obtenemos que
$$\align
G^{00} &=-\frac{\Delta^2}{2}h^T, \tag 4.28,a\\
G^{0i} &=\frac{\Delta}{2}\ep_{ij}\partial_jn^T, \tag 4.28,b\\
G^{ij} &=-\frac{1}{2}(\delta_{ij}\Delta -\partial_i\partial_j)n, \tag 4.28,c
\endalign
$$
donde notamos que las partes sensibles a transformaciones de calibre 
$n^L,h^{TL}$ y $h^L$ no intervienen en la soluci\'on de (4.27). Tomamos 
$n^L=h^{TL}=0$ y $h^L=h^T$, llegamos a que 
$$\align
h_{\ov{00}}(r) &=\kappa (m+\mu\sigma )Y(\mu r),\tag 4.29,a\\
h_{\ov{0i}}(r) &= \ep_{ij}\partial_j[-\kappa \frac{(m+\mu\sigma )}{\mu}
(Y(\mu r)-C(\mu r))],\tag 4.29,b\\
h_{\ov{ij}}(r) &=\delta_{ij}[\kappa (m+\mu\sigma )Y(\mu r)-2\kappa mC(\mu r)].
\tag 4.29,c
\endalign
$$
Observamos que (4.29) s\'olo constituye la generalizaci\'on uniforme del caso 
vectorial ( la TM vectorial ) cuando $\sigma =0(g=0)$. En el caso especial 
que $m+\mu\sigma =0$ no hay posibilidad de comportamiento any\'onico
$(h_{\ov{0i}}=0)$ y la  soluci\'on corresponde a la ge\-ne\-rada por una
part\'{\i}cula masiva sin spin  en la teor\'{\i}a de Einsten linealizada [29].
As\'{\i}, puede decirse que  el efecto del t\'ermino CS y del $T^{0i}$ se
cancelan mutuamente.

El par\'ametro de comportamiento any\'onico es para un c\'{\i}rculo de radio 
$R$
$$
\alpha_{TM}(R)=\frac{\kappa m\kappa (m+\mu\sigma )}{\mu}(1-\mu RK_1(\mu R)), \tag 4.30
$$
que difiere del ``correspondiente'' caso vectorial (la TM vectorial), y 
observamos que $\alpha_{TM}(R)=0$ (para cualquier contorno) si 
$m+\mu\sigma =0$. La analog\'{\i}a con el modelo TM vectorial se da como 
dijimos, cuando $\sigma =0(g=0)$ [28,29,66].

\vskip 5mm
\noi
{\bf 4.3.- Par\'ametro de comportamiento any\'onico de la teor\'{\i}a AD y de la
teor\'{\i}a de  Einstein autodual}
\vskip 3mm

Ya vimos para el caso vectorial que cuando hay rotura de simetr\'{\i}a el
par\'ametro de comportamiento any\'onico se anula. Con el t\'ermino TCS
li\-nea\-lizado podemos  considerar dos posibilidades: una cuando tenemos adem\'as
de $S_{TCS}$ al  t\'ermino de Fierz-Pauli (teor\'{\i}a autodual), y otra si
adem\'as tenemos el  de Einstein que corresponder\'{\i}a a la gravedad de
Einstein autodual. Ambas  teor\'{\i}as han perdido la invariancia de calibre y
veremos que el  par\'ametro de comportamiento any\'onico es nulo.

Si consideramos la acci\'on $AD$, partir\'{\i}amos de (4.6) sum\'andole el 
t\'ermino de Fierz-Pauli $\sim m^2(hh-hh)$. Las ecuaciones de mo\-vi\-mien\-to para 
$h_{mn}$, son
$$
\mu \ep^{prs}\partial_rh_s{}^n-M^2(h^{np}-\eta^{pn}h_l{}^l)=-\kappa T^{pn}.\tag 4.31
$$
El m\'etodo a seguir es el ya expuesto con $T^{pn}$ como en (4.10) y la 
descomposici\'on (4.11). Obtenemos 
$$\align
h_{00}(r) &=\frac{\kappa \sigma}{2\mu}\delta^{(2)}(\ovr{r})+\frac{\kappa 
m_{{}_{AD}}}{2\mu}(m-m_{{}_{AD}}\sigma )Y(m_{{}_{AD}}r), \tag 4.32,a\\
h_{i0}(r)&=h_{0i}(r) =\frac{\kappa}{2\mu}(m-m_{{}_{AD}}\sigma )\ep_{ij}
\partial_jY(m_{{}_{AD}}r),\tag 4.32,b\\
h_{ij}(r) &=\frac{\kappa
\sigma}{2\mu}\delta^{(2)}(\ovr{r})\delta_{ij}+\frac{\kappa  m_{{}_{AD}}}{2\mu}
(m-m_{{}_{AD}}\sigma )\Bigl(\delta_{ij}-\frac{\partial_i\partial_j}
{m^2_{{}_{AD}}}\Bigr)
Y(m_{{}_{AD}}r), \tag 4.32,c
\endalign
$$
y como $h_{mn}=h_{nm}$, resultar\'a que $h_{\ov{mn}}=2h_{mn}$. En (4.32) 
$m_{AD}\equiv M^2/\mu$ corresponde a la masa de las excitaciones de la 
teor\'{\i}a autodual. El par\'ametro de comportamiento any\'onico ser\'a 
para un c\'{\i}rculo de radio $R$
$$
\alpha_{AD}(R)=\frac{\kappa m\kappa (m-m_{AD}\sigma)}{\mu}m_{AD}RK_1(m_{AD}R),
\tag 4.33 $$
que tiende a cero cuando $R\to \infty$. Cuando $m-m_{AD}\sigma =0$,
$\alpha_{AD}$ es id\'enti\-ca\-men\-te cero. La analog\'{\i}a con el caso
vectorial se obtiene si se identifica $km \leftrightarrow q$ y $k\sigma
\leftrightarrow g$.

Para la teor\'{\i}a de Einstein autodual partir\'{\i}amos del sitema
$$\align
(\frac{1}{2}\ep^{pmn}\ep^{srl}-\ep^{pml}\ep^{srn})&\partial_m\partial_rh_{sl}
-\mu\ep^{prs}\partial_sh_r{}^n+\\
&-M^2(h^{np}-\eta^{pn}h_l{}^l)=-\kappa T^{pn}.\tag 4.34
\endalign
$$
Las masas de las excitaciones, como vimos en el Cap\'{\i}tulo V son 
$(\ep =\mu/M)$
$$
m_\pm =\frac{M\ep}{2}\Bigl(\sqrt{1+\frac{4}{\ep^2}}\pm 1\Bigr).\tag 4.35
$$

Siguiendo el mismo proceso y descrito, obtenemos, de nuevo, que 
$h_{mn}=h_{nm}$ y
$$\align
h_{\ov{00}} &=\frac{\kappa}{m_++m_-}[m_+(m-m_+\sigma )Y(m_+r)+\\
&\ \ \ \ \ \ \ \ \ \ \ \ \ \ \ \ \ \ \ \ \ \ \ +m_-(m+m_-\sigma )
Y(m_-r)], \tag 4.36,a\\
h_{\ov{0i}} &=\frac{\kappa}{m_++m_-}\ep_{ij}\partial_j[(m-m_+\sigma )
Y(m_+r)-\\
&\ \ \ \ \ \ \ \ \ \ \ \ \ \ \ \ \ \ \ \ \ \ \ -(m+m_-\sigma )Y(m_-r)],\tag 4.36,b\\
h_{\ov{ij}}&=\frac{\kappa}{m_++m_-}[m_+(m-m_+\sigma )\Bigl(\delta_{ij}
-\frac{\partial_i\partial_j}{m^2_+}\Bigr)Y(m_+r)+\\
&\ \ \ \ \ \ \ \ \ \ \ \ \ \ \ \ \ \ \ +m_-(m+m_-\sigma )
\Bigl(\delta_{ij}-\frac{\partial_i\partial_j}{m^2_+}\Bigr)
Y(m_-r)],\tag 4.36,c\\
\endalign
$$
El par\'ametro de comportamiento any\'onico, para un c\'{\i}rculo de radio $R$,
es $$\align
\alpha_{E_{(AD)}}(R)=\frac{\kappa m}{m_++m_-}[\kappa m_+&R(m-m_+\sigma
)K_1(m_+R)+\\ 
-&\kappa m_-R(m+m_-\sigma )K_1(m_-R)],\tag 4.37
\endalign
$$
que tiende a cero si hacemos tender $R\to \infty$.

El resultado (4.36) pod\'{\i}amos obtenerlo directamente teniendo en cuenta 
el hecho que el propagador de la acci\'on de Einstein autodual es la 
combinaci\'on lineal de dos propagadores autoduales. Para esto tomamos 
$M=\mu $ en (4.32) y tomamos esto como el resultado para la teor\'{\i}a $AD$ 
con helicidad $+2$. Lo llamamos $h^{(+)}_{mn}(M)$. Es inmediato notar que, 
en virtud de la ecuaci\'on (2.18) del Cap\'{\i}tulo V, para la teor\'{\i}a 
autodual
$$
h_{mn}=\frac{1}{m_++m_-}[m_+h^{(+)}_{mn}(m_+)+m_-h^{(-)}_{mn}(m_-)],\tag 4.38
$$
que corresponde a las ecuaciones (4.36).

\vskip 3mm
\noi
{\bf 4.4 La teor\'{\i}a autodual con acoplamiento no minimal}

\vskip 3mm

En virtud de la analog\'{\i}a que existen entre las teorias de spin 1 y 
spin 2, probamos que sucede si miramos el comportamiento any\'onico de la 
teor\'{\i}a $AD$ con una fuente de la forma
$$
\wt{T}^{mn}=-m_{AD}\ep^{nls}\partial_l(-\Delta )^{-1}T_s{}^m, \tag 4.39
$$
con $T^{mn}$ igual que en (4.10). Esta forma particular de fuente la 
escogemos en analog\'{\i}a con eso vectorial (ecuaci\'on (3.24)). Observamos 
que $\wt{T}^{mn}$ no es sim\'etrica.

Partimos de (4.31) con la guente (4.39) y $m_{AD}=\frac{M^2}{\mu} =\mu$, 
consiguiendo en principio que $h^{TL}$, $V$, $n^L-v^T$ y $n^L+v^T$ son nulos. 
Las ecuaciones restantes son
$$\align
-m_{AD}\Delta h^T+\Delta (n^T+^L)-m_{AD}\Delta h^L &=\frac{\kappa
m_{AD}\sigma}{2} \delta^{(2)}(\ovr{r}),\tag 4.40,a\\
m_{AD}(n^T+v^L)-\Delta h^T &=\kappa mC(m_{AD}r),\tag 4.40,b\\
m_{AD}(n^T-v^L)-n &=0,\tag 4.40,c\\
m_{AD} h^L+(n^T-v^L)+m_{AD}(-\Delta )^{-1}n &=-\frac{\kappa \sigma}{2}C(m_{AD}r),
\tag 4.40,d\\
h^T+(-\Delta )^{-1}n &= 0,\tag 4.40,e
\endalign
$$
de donde

$$\align
n(r) &=\frac{\kappa m_{AD}(m-m_{AD}\sigma )}{2\mu}Y(m_{AD}r),\tag 4.41,a\\
h^T(r) &=-\frac{\kappa (m-m_{AD}\sigma )}{2\mu m_{AD}}(C(m_{AD}r)-Y(m_{AD}r)),
\tag 4.41,b\\
h^L(r) &=-\frac{\kappa m}{2\mu m_{AD}}C(m_{AD}r),\tag 4.41,c\\
n^T(r) &=\frac{\kappa m_{AD}(m-m_{AD}\sigma )}{2\mu}Y(m_{AD}r)-\frac{km}{2\mu}
C(m_{AD}r),\tag 4.41,d\\
v^L(r) &=-\frac{\kappa m}{2\mu}C(m_{AD}r),\tag 4.41,e
\endalign
$$

Las soluciones (4.41) son iguales a las de la teoria $VCS$ si hacemos la 
identificaci\'on $m_{AD}\to -\mu$, $\mu \to -\mu$ y fijamos a 
$\mu h^L=-v^L$. El comportamiento any\'onico, a partir de (4.41) es, para 
un c\'{\i}rculo de radio $R$
$$
\wt{\alpha}_{AD}(R)=-\frac{\kappa^2m^2}{\mu}+\frac{\kappa m}{\mu}k(m-m_{AD}\sigma
) m_{AD}R K_1(m_{AD}R).\tag 4.42 
$$
Este es igual al de la teor\'{\i}a $TCS$ asint\'oticamente, observamos 
que esta igualdad de comportamientos any\'onicos no depende del t\'ermino 
inercial, s\'olo del t\'ermino $TCS$. $\wt{\alpha}_{AD}(R)=\alpha_{VCS}(R)$ 
(e\-cua\-ci\'on (4.24)) si hacemos la identificaci\'on antes se\~nalada.
Conclu\'{\i}mos entonces, que la teor\'{\i}a $VCS$ en un calibre y acoplamiento 
minimal con la fuente (4.10) corresponde a la teor\'{\i}a autodual con un 
acoplamiento no minimal. 

\newpage

\pageno=100
\headline={\ifnum\pageno=100\hfil\else\hss\tenrm \folio\ \fi}

$\ $
\vskip 1cm
\centerline{\catorcerm Cap\'{\i}tulo {\catorcebf VII}}
\vskip 1cm

\centerline{\dseis CONCLUSIONES}
\vskip 2cm

En este trabajo hemos introducido y analizado nuevas teor\'{\i}as en 2+1 
dimensiones. Una de estas es la Gravedad masiva Vectorial de Chern-Simons. 
Esta tiene la acci\'on (Cap\'{\i}tulo IV) [43,47]
$$
S_{VCS}=S_E\pm \mu S_{TCS}, \tag 6.1
$$
donde $S_E$ es la acci\'on de Einstein y $S_{TCS}$ es el t\'ermino de $CS$ 
tri\'adico. $S_{VCS}$ representa a una teor\'{\i}a curva que propaga una 
excitaci\'on masiva de helicidad $2\mu /|\mu |$. Tiene la ventaja respecto 
a la teor\'{\i}a $TM$ existente, que es de segundo orden. Sin embargo, no 
tiene invariancia Lorentz local; aunque mantiene la in\-va\-rian\-cia bajo 
transformaciones de coordenadas. A nivel linealizado hicimos un estudio 
din\'amico bastante amplio, que no hab\'{\i}a sido realizado ya que, 
$S_{VCS}^l$, s\'olo hab\'{\i}a aparecido como una acci\'on intermedia entre 
la acci\'on maestra de spin 2 y la teor\'{\i}a autodual [17]. Se analiz\'o su 
espectro f\'{\i}sico covariantemente, se hall\'o su acci\'on reducida y se 
analiz\'o parcialmente con el formalismo 
can\'onico. Como posibles ampliaciones al estudio de esta teor\'{\i}a queda 
el an\'alisis can\'onico de la teor\'{\i}a curva, as\'{\i} como la b\'usqueda 
de soluciones exactas a sus ecuaciones de movimiento.

Dentro de los resultados resaltantes de esta t\'esis est\'a tambi\'en el 
hecho de que los fen\'omenos f\'{\i}sicos que se encuentran en las 
teor\'{\i}as de spin 1 se repiten uniformemente en las teor\'{\i}as de spin 2:

\noi
1) Tenemos una teor\'{\i}a invariante bajo $P$ y $T$ que describe a dos 
excitaciones masivas de igual masa y helicidades opuestas. Para spin 1 
est\'a la teor\'{\i}a de Proca, para spin 2 la de Fierz-Pauli. La 
``ra\'{\i}z'' de sus ecuaciones de movimiento nos proporciona la 
``condici\'on de autodualidad'' que verifican las partes din\'amicas del 
campo matriz.

\noi
2) Tenemos una teor\'{\i}a autodual $(AD)$ para spin 1 y 2, cuyas ecuaciones 
de movimiento son justamente la condici\'on de autodualidad (sobre 
$a_m^T$ y $h^{Tt}_{mn}$) ya mencionadas. Estas teor\'{\i}as $AD$ violan $P$ y 
$T$, no son sensibles a transformaciones de calibre, y 
describen una excitaci\'on masiva de masa $m$ y helicidad +1 \'o $-$1 (+2 \'o
$-$2) dependiendo del signo con que aparece $m$ en la acci\'on. Estas  acciones
autoduales tambi\'en existen para spines altos y conjeturamos que  existen para
cualquier spin entero [19,20]. La condici\'on de  autodualidad constituye una
realizaci\'on de la ecuaci\'on de Pauli-Lubanski que  verifican las distintas
representaciones irreducibles del grupo de  Poincar\'e en 2+1 dimensiones [67].

\noi
3) Existe una teor\'{\i}a masiva invariante de calibre de segundo orden. Para 
spin 1 es la teor\'{\i}a $TM$ y para spin 2 la gravedad $VCS$ linealizada. 
Estas teor\'{\i}as son equivalentes a las teor\'{\i}as autoduales como 
teor\'{\i}as libres [16,18,31,44] (Cap\'{\i}tulo II y III). Esto ocurre 
tambi\'en con teor\'{\i}as de spin 3 [20]. Puede probarse, tambi\'en, que las
teor\'{\i}as son equivalentes can\'onicamente en el sentido que una corresponde
a la otra con el calibre fijado ([36,44] y Cap\'{\i}tulo III). Cuando hay
fuentes externas las teor\'{\i}as deben estar acopladas de manera distinta con
la fuente para mirar su equivalencia. 

\noi
4) Las teor\'{\i}as invariantes de calibre de segundo orden pueden sufrir 
un proceso de rotura de simetr\'{\i}a. Para spin 1 la 
teor\'{\i}a con simetr\'{\i}a rota es la de Proca-Chern-Simons
$$
S_{PCS}=S_{Maxwell}-\mu S_{CS}-m^2S_P, \tag 6.2,a
$$
para spin 2 es la de Einstein autodual
$$
S_E^{(AD)}=S_E-\mu S_{TCS}-m^2S_{FP}. \tag 6.2,b
$$
Ambas teor\'{\i}as vimos que describen 2 excitaciones masivas, au\-to\-dua\-les, 
de masas distintas
$$
m_\pm =\frac{m\ep}{2}(\sqrt{1+\frac{4}{\ep^2}}\pm 1), \tag 6.3
$$
y tienen energ\'{\i}a definida positiva (Cap\'{\i}tulo V) [34,51,52,54,56]. 
Este resultado tambi\'en se verifica con la teor\'{\i}a invariante de calibre 
de spin 3.

\noi
5) Es posible implementar din\'amicamente estad\'{\i}stica y spin fraccional. 
Para spin 1 el t\'ermino responsable de esto es el de $CS$ vectorial. 
Para spin 2 encontramos que es el de $TCS$ linealizado (Cap\'{\i}tulo VI) 
[66]. Este cuadro se mantiene asint\'oticamente en presencia del t\'ermino de 
Maxwell, para el caso de spin 1 [29,30]; y del t\'ermino de Einstein, para 
el caso de spin 2 [66]. Cuando hay acoplamiento minimal y se ha roto la 
simetr\'{\i}a, se pierde el comportamiento any\'onico. \'Este se mantiene si 
el acoplamiento es no minimal (Cap\'{\i}tulo VI).

Dentro de estas comparaciones incluimos a la teor\'{\i}a $TM$ li\-ne\-a\-li\-za\-da, 
encontrando que no es posible romper la invariancia de calibre de 
$S_{TM}^{2,l}$ parcial o totalmente (Cap\'{\i}tulo V) ya que la teor\'{\i}a 
resultante tiene energ\'{\i}a no acotada inferiormente. El
par\'ametro de comportamiento any\'onico no es igual al caso de spin 1 y en
determinado casos  $(m+\mu\sigma =0)$ no lo hay (Cap\'{\i}tulo VI) [66].

Finalmente queda como pregunta abierta la implementaci\'on de 
estad\'{\i}stica fraccionaria a nivel covariante curvo.

\newpage

$\ $

\pageno=103
\headline={\ifnum\pageno=103\hfil\else\hss\tenrm \folio\ \fi}

\vskip 1cm

\item{}{\bf{\catorcebf Referencias}}
\vskip 5mm
\item{[1]}D. Gross, R. D. Pisarski y L. G. Yaffe, Rev. Mod. Phys. {\bf 53} 
(1981) 43.
\item{[2]}G. 'tHooft, {\it ``Dimensional Reduction in Quantum Gravity''}, 
Preprint THU-93/26.
\item{[3]}F. Wilczek, Phys. Rev. Lett. {\bf 48} (1982); {\bf 49} (1982) 957.
\item{[4]}R. B. Laughlin, Phys. Rev. Lett. {\bf 60} (1988) 2677.
\item{[5]}Y. H. Chen, F. Wilczek, E. Witten y B. I. Halperin, Int. J. Mod. 
Phys. {\bf B39} (1989) 9679.
\item{[6]}R. Jackiw, Nuc. Phys. {\bf B252} (1985) 343.
\item{[7]}B. Binegar, J. Math. Phys. {\bf 23} (1982) 1511.
\item{[8]}R. Jackiw y V. P. Nair, Phys. Rev. {\bf D43} (1991) 1933.
\item{[9]}S. Deser y R. Jackiw, Phys. Lett. {\bf B263} (1991) 431.
\item{[10]}J. Fr\"ohlich y P. A. Marchetti, Commun. Math. Phys. {\bf 121} 
(1989) 177.
\item{[11]}S. Doplicher, R. Haag y J. E. Roberts, Commun. Math. Phys. 
{\bf 13} (1969) 1; {\bf 15} (1969) 173; {\bf 23} (1971) 199; {\bf 35} (1974) 
49.
\item{[12]}P. de Sousa Gerbert, Nuc. Phys. {\bf B346} (1990) 440.
\item{[13]}J. Schonfeld, Nuc. Phys. {\bf B185} (1981) 157.
\item{[14]}S. Deser, R. Jackiw y S. Tempelton, Phys. Rev. Lett. {\bf 48} 
(1982) 975; Ann. Phys. {\bf 140} (1982) 372; (E) {\bf 185} (1988) 406.
\item{[15]}P. K. Townsend, K. Pilch y P. van Nieuwenhuizen, Phys. Lett. 
{\bf B136} (1984) 38.
\item{[16]}S. Deser y R. Jackiw, Phys. Lett. {\bf B139} (1984) 371.
\item{[17]}C. Aragone y A. Khoudeir, Phys. Lett. {\bf B173} (1986) 141.
\item{[18]}C. Aragone y A. Khoudeir, {\it ``Quantum Mechanics of Fundamental 
Systems 1''}, Ed. C. Teitelboim, Plenum Press, N.Y. (1988) p.17.
\item{[19]}C. Aragone y A. Khoudeir, por aparecer en la Revista Mexicana de 
F\'{i}sica (1993)
\item{[20]}A. Khoudeir, Tesis de Doctorado, U.S.B. (1993); C. Aragone y A. 
Khoudeir, {\it ``Massive Triadic Chern-Simons spin-3 theory''} por aparecer 
en los procedings del SILARG VIII, World Scientific (1993).
\item{[21]}D. Arovas, R. Schrieffer, F. Wilczek y A. Zee, Nuc. Phys. 
{\bf B251} (1985) 117.
\item{[22]}S. Forte, Rev. Mod. Phys. {\bf 64} (1992) 193.
\item{[23]}R. Jackiw, Ann. Phys. {\bf 201} (1990) 83.
\item{[24]}E. Dzyaloshinski, A. M. Polyakov y P. B. Wiegman, Phys. Lett. 
{\bf A127} (1988) 112.
\item{[25]}X. G. Wen y A. Zee., Nucl. Phys. Proc. Suppl. {\bf 15} 
(1990) 135.
\item{[26]}B. De Witt, Phys. Rev. Lett. {\bf 16} (1966) 1902.
\item{[27]}J. S. Dowker y J. A. Roche, Proc. Phys. Soc. {\bf 92} (1967) 1.
\item{[28]}M. Ortiz, Nuc. Phys. {\bf B375} (1991) 127.
\item{[29]}S. Deser, Phys. Rev. Lett. {\bf 64} (1990) 611.
\item{[30]}S. Deser y J. Mc. Carthy, Nuc. Phys. {\bf B344} (1990) 747; S. 
Deser, Class. Quantum Grav. {\bf 9} (1992) 61.
\item{[31]}S. Deser y J. Mc Carthy, Phys. Lett. {\bf B245} (1990) 441; (A) 
{\bf B248} (1990) 473.
\item{[32]}X. G. Wen y A. Zee, J. Phys. France {\bf 50} (1989) 1623.
\item{[33]}D. Boyanovsky, Phys. Rev. {\bf D42} (1990) 1179.
\item{[34]}S. K. Paul y A. Khare, Phys. Lett. {\bf B171} (1986) 244.
\item{[35]}S. Deser, En {\it College Park 1993, Directions in 
General Relativity}, Vol. 2, p. 114 .
\item{[36]}R. Gianvittorio, A. Restuccia y J. Stephany, Mod. Phys. Lett. 
{\bf A6} (1991) 2121.
\item{[37]}C. R. Hagen, Phys. Rev. Lett. {\bf 58} (1987) 1074.
\item{[38]}S. Deser y R. Jackiw, Phys. Rev. Lett. {\bf 59} (1987) 1981; 
C. R. Hagen, Phys. Rev. Lett. {\bf 59} (1987) 1982.
\item{[39]}S. M. Latinsky y D. P. Sorokin, {\it ``On a non-minimal gauge 
interaction of scalar fields in D=2+1 Chern-Simons-Maxwell theory''}, Preprint 
KFTI 91-92, Kharkov Institute of Physics and Technology.
\item{[40]}S. Deser, R. Jackiw y G. 'tHooft, Ann. Phys. {\bf 152} (1984) 220.
\item{[41]}S. Giddins, J. Abbot y K. Kuchar, Gen. Rel. Grav. {\bf 16} (1984) 
751.
\item{[42]}J. R. Gott III y M. Alpert, Gen. Rel. Grav. {\bf 16} (1984) 243.
\item{[43]}C. Aragone, P. J. Arias y A. Khoudeir, {\it ``Massive Vector 
Chern-Simons Gravity''}, Preprint SB/F/92/192, U.S.B.(Por aparecer en Nuovo 
Cimento {\bf B}) (Nuov. Cim. {\bf 109B} (1994) 303).
\item{[44]}P. J. Arias y J. Stephany, {\it ``Gauge Invariance and Second 
class constraints in 3-D linearized massive gravity''}, Preprint SB/F/93/210, 
U.S.B. (J. Math. Phys. {\bf 36} (1995) 1868).
\item{[45]}P. Senjanovic, Ann. Phys. {\bf 100} (1976) 227.
\item{[46]}R. Jackiw, en {\it ``Relativity and Gravitation: Classical and 
Quantum''}, Pro\-cee\-dings del SILARG VII, Eds. J.C. D'Olivo et al., World 
Scientific (1991) p. 74.
\item{[47]}C. Aragone, P. J. Arias y A. Khoudeir, {\it ``Light-front dynamics 
of Massive Vector Chern-Simons gravity''}, proceedings 
del SILARG VIII, World Scientific (1993), p.523, hep-th/9309132.
\item{[48]}Y. Hosotani, Phys. Lett. {\bf B319} (1993) 332.
\item{[49]}G. 'tHooft, Commun. Math. Phys. {\bf 117} (1988) 685.
\item{[50]}E. W. Mielke y P. Baekler, Phys. Lett. {\bf A156} (1991) 399.
\item{[51]}C. Aragone y P. J. Arias, Mod. Phys. Lett. {\bf A5} (1990) 1651.
\item{[52]}C. Aragone, P. J. Arias y A. Khoudeir, en {\it ``Relativity and 
Gravitation: Classical and Quantum''}, Proceedings del SILARG VII, Eds. J.C. 
D'Olivo et al., World Scientific (1991), p. 437.
\item{[53]}C. Itzykson y J. B. Zuber, {\it ``Quantum Field Theory''}, Ed. Mc 
Graw Hill International, p. 162.
\item{[54]}C. Aragone, P. J. Arias y A. Khoudeir, {\it ``Spontaneously broken 
Einstein Selfdual massive spin two theory''}, Preprint SB-F-162, U.S.B..
\item{[55]}C. Aragone, P. J. Arias y A. Khoudeir, {\it ``Two  Gravitationally 
Chern-Simons terms are too many''}, hep-th/9309149.
\item{[56]}C. Aragone, P. J. Arias y A. khoudeir, {\it ``On the spontaneous 
breakdown of topological massive gravities''}, Preprint SB/F/93/202, U. S. B. 
(Nuov. Cim. {\bf 112B} (1997) 63).
\item{[57]}G. Morandi, {\it ``The role of topology in Classical and Quantum 
Physics''}, Lecture Notes in Physics, m. 7, Springer Verlag.
\item{[58]}A. P. Balachandran, Int. J. Mod. Phys. {\bf B5} (1991) 2585.
\item{[59]}J. P. Lupi, Trabajo de grado, U.S.B. (1992).
\item{[60]}J. M. Leinaas y J. Myrheim, Il Nuovo Cimento, {\bf 37} (1977) 1.
\item{[61]}Y. S. Wu, Phys. Rev. Lett. {\bf 52} (1984) 2108.
\item{[62]}M. Laidlaw, C.M. deWitt Phys. Rev. {\bf D3}(1971) 1375.
\item{[63]}J. S. Dowker, J. Phys. {\bf A5} (1972) 936.
\item{[64]}L. Schulman, Phys. Rev. {\bf 176} (1968) 1558.
\item{[65]}R. Mackenzie y F. Wilczek, Int. J. Mod. Phys. {\bf A3} (1988) 2827.
\item{[66]}C. Aragone y P. J. Arias, {\it ``More gravitational anyons'''}, 
proceedings del SILARG VIII, World Scientific (1993), p. 553, hep-th/9309131.
\item{[67]}T. Gonera y P. Kosinski, Phys. Lett. {\bf B268} (1991) 81.

\newpage

$\ $

\pageno=107
\headline={\ifnum\pageno=107\hfil\else\hss\tenrm \folio\ \fi}

\vskip 1cm

\centerline{\catorcerm AP\'ENDICE A}

\vskip 1cm

\centerline{\dseisbf CONVENCIONES PARA TEOR\'IAS DE}

\vskip 3mm

\centerline{\dseisbf GRAVEDAD CURVA Y LINEALIZADA}

\vskip 2cm

\noi
{\bf A.1 Transformaciones bajo difeomorfismos}
\vskip 3mm
Ante un cambio de coordenadas

$$
x\to x'= x-\xi(x), \tag A.1
$$
los objetos quedan definidos por su comportamiento ante estas 
transformaciones:

\noi
a) Un escalar cambia como

$$
\delta f(x)=\xi^m\partial_m f, \tag A.2
$$
\noi
b) Para vectores covariantes $U_m$ y contravariantes $V^m$

$$\align
\delta U_m &= \xi^n\partial_nU_m+(\partial_m\xi^n)U_n, \tag A.3,a\\
\delta V^m&=\xi^n\partial_nV^m-\partial_n\xi^mV^n, \tag A.3,b
\endalign
$$
\noi 
c) Un tensor general cambia como 

$$\align
\delta T^{m_1\cdots m_j}{}_{n_1\cdots n_j} &= \xi^n\partial_n
T^{m_1\cdots m_j}{}_{n_1\cdots n_j}+\partial_{n_1}\xi^p
T^{m_1\cdots}_{pn_2\cdots}+\\ 
&\ \ \ \ + \partial_{n_2}\xi^pT^{m_1\cdots}_{n_1p\cdots}+\cdots \tag A.3,c
\endalign
$$
\noi
d) Una densidad de peso $p$

$$
\delta h=\xi^n\partial_nh-p\partial_n\xi^nh, \tag A.4
$$

\noi
e) Para una densidad vectorial $k^m$ de peso $p$

$$
\delta k^m=\xi^n\partial_nk^m-(\partial_n\xi^m)k^n-p\partial_n\xi^nk^m, 
\tag A.5
$$
\vskip 5mm
\noi
{\bf A.2 Derivadas covariantes, conexiones, tensores de Riemmann, Ricci y 
Einstein}
\vskip 3mm
Las derivadas covariantes de un vector covariante o con\-tra\-va\-rian\-te son

$$\align
\Cal{D}_mU_n &=\partial_mU_n-\Gamma_{mn}^lU_l, \tag A.6,a\\
\Cal{D}_mV^n &=\partial_mV^n+\Gamma_{ml}^nV^l, \tag A.6,b
\endalign
$$
las cuales transforman bajo difeomorfismos como vectores. $\Gamma_{mn}{}^l$ 
no es un buen tensor, pero la torsi\'on si lo es

$$
T_{mn}{}^l\equiv \Gamma_{mn}^l-\Gamma_{nm}^l. \tag A.7
$$
El tensor de curvatura $R_{mnl}{}^s$ se define por el conmutador entre 
derivadas covariantes

$$
[\Cal{D}_m,\Cal{D}_n]V^l=R_{mns}{}^lV^s-T_{mn}{}^s\Cal{D}_sV^l, \tag A.8
$$
donde

$$
R_{mns}{}^l=\partial_m\Gamma_{ns}^l+\Gamma_{mr}^l\Gamma_{ns}^r-
(m\leftrightarrow n).\tag A.9
$$
Las identidades de Jacobi para los conmutadores permite hallar las 
identidades de Bianchi para $T_{mn}{}^l$ y $R_{mnl}{}^s$

$$\align
&\Cal{D}_rR_{mns}{}^l+T_{mn}{}^tR_{trs}{}^l+
\text{(C\'{\i}clicos en $r,m,n$)}=0,\tag A.10,a\\
&R_{mnr}{}^t-\Cal{D}_rT_{mn}{}^t-T_{mn}{}^sT_{sr}{}^t+
\text{(C\'{\i}clicos en $r,m,n$)}=0\tag A.10,b
\endalign
$$

Con el tensor de curvatura o tensor Riemmann se define el tensor de Ricci 
como la contracci\'on

$$
R_{mn}\equiv R_{lmn}{}^l=R_{nm},\tag A.11,a
$$
el cual al contraerlo, da el escalar de curvatura

$$
R\equiv R_m{}^m. \tag A.11,b
$$
En $d$ dimensiones el tensor de curvatura tiene $(1/12)d^2(d^2-1)$ 
componentes independientes y el de Ricci $(1/12)d(d+1)$.

Con $R_{mn}$ y $R$ se define el tensor Einstein como 

$$
G_{mn}=R_{mn}-\frac{1}{2}g_{mn}R \tag A.12
$$
el cual es covariantemente conservado, en virtud de las identidades de 
Bianchi (A.10)

$$
\Cal{D}_mG^{mn}=0.\tag A.13
$$
\vskip 5mm
\noi
{\bf A.3 Particularidades en d=2+1}
\vskip 3mm

En tres dimensiones, el tensor de Einstein y de Riemmann tiene el mismo 
n\'umero de componentes independientes. As\'{\i} que uno puede expresarse 
en fun\-ci\'on del otro. De hecho

$$
R_{mnl}{}^s=-\ep_{mnt}\ep^s{}_{pl}G^{tp}. \tag A.14
$$
As\'{\i}, no es posible definir el tensor de Weyl de manera usual. Sin 
embargo tenemos un tensor conforme. Este es el tensor de Cotton

$$
C^{mn}\equiv \frac{1}{\sqrt{-g}}\ep^{mpl}\Cal{D}_p\wt{R}_l{}^n, \tag A.15
$$
el cual es sim\'etrico, tiene traza nula y es covariantemente conservado. En 
(A.15)

$$
\wt{R}_m{}^n=R_m{}^n-\frac{1}{4}\delta_m{}^nR \tag A.16
$$
\vskip 5mm

\noi
{\bf A.4 Lenguaje de las Tr\'{\i}adas}
\vskip 3mm
Resulta conveniente, en ocasiones, remitirse al espacio tangente de la 
va\-rie\-dad. Ah\'{\i}, introducimos las tetradas (vielbeins), $e_a$, las 
cuales tienen componentes curvil\'{\i}neas $e_a{}^m$, sus duales, $e^a$, 
tendr\'an componentes $e_m{}^a$. As\'{\i}

$$
e_a{}^me_m{}^b=\delta_a{}^b.\tag A.17
$$

Referimos a los \'{\i}ndices ``de mundo'' con las letras intermedias del 
alfabeto y a los \'{\i}ndices del espacio tangente con las primeras letras 
del alfabeto.

En el espacio tangente definimos las transformaciones que conservan la norma 
$V_aU^a$

$$
\delta V^a=V^bX_b{}^a,\ \ \ \delta U_a=-X_b^aU_a, \tag A.18
$$
y respecto a estas transformaciones tenemos las derivadas co\-va\-rian\-tes

$$\align
D_mU_a &=\partial_mU_a-\omega_{ma}{}^bU_b, \tag A.19,a\\
D_mV^a &=\partial_mV^a+V^b\omega_{mb}{}^a. \tag A.19,b
\endalign
$$
$\omega_{ma}{}^b$ transforma como una buena conexi\'on

$$
\delta \omega_{mb}{}^a=-D_mX_b{}^a \tag A.20
$$

Para objetos mixtos, como por ejemplo $e_m{}^a$, la derivada co\-va\-rian\-te 
total es

$$
\Cal{D}_me_n{}^a=D_me_n{}^a-\Gamma_{mn}^le_l{}^a.\tag A.21
$$
Por consistencia entre las definiciones se pide que $\Cal{D}_m$ $ e_m{}^a=0$. 
As\'{\i}

$$
D_me_n{}^a=\Gamma_{mn}^le_l{}^a, \tag A.22
$$
Por lo tanto

$$\align
T_{mn}{}^le_l{}^a &=(D_me_n{}^a-D_ne_m{}^a)e_a{}^l\\
&\equiv T_{mn}{}^a e_a{}^l\tag A.23
\endalign
$$

Cuando $T_{mn}{}^a=0$ es posible obtener $\omega_m{}^a$ en funci\'on de 
$e_m{}^a$. Podemos, tambi\'en, definir el an\'alogo al tensor de curvatura. 
Este es $R_{mna}{}^b=\partial_m\omega_{na}{}^b-\partial_n\omega_{ma}{}^b+
\omega_{na}{}^c\omega_{mc}{}^b-\omega_{nb}{}^c\omega_{mc}{}^a$, y

$$
[D_m,D_n]V^a=V^bR_{mnb}{}^a.\tag A.24
$$
Resulta que

$$
R_{mna}{}^b=R_{mnl}{}^se_s{}^be^l{}_a \tag A.25
$$

En dimensi\'on $2+1$ llamamos al vielbein, dreibein (tr\'{\i}ada). En esta 
dimensi\'on podemos definir ``duales'' a los $R_{mna}{}^b$, 
$\omega_{ma}{}^b$, $\cdots$

$$\align
\omega_m{}^a &\equiv \frac{1}{2}\ep^{abc}\omega_{mbc} \tag A.26,a\\
R_{mn}^*{}^a &=\frac{1}{2}\ep^{abc}R_{mnbc},\tag A.26,b
\endalign
$$
Y, a\'un mas

$$\align
R^{**pa} &=\frac{1}{2}\ep^{pmn}R_{mn}^*{}^a\\
&=\frac{1}{4}\ep^{pmn}\ep^{abc}R_{mnbc}.\tag A.26,c
\endalign
$$

El determinante de dreibein es

$$
e=-\frac{1}{3!}\ep^{pmn}\ep_{abc}e_p{}^ae_m{}^be_n{}^c, \tag A.27
$$
as\'{\i}

$$
e_a{}^p=\frac{1}{2e}\ep^{pmn}\ep_{abc}e_m{}^be_n{}^c. \tag A.28
$$
Puede mostrarse que 

$$
R^{**pa}e_a{}^r=eG^{pr}. \tag A.29
$$
Por \'ultimo, la soluci\'on a (A.23) es

$$
e\omega_m{}^a=(e_{mb}e_p{}^a\ep^{prs}\partial_re_s{}^b-\frac{1}{2}e_m{}^a
e_{pb}\ep^{prs}\partial_re_s{}^b)\tag A.30
$$
\vskip 5mm

\noi
{\bf A.5 La acci\'on de Einstein y su linealizaci\'on}
\vskip 3mm
La acci\'on de Einstein

$$
S_E=-\frac{1}{2k^2}\left<\sqrt{-g}R\right> ,\tag A.31
$$
se escribe, alternativamente, como

$$
S_E=\frac{1}{2k^2}\left<e_{pa}\ep^{pmn}R_{mn}^*{}^a\right> .\tag A.32
$$ 

La prescripci\'on al linearizar es $e_m{}^a=\delta_m{}^a+kh_m{}^a$, 
obtenemos entonces

$$
S_E^l=\frac{1}{2}\left<2h_{pa}\ep^{pmn}\partial_m\omega_n{}^a-
(\omega_{ma}\omega^{am}-\omega_m{}^m\omega_a{}^a)\right> .\tag A.33
$$

Al linearizar la m\'etrica $(g_{mn}=\eta_{mn}+kh_{mn}^{(s)})$ el tensor de 
Einstein es

$$\align
G^{mn}&=\frac{1}{2}(-\sq h^{(s)mn}+\partial^m\partial_lh^{(s)ln}+
\partial^n\partial_lh^{(s)lm}+\\ 
&\ \ \ \ \ \ \ -\partial^m\partial^nh^{(s)r}_r+\eta^{mn}
(\sq h^{(s)}-\partial_r\partial_s h^{rs}))\tag A.34\\
&=-\frac{1}{2}\ep^{mrs}\ep^{nlp}\partial_r\partial_lh_{sp}^{(s)}\tag A.35
\endalign
$$
A nivel linealizado (A.30) se escribe como 

$$
\omega_m{}^a=-\frac{1}{2}\delta_m{}^a\ep^{pnr}\partial_ph_{nr}+
\ep^{anr}\partial_nh_{rm} \tag A.36
$$
que podemos sustituir en (A.33) y obtenemos

$$
S_E=-\frac{1}{4}<h_{pa}^{(s)}G^{pa}> \tag A.37
$$
con

$$
h^{(s)}_{pa}=h_{pa}+h_{ap} \tag A.38
$$

\newpage

$\ $

\pageno=114
\headline={\ifnum\pageno=114\hfil\else\hss\tenrm \folio\ \fi}

\vskip 1cm

\centerline{\catorcerm AP\'ENDICE B}

\vskip 1cm

\centerline{\dseisbf OPERADORES DE PROYECCI\'ON Y}

\vskip 3mm

\centerline{\dseisbf TRANSFERENCIA PARA TEOR\'IAS}

\vskip 3mm

\centerline{\dseisbf DE SPIN 2 EN D=2+1}

\vskip 2cm

Para el c\'alculo de los propagadores de teor\'{\i}as de gravedad 
linealizada, resulta de gran utilidad el uso de los operadores de 
proyecci\'on. Estos proyectan al campo $h_{mn}$ en sus distintas partes 
irreducibles, lo cual convierte el problema de hallar el propagador en uno 
netamente algebraico.

Adem\'as de estos operadores de proyecci\'on, los cuales como veremos 
constituyen una descomposici\'on de la ``unidad", necesitamos los operadores 
de transferencia. Estos tienen la propiedad de transferir una componente 
irreducible, de $h_{mn}$, a otra de igual spin. Especializamos a D=2+1 lo 
ya hecho en D=3+1 [1,2].

Para campos vectoriales, la construcci\'on de los operadores de proyecci\'on, 
es un pro\-blema trivial. Adem\'as de que no es necesario introducir 
operadores de transferencia.

Introducimos, primeramente los operadores 

$$
\wh{\partial}^m\equiv \frac{\partial^m}{\sq^{1/2}},\tag B.1
$$
los cuales cumplen 

$$
\wh{\partial}_m\wh{\partial}^m=1.\tag B.2
$$
La acci\'on de $\wh{\partial}_m$ se entiende en el espacio de Fourier, ya que 
si 

$$
\varphi (x)=\frac{1}{(2\pi )^3}\int d^3ke^{-ik^nx_n}\varphi (k),
$$
entonces

$$
\wh{\partial}_m\varphi (x)=-\frac{1}{(2\pi )^3}\int d^3k
\frac{e^{-ik^nX_n}ik_m\varphi (k)}{(k^lk_l)^{1/2}}.\tag B.3
$$

Introducimos el proyector transverso

$$
P_m{}^n\equiv \delta_m{}^n+\omega_m{}^n,\tag B.4
$$
donde $\omega_m{}^n=\wh{\partial}_m\wh{\partial}^n$. El proyector 
$P_m{}^n$ manda a cualquier campo vectorial $V_n$ en su parte transversa 

$$
P_m{}^nV_n\equiv V_m^T\ ;\ \wh{\partial}^mV_m^T=0.\tag B.5
$$
Como es sabido la parte de spin 1 de un campo vectorial est\'a en su parte 
transversa $V_m^T$. Esta, a su vez, tiene dos helicidades posibles. Podemos 
proyectar a $V_m^T$en estas, con los proyectores $P_{\pm m}{}^n$, definidos 
por 

$$
P_{\pm m}{}^n\equiv \frac{1}{2}(P_m{}^n\pm\xi_m{}^n),\tag B.6
$$
donde hemos introducido el operador

$$
\xi_m{}^n\equiv \ep_m{}^{rn}\wh{\partial}_r,\tag B.7
$$
el cual es como la "ra\'{\i}z" de $P_m{}^n$ y adem\'as es sensible a paridad. 
As\'{\i} que los proyectores $P_{\pm m}{}^n$ son sensibles, tambi\'en, a 
paridad.

Los operadores $P_m{}^n$, $P_{\pm m}{}^n$ y $\xi_m{}^n$, verifican el 
\'algebra

$$\align
P_m{}^nP_n{}^l &=P_m{}^l,\tag B.8,a\\
\xi_m{}^n\xi_n{}^l &=P_m{}^l,\tag B.8,b\\
P_m{}^n\xi_n{}^l &=\xi_m{}^nP_n{}^l=\xi_m{}^l\tag B.8,c\\
P_{\pm m}{}^nP_n{}^l &=P_m{}^nP_{\pm n}{}^l=P_{\pm m}{}^l,\tag B.8,d\\
P_{\pm m}{}^n\xi_n{}^l &=\xi_m{}^nP_{\pm n}{}^l=\pm P_{\pm m}{}^l,\tag B.8,e
\endalign
$$
La descomposici\'on de la unidad para campos vectoriales es entonces

$$
1_m{}^n=P_{+m}{}^n+P_{-m}{}^n+\omega_m{}^n,\tag B.9
$$
donde $\omega_m{}^n$ es el proyector en la parte de spin 0. Podemos pasar 
ahora a los proyectores para las distintas partes de un objeto de 2 
\'{\i}ndices, $h_{mn}$.

Hacemos una descomposici\'on primaria de la unidad, 
$1_{mn}{}^{ls}\equiv \delta_m{}^l\delta_n{}^s$, en sus partes sim\'etrica y 
antisim\'etrica.

$$
1=S+A,\tag B.10
$$
donde

$$\align
S_{mn}{}^{ls} &=\frac{1}{2}(\delta_m{}^l\delta_n{}^s+\delta_m{}^s\delta_n{}^l)
,\tag B.11,a\\
A_{mn}{}^{ls} &=\frac{1}{2}(\delta_m{}^l\delta_n{}^s-\delta_m{}^s\delta_n{}^l)
.\tag B.11,b
\endalign
$$

La parte de spin 2 de $h_{mn}$ est\'a en la componente sim\'etrica, 
transversa y sin traza $h_{mn}{}^{Tt}$
$(h_{mn}{}^{Tt}=h_{nm}{}^{Tt},\eta^{mn}h_{mn}{}^{Tt}=\wh{\partial}^m
h_{mn}{}^{Tt}=0)$. Su parte de spin 1 se encuentra al tomar divergencia 
respecto a alg\'un \'{\i}ndice, y la parte de spin 0 al tomar traza, o la 
doble divergencia. Siguiendo estos lineamientos tenemos que las 9 componentes 
que originalmente tienen $h_{mn}$, de las cuales extraemos 6 con $S$ y 3 con 
$A$, quedar\'an repartidas as\'{\i}:

$$
S:\ 6\ {\text{en la parte sim\'etrica}}
\cases
2 &{\text{de spin 2}}\\
2 &{\text{de spin 1}}\\
2 &{\text{de spin 0}}
\endcases
$$

$$
A:\ 3\ {\text{en la parte antisim\'etrica}}
\cases
2 &{\text{de spin 1}}\\
1 &{\text{de spin 0}}
\endcases
$$

As\'{\i} para la parte sim\'etrica, tenemos que los proyectores que nos 
extraen las partes de spin 2, spin 1 y spin 0, de 
$S_{mn}{}^{ls}h_{ls}=h_{mn}^{(s)}$, son respectivamente

$$\align
P^2_{S\ mn}{}^{ls} &=\frac{1}{2}(P_m{}^lP_n{}^s+P_m{}^sP_n{}^l-P_{mn}P^{ls}),
\tag B.12,a\\
P^1_{S\ mn}{}^{ls} &=\frac{1}{2}(P_m{}^l\omega_n{}^s+P_m{}^s\omega_n{}^l+
P_n{}^l\omega_m{}^s+P_n{}^s\omega_m{}^l),\tag B.12,b\\
P^0_{S\ mn}{}^{ls} &=\frac{1}{2}P_{mn}P^{ls},\tag B.12,c\\
P^0_{W\ mn}{}^{ls} &=\omega_{mn}\omega^{ls},\tag B.12,d
\endalign
$$
Observamos que por construcci\'on $P^2_S$ es transverso y sin traza, 
requerimiento exigido a la parte de spin 2. Para $P^1_S$ vemos que es una 
construcci\'on sim\'etrica donde primero se toma la divergencia y luego se 
proyecta ``lo que queda" en su parte transversa. Finalmente, para las 
componentes de spin 0, $P^0_W$ toma la doble divergencia y $P^0_S$ est\'a 
relacionado con la traza. Una manera alterna para las componentes de spin 0 
es construir el proyector que extrae la traza

$$
T_{mn}{}^{ls}\equiv \frac{1}{3}\eta_{mn}\eta^{ls},\tag B.13
$$
y luego exigiendo que $S=P^2_S+P^1_S+T+R$, queda perfectamente identificado 
el proyector que estar\'a relacionado con la doble divergencia, de manera 
an\'aloga como est\'a relacionado $P^0_S$ con la traza, de $h_{mn}^{(s)}$. 
As\'{\i} [2]

$$
R_{mn}{}^{ls}=\frac{1}{6}Q_{mn}Q^{ls},\tag B.14
$$
donde

$$
Q_{mn}\equiv \eta_{mn}-3\wh{\partial}_m\wh{\partial}_n.\tag B.15
$$
Nosotros tomamos (B.12) como la descomposici\'on de $S$. Es decir

$$
S=P^2_S+P^1_S+P^0_S+P^0_W.\tag B.16
$$

La descomposici\'on de $A$, se obtiene de forma an\'aloga. Tenemos que los 
proyectores en las partes de spin 1 y spin 0, son entonces 

$$\align
P^1_{E\ mn}{}^{ls} &=\frac{1}{2}(P_m{}^l\omega_n{}^s-P_m{}^s\omega_n{}^l-
P_n{}^l\omega_m{}^s+P_n{}^s\omega_m{}^l),\tag B.17,a\\
P^0_{B\ mn}{}^{ls} &=\frac{1}{2}(P_m{}^lP_n{}^s-P_m{}^sP_n{}^l).\tag B.17,b
\endalign
$$
En particular $P_B^0$ puede reescribirse como

$$
P^0_{B\ mn}{}^{ls} =-\frac{1}{2}\xi_{mn}\xi^{ls}.\tag B.17,a
$$
concluimos, entonces, que la unidad se descompone como

$$\align
1 &=S+A\\
&=(P^2_S+P^1_S+P^0_S+P_W^0)+(P^1_E+P_B^0).\tag B.18
\endalign
$$

La proyecci\'on en cada una de las partes de $h_{mn}$ puede, todav\'{\i}a, 
particularizarse mas. Esto se debe a que cada componente de spin distinto de 
0 tiene dos helicidades posibles. Adem\'as, si tomamos la ``traza'' de 
$P^2_S$, $P^1_{S}$, $P^1_E$ (i.e. traza de $P_{mn}{}^{ls}$ es 
$P_{mn}{}^{mn}$), estas dan 2, lo que corresponde a la ``dimensi\'on'' del 
subespacio en que se proyecta. En particular para la parte sim\'etrica, 
transversa y sin traza ( o de spin 2) reconocemos sus dos partes [3]

$$
h_{mn}^{(\pm )Tt}=\frac{1}{2}[h_{mn}^{Tt}\pm\frac{1}{2}(\xi_m{}^r\delta_n{}^s+
\xi_n{}^s\delta_m{}^r)h_{rs}^{Tt}].\tag B.19
$$
As\'{\i}, tenemos que 

$$
P^2_{\pm Smn}{}^{ls}=\frac{1}{4}[P_{\pm m}{}^lP_n{}^s+P_{\pm m}{}^s
P_n{}^l+P_{\pm n}{}^lP_m{}^s+P_{\pm n}{}^sP_m{}^l-P^{mn}P_{ls}],\tag B.20
$$
y ya que $P_m{}^n=P_{+m}{}^n+P_{-m}{}^n$, es f\'acil convencerse que 

$$\align
P^1_{\pm Smn}{}^{ls} &=\frac{1}{2}(P_{\pm m}{}^l\omega_n{}^s+P_{\pm m}{}^s
\omega_n{}^l+P_{\pm n}{}^l\omega_m{}^s+P_{\pm n}{}^s\omega_m{}^l),
\tag B.21,a\\
P^1_{\pm Emn}{}^{ls} &=\frac{1}{2}(P_{\pm m}{}^l\omega_n{}^s-P_{\pm m}{}^s
\omega_n{}^l-P_{\pm n}{}^l\omega_m{}^s+P_{\pm n}{}^s\omega_m{}^l),
\tag B.21,b
\endalign
$$
Adem\'as

$$\align
P^2_S &=P^2_{+S}+P^2_{-S},\tag B.22,a\\
P^1_S &=P^1_{+S}+P^1_{-S},\tag B.22,b\\
P^1_E &=P^1_{+E}+P^1_{-E},\tag B.22,c
\endalign
$$
As\'{\i} en dimensi\'on $D=2+1$ tenemos un proyector para cada una de las 
componentes irreducibles de $h_{mn}$. Cada uno de estos proyectores es 
ortogonal al otro, como se mostrar\'a mas adelante.

A pesar de la completitud de estos proyectores, antes cons\-truidos, no es 
posible cons\-truir con ellos a todo operador diferencial que act\'ue sobre un 
objeto de 2 \'{\i}ndices. Existen otros operadores, u operadores de 
transferencia, que mandan una parte de spin y helicidad definida en otra de 
igual spin y helicidad. Estos son $P^0_{SW},P^0_{WS},P^0_{BW},P^0_{WB},
P^0_{SB},P^0_{BS},P^1_{ES}$ y $P^1_{SE}$, donde la letra sub\'{\i}ndice 
indica primero el sector de llegada y segundo el de partida. Pedimos que 
adem\'as cumplan $P^0_WP^0_{WS}=P^0_{WS}$, $P^0_{WS}P^0_{SW}=P^0_S$, 
$P^0_{SW}P^0_{WS}=P^0_S$, $P^1_{SE}P^1_{ES}=P^1_S$ etc.. 
Obteniendo

$$\align
P^0_{SWmn}{}^{ls} &=\frac{1}{\sqrt{2}}P_{mn}\omega^{ls},\tag B.23,a\\
P^0_{WSmn}{}^{ls} &=\frac{1}{\sqrt{2}}\omega_{mn}P^{ls},\tag B.23,b
\endalign
$$

$$\align
P^0_{SBmn}{}^{ls} &=\frac{1}{2}P_{mn}\xi^{ls},\tag B.23,c\\
P^0_{BSmn}{}^{ls} &=-\frac{1}{2}\xi_{mn}P^{ls},\tag B.23,d\\
P^0_{WBmn}{}^{ls} &=\frac{1}{\sqrt{2}}\omega_{mn}\xi^{ls},\tag B.23,e\\
P^0_{BWmn}{}^{ls} &=-\frac{1}{\sqrt{2}}\xi_{mn}\omega^{ls},\tag B.23,f\\
P^1_{ESmn}{}^{ls} &=\frac{1}{2}(P_m{}^l\omega_n{}^s+P_m{}^s\omega_n{}^l-
P_m{}^l\omega_n{}^s-P_m{}^s\omega_n{}^l),\tag B.23,g\\
P^1_{SEmn}{}^{ls} &=\frac{1}{2}(P_m{}^l\omega_n{}^s-P_m{}^s\omega_n{}^l+
P_n{}^l\omega_m{}^s-P_n{}^s\omega_m{}^l),\tag B.23,h
\endalign
$$
Adem\'as $P^1_{SE}$ y $P^1_{ES}$ tienen sus respectivas versiones 
$P^1_{\pm SE}$ y $P^1_{\pm ES}$, con $P^1_{SE}=P^1_{+SE}+P^1_{-SE}$ y 
$P^1_{ES}=P^1_{+ES}+P^1_{-ES}$, donde en vez de $P_m{}^n$ ponemos 
$P_{\pm m}{}^n$. Para los sectores, de spin 2 (partes + y -) no hay, por 
definici\'on, operadores de transferencia.

Si llamamos $P_I^\alpha$ a los proyectores, con $\alpha =0,1,2$ e 
$I = S, W, E$ o $B$, seg\'un el caso, y $P_{IJ}^\alpha$ a los operadores de 
transferencia, tendremos que se verifica el siguiente \'algebra 

$$\align
P_I^\alpha P_J^\beta &=\delta^{\alpha\beta}\delta_{IJ}P_J^\beta ,\tag B.24,a\\
P_I^\alpha P_{JK}^\beta &=\delta^{\alpha\beta}\delta_{IJ}P_{IK}^\beta 
,\tag B.24,b\\
P_{IJ}^\alpha P_K^\beta &=\delta^{\alpha\beta}\delta_{JK}P_{IK}^\beta 
,\tag B.24,c\\
P_{IJ}^\alpha P_{KL}^\beta &=\delta^{\alpha\beta}\delta_{JK}P_{IL}^\beta 
\ \ si\ \ I\neq L,\tag B.24,d\\
P_{IJ}^\alpha P_{KI}^\beta &=\delta^{\alpha\beta}\delta_{JK}P_I^\beta 
,\tag B.24,e
\endalign
$$
donde no se aplica la convenci\'on de suma de Einstein. Para los proyectores 
y operadores con versiones + y - estas reglas se cumplen entre proyectores u 
operadores de igual ``helicidad". 
\vskip 5mm

\noi
{\bf El proyector transverso respecto al 1$^{er}$ \'{\i}ndice}. 
\vskip 3mm

Para el c\'alculo de algunos propagadores necesitamos escoger el calibre 
$\wh{\partial}^mh_{mn}=0$. Esto es equivalente a pedir que 

$$
\omega^{ml}\eta^{ns}h_{ls}=0.\tag B.25
$$
Resulta que 

$$
\omega_m{}^l\delta_n{}^s=(P^0_W+\frac{1}{2}(P^1_S+P^1_E-P^1_{SE}-
P^1_{ES}))_{mn}{}^{ls}.
\tag B.26
$$

Como $h=(P^2_S+P^1_S+P^0_S+P^0_W+P^1_E+P^0_B)h$, si definimos 
$h^T\equiv \{h_{mn}\ t.q.\ \partial^mh_{mn}=0\}$, tendremos que 

$$\align
h^T &=(P^2_S+P^0_S+P^0_B+\frac{1}{2}(P^1_{SE}+P^1_{ES}+P^1_S+P^1_E))h\\
& \equiv T_{mn}{}^{ls}h_{ls},\tag B.27
\endalign
$$
y $T_{mn}{}^{ls}$ es un proyector, con las caracter\'{\i}sticas exigidas 

$$\align
T_{mn}{}^{ls} &=P_m{}^lP_n{}^s+P_m{}^l\omega_n{}^s,\tag B.28,a\\
T_{mn}{}^{ls}T_{ls}{}^{pr} &=T_{mn}{}^{pr}\ \ ;\ \ 
\wh{\partial}^mT_{mn}{}^{ls}=0.
\tag B.28,b
\endalign
$$

Si $h_{mn}$ fuera adem\'as sim\'etrico, el que sea transverso respecto a un 
\'{\i}ndice implicar\'{\i}a, que lo es respecto al otro. La construcci\'on 
del proyector es mas sencilla y resulta ser 

$$\align
T^{(S)} &=P_S^2+P_S^0 \tag B.29,a\\
T^{(S)}_{mn}{}^{ls} &=\frac{1}{2}(P_m{}^lP_n{}^s+P_m{}^sP_n{}^l).\tag B.29,b
\endalign
$$
Observamos que $tr(T)=6$ y $tr(T^{(s)})=3$, como debe ser.
\vskip 1cm
\item{}{\bf Referencias}
\item{[1]}K. J. Barnes, J. Math. Phys. {\bf 6} (1965) 788; P. van
Nieuwenhuizen,  Nuc. Phys. {\bf B60} (1973) 478.
\item{[2]}R. J. Rivers, Nuov. Cim. {\bf 24} (1964) 386.
\item{[3]}C. Aragone y A. Khoudeir, Phys. Lett. {\bf B173} (1986) 141.

\bye